\documentclass[11pt,a4paper]{article}
\usepackage{float}
\usepackage{jheppub}
\usepackage[T1]{fontenc}
\usepackage{lmodern,microtype,booktabs,mathtools,bm}
\usepackage{tikz}
\usepackage{placeins,needspace}
\usetikzlibrary{arrows.meta,positioning,calc}
\providecommand{\quivercircleloop}[5][particle loop]{%
  \begingroup
  \pgfmathsetmacro{\quiverLoopOffset}{#3+#5}%
  \pgfmathsetmacro{\quiverLoopStart}{#4+150}%
  \draw[#1]
    ($(#2)+(#4:\quiverLoopOffset)+(\quiverLoopStart:#3)$)
    arc[start angle=\quiverLoopStart,delta angle=-300,radius=#3];%
  \endgroup
}

\usepackage{etoolbox}
\usepackage{todonotes}
\makeatletter
\patchcmd{\maketitle}{\PackageWarningNoLine{\jname}{E-mails are missing.\MessageBreak Plese use \protect\emailAdd\space macro to provide e-mails.}}{}{}{}
\makeatother
\hypersetup{pdftitle={Reconstructing kink and particle quivers on a defect circle},pdfauthor={},
  colorlinks=true,linkcolor=blue!45!black,citecolor=blue!45!black,urlcolor=blue!45!black}
\newcommand{\C}{\mathcal C}
\newcommand{\Hc}{\mathcal H}
\newcommand{\V}{\mathcal V}

\newcommand{\ii}{i}
\newcommand{\ee}{\mathrm e}
\newcommand{\dd}{\mathrm d}
\newcommand{\Tr}{\operatorname{Tr}}

\newcommand{\Hom}{\operatorname{Hom}}

\newcommand{\ket}[1]{\left\lvert #1\right\rangle}
\newcommand{\bra}[1]{\left\langle #1\right\rvert}
\newcommand{\F}[6]{\bigl(F^{#1#2#3}_{#4}\bigr)_{#5#6}}
\newcommand{\qnum}[1]{\left(#1\right)_{\zeta}}
\newcommand{\sixj}[6]{\begin{Bmatrix}#1&#2&#3\\#4&#5&#6\end{Bmatrix}_{\!\zeta}}

\preprint{USTC-ICTS/PCFT-26-91}
\title{Reconstructing massive quivers on a defect circle}
\author{Ting-Kai Chen$^{a}$}
\author{and Jin Chen$^{a,\,b,\,\dagger}$}

\affiliation{$^{a}$Department of Physics, Xiamen University, Xiamen, 361005, China}
\affiliation{$^{b}$Peng Huanwu Center for Fundamental Theory, Hefei, Anhui 230026, China}

\emailAdd{elysianchan13@gmail.com}
\emailAdd{zenofox@gmail.com}

\abstract{
We study the kink and particle spectra for the massive $\phi_{2,1}$-deformations of minimal models $\mathcal M_{k+1,k+2}$ via truncated conformal space approach (TCSA) on a defect circle. The massive $\phi_{2,1}$-deformation preserves a set of non-invertible symmetries, furnishing a fusion category $\mathcal C_k\simeq {\rm SO}(3)_k$. These non-invertible symmetries are spontaneously broken, and thus label the vacua and organize the kinks and particles into their multiplets in terms of quiver diagrams. We employ a finite-volume truncated Hamiltonian approach on defect Hilbert spaces, $\mathcal H_{X_i}$ for $X_i\in\mathcal C_k$, to study these quivers in the IR theories. The symmetry defect can then close an individual kink on a circle.
We first establish the relation between states of definite fractional momentum on the defect circle and physical kink and particle states labelled by their vacuum pairs.
Using form factors of local fields, we independently measure the squared amplitudes between the states of the two bases in defect TCSA.
Combining these measurements with defect TCSA spectra, we reconstruct the quiver diagrams for $k=3,4,5$ without assuming them in advance.
In particular, for $k=4$ we reconstruct the distinct quivers of the ordered and disordered phases. For $k=5$, we identify the quivers of both the light and heavy multiplets.
}
\begin{document}
\begin{NoHyper}
\note[$\dagger$]{Corresponding author.}
\end{NoHyper}
\maketitle
\section{Introduction and summary}
\label{sec:intro}

Symmetry has been always playing a central role in study of quantum field theories, as it can provide many valuable non-perturbative constraints even when the dynamics of the systems are strongly coupled. It organizes states in a multiplet of the same energy, constrains correlation functions and scattering amplitudes, and characterizes the different phases of the theory through its realization on the vacuum. In recent years, this organizing principle has been extended by describing symmetries in terms of topological operators\cite{Gaiotto:2014kfa,McGreevy:2022oyu,Shao:2023gho,Luo:2023ive}. Ordinary internal symmetries are represented by invertible topological defects, whose composition follows the group law. More generally, the fusion of two topological defects can give a sum of defects~\cite{Frohlich:2004ef,Choi:2021kmx,Bhardwaj:2022yxj,Kaidi:2022cpf}. Such non-invertible symmetries provide additional constraints on quantum field theories and their renormalization group flows~\cite{CLSWY,Kaidi:2023maf}.

In two dimensions, the symmetry defects are topological lines, known as topological defect lines (TDLs), commuting with the Hamiltonian, and their fusion rules and junctions form a fusion category~\cite{Bhardwaj:2017xup,CLSWY}. When the theory is gapped and the symmetry is spontaneously broken, these lines act on the degenerate vacua and on the excitations interpolating between them~\cite{Thorngren:2019iar,Bhardwaj:2023idu}. This action can enforce mass degeneracies between particles and solitons \cite{CGH,Chen:2025qub,Chen:2026wyn}, which are organized into multiplets of a strip algebra~\cite{CHO}. The physical content of such a representation is conveniently displayed by a quiver: its vertices label the vacua, an oriented edge between distinct vertices describes a kink, and a loop describes a particle above a fixed vacuum. The massive $\phi_{2,1}$ deformations of unitary Virasoro minimal models provide a concrete setting in which to study these multiplets~\cite{LMC,Colomo:1991gw,Ellem:1997vz,MM,MussardoTakacs2009}. In this note, we consider the $\phi_{2,1}$ deformations of $\mathcal M_{k+1,k+2}$, focusing on $k=3,4,5$, for which the preserved symmetry category is given by $\C=\mathrm{SO}(3)_k$. The quiver representations of the associated strip algebra $\mathbf{Str}(\mathcal C)$ specify the possible kink and particle multiplets. However, a given fusion category can admit several such representations, whereas the masses and the representations actually realized by stable excitations are model dependent, and are determined solely by the infrared dynamics. This raises two questions we will address in this paper: Which quiver representations are realized in the low energy spectrum of each massive flow? Can we reconstruct these quivers directly from finite volume data without specifying their arrows \emph{a priori}?

We investigate these questions using the truncated conformal space approach (TCSA), which starts from conformal data in the ultraviolet and diagonalizes the perturbed Hamiltonian in a truncated Hilbert space~\cite{YZ,Yurov:1991my,LMC,Martins:1991jj,Konik:2007cb,James:2017cpc,CFTCSA}. To resolve the kink components, the choice of Hilbert space is essential~\cite{Dorey:1997yg,Feverati:1998va}. A kink approaches different vacua at the two ends of the spatial line, whereas periodic boundary conditions require the vacuum to return to itself around the circle. Consequently, the periodic spectrum contains neutral particles and combinations such as a kink and an antikink~\cite{Lepori:2009ip,Lencses:2014tba}, while an individual kink, however, must be accessed in another sector connecting distinct boundary conditions~\cite{Mong}. Earlier TCSA studies have resolved kink spectra using $\mathbb Z_2$ twisted boundary conditions~\cite{LMC} and non-diagonal sectors of the $\mathbb Z_3$ parafermion theory~\cite{Mong}. A more general TCSA approach on defect Hilbert spaces twisted by a non-invertible TDL has been introduced in~\cite{DLY} to study both flows between conformal defects~\cite{Kormos:2009sk} and finite volume spectra under combined bulk and defect perturbations in the Lee-Yang model. 

\begin{figure}[t]
\centering
\input{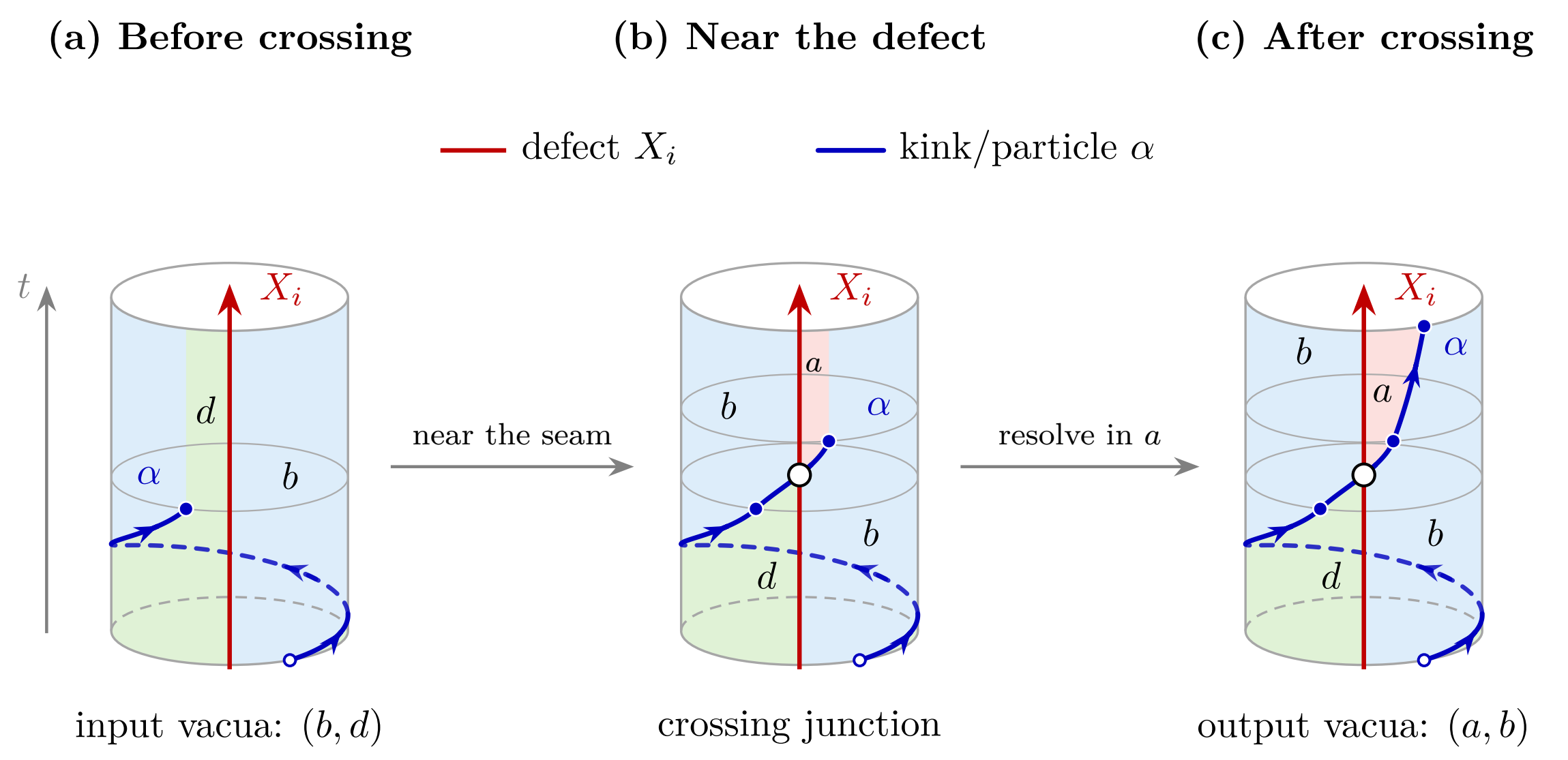}%

\caption{Successive stages of a kink/particle worldline on a spacetime
cylinder. Red denotes the symmetry defect $X_i$ and blue a kink/particle worldline in the multiplet labelled by $\alpha$.}
\label{fig:recoupling}
\end{figure}

In this paper, we further generalize the defect TCSA approach to the massive deformed CFTs with multiple vacua from spontaneously breaking fusion categories.
We  construct Hilbert spaces twisted by spontaneously broken TDLs preserved under the $\phi_{2,1}$-deformations in minimal models, and numerically study the kink and particle spectra therein. More concretely, to study individual kink states on a circle, we formulate TCSA on the defect Hilbert spaces $\Hc_{X_i}$ associated with the preserved topological lines $X_i\in\C$. A line inserted along the time direction supplies a junction at which the two vacuum regions of a single kink can be joined, so that the kink becomes a state on such a defect circle. Such defect Hilbert space $\mathcal H_{X_i}$, after reaching deep IR, will contain all allowable kink or particle states $|ab\rangle$ so long as the two vacua $a$ and $b$ can be connected by the defect $X_i$-line. This construction therefore makes it approachable to measure the kink and particles in a single non-invertible $X_i$-multiplet by default. Nevertheless, the states directly measured in TCSA are eigenstates of energy and fractional momentum. We therefore establish the relation between the eigenstates of momentum on such a circle and the physical kink and particle components of the possible massive quivers predicted by the strip algebra $\mathbf {Str}(\mathcal C)$: The momentum eigenstates have definite fractional momentum, while a physical kink or particle on the line has definite vacua on its two sides. We establish the transformation between these two bases from the associativity data of the preserved $X_1$-line and the corresponding vacua. When an kink or a particle state is transported once around the defect circle through the closing junction, the $F$-symbol associated with the $X_1$-line will relate two kink or particle states with different allowable vacua labels. For the $X_1$ multiplet on a circle closed by the $X_1$ line in the regular vacuum module, the transport matrix is given by
\begin{align}
	T^{X_1}_{(a,b),(c,d)}=\delta_{bc}\F{1}{b}{1}{b}{a}{d}\,,
	\label{eq:introT}
\end{align}
where $(a,b)$ and $(c,d)$ denote the outgoing and incoming
vacuum pairs, and the label $1$ in the $F$-symbol denotes $X_1$,
as depicted in Figure~\ref{fig:recoupling}. On the other hand, the eigenvalues $\ee^{2\pi\ii q}$ from TCSA measurements determine the fractional momentum shifts $q$, giving $p=\frac{2\pi(n+q)}{L}$ on a circle of circumference $L$, with $n\in\mathbb Z$. We call the transport eigenchannels with definite fractional momentum shifts the \emph{momentum basis}, and the physical kink and particle components closed through the defect with fixed adjacent vacua the \emph{circle basis}. The unitary matrix $U$ relates these two internal bases. In particular, this construction resolves the different vacuum compositions of states that have the same fractional momentum. We determine this composition numerically using local probes that distinguish the vacua on the two sides of the defect. Their TCSA matrix elements allow us to extract the relative squared amplitudes of the kink and neutral particle components.

To reconstruct the quiver of the multiplet, we take the vacua identified on the periodic circle as its vertices. We then fix a defect $X_i$ and measure the probability of each ordered vacuum pair $(a,b)$. If $X_i$ admits a single junction between $a$ and $b$, each physical component with these vacua contributes one state on the defect circle. We further sum the measured probability of finding $(a,b)$ over a complete momentum basis. A sum close to zero indicates that there is no arrow from $a$ to $b$, whereas a sum close to one gives one arrow, which is a kink from $a$ to $b$ if the vacua differ and a particle above $a$ if they coincide. Finally the number of particles at the same mass in the periodic spectrum provides an independent check on the total number of loops.

In every phase studied here we find a connected quiver realizing $R_1$, the multiplet labelled by the spin one generator $X_1=\mathcal L_{(1,3)}$ of $\mathrm{SO}(3)_k$.
For $k=3$, this multiplet consists of a kink, its antikink and a neutral particle, all with the same mass.
For $k=4$, the perturbing operator is $\mathbb Z_2$ even, and the two signs of the coupling lead to an ordered phase at $\lambda<0$ and a disordered phase at $\lambda>0$.
In the ordered phase, four kinks connect three vacua and the middle vacuum supports a particle. In the disordered phase, a kink and its antikink connect two vacua, each of which supports a particle.
These two realizations of $R_1$ are representations of the strip algebras associated with the respective vacuum modules.
For $k=5$, the light quiver contains four kinks and two particle loops, one at vacuum $1$ and one at vacuum $2$.
We also identify a heavier $R_0$ multiplet of three neutral particles of equal mass, one above each vacuum.
These results are consistent with the exact scattering theory of the corresponding integrable models~\cite{Klassen:1992qy,DoreyPocklingtonTateo2003,DoreyPocklingtonTateo2002II}.
In particular, the heavier particles at $k=5$ are bound states of the light excitations, with mass $m_H=2m_L\cos\left(\frac{\pi}{30}\right)$, where $m_L$ is the light multiplet mass.
Table~\ref{tab:spectrum-summary} summarizes the quivers and their kink and particle content. In a forthcoming work, we further extend this approach to the massive
$\phi_{2,1}$-deformations of exceptional minimal models with
$E_6$, $E_7$ and $E_8$ modular invariants~\cite{Chen:ExceptionalMassiveQuivers}.

\begin{table}[htbp]
\centering
\begingroup
\providecommand{\quivercircleloop}[5][particle loop]{%
  \begingroup
  \pgfmathsetmacro{\quiverLoopOffset}{#3+#5}%
  \pgfmathsetmacro{\quiverLoopStart}{#4+150}%
  \draw[#1]
    ($(#2)+(#4:\quiverLoopOffset)+(\quiverLoopStart:#3)$)
    arc[start angle=\quiverLoopStart,delta angle=-300,radius=#3];%
  \endgroup
}
\small
\setlength{\tabcolsep}{5pt}
\renewcommand{\arraystretch}{1.12}
\tikzset{
  spectrum vertex/.style={circle,fill=black,inner sep=1.35pt},
  spectrum arrow/.style={draw=cyan!75!black,
    -{Stealth[length=1.45mm,width=1.05mm]},line width=.60pt}
}
\newcommand{\spectrumquiver}[3]{%
  \begin{tikzpicture}[x=.80cm,y=.80cm,
    baseline={([yshift=-.5ex]current bounding box.center)}]
    \ifnum#1=2
      \path[use as bounding box] (-.28,-.28) rectangle (2.60,.28);
      \node[spectrum vertex] (a) at (.5,0) {};
      \node[spectrum vertex] (b) at (1.5,0) {};
    \else
      \path[use as bounding box] (-.28,-.24) rectangle (2.60,.74);
      \node[spectrum vertex] (a) at (0,0) {};
      \node[spectrum vertex] (b) at (1,0) {};
      \node[spectrum vertex] (c) at (2,0) {};
    \fi
    \ifnum#2=1
      \draw[spectrum arrow] ($(a)+(.13,.04)$)
        .. controls ($(a)+(.30,.22)$) and ($(b)+(-.30,.22)$)
        .. ($(b)+(-.13,.04)$);
      \draw[spectrum arrow] ($(b)+(-.13,-.04)$)
        .. controls ($(b)+(-.30,-.22)$) and ($(a)+(.30,-.22)$)
        .. ($(a)+(.13,-.04)$);
      \ifnum#1=3
        \draw[spectrum arrow] ($(b)+(.13,.04)$)
          .. controls ($(b)+(.30,.22)$) and ($(c)+(-.30,.22)$)
          .. ($(c)+(-.13,.04)$);
        \draw[spectrum arrow] ($(c)+(-.13,-.04)$)
          .. controls ($(c)+(-.30,-.22)$) and ($(b)+(.30,-.22)$)
          .. ($(b)+(.13,-.04)$);
      \fi
    \fi
    \foreach \vertex/\direction in {#3}{%
      \def\spectrumLoopClearance{.08}%
      \ifnum#2=1\relax
        \ifnum\direction=90\relax
          \def\spectrumLoopClearance{.18}%
        \fi
      \fi
      \quivercircleloop[spectrum arrow]{\vertex}{.23}{\direction}{\spectrumLoopClearance}%
    }%
  \end{tikzpicture}%
}
\begin{tabular*}{\textwidth}{@{\extracolsep{\fill}}ccccc@{}}
\toprule
$k$ & Preserved fusion category & Phase & Quiver &
\shortstack{Kinks, particles\\$(n_K,n_P)$} \\
\midrule
$3$ & $\mathrm{SO}(3)_3\simeq\mathrm{Fib}$ & --- &
$R_1\!:$\quad\spectrumquiver{2}{1}{b/0} & $(2,1)$ \\
\midrule
$4$ & $\mathrm{SO}(3)_4\simeq\mathrm{Rep}(S_3)$ &
\shortstack{Ordered\\$\lambda<0$} &
$R_1\!:$\quad\spectrumquiver{3}{1}{b/90} & $(4,1)$ \\
\addlinespace[2pt]
 & & \shortstack{Disordered\\$\lambda>0$} &
$R_1\!:$\quad\spectrumquiver{2}{1}{a/180,b/0} & $(2,2)$ \\
\midrule
$5$ & $\mathrm{SO}(3)_5$ & --- &
$R_1\!:$\quad\spectrumquiver{3}{1}{b/90,c/0} & $(4,2)$ \\
\addlinespace[2pt]
 & & &
$R_0\!:$\quad\spectrumquiver{3}{0}{a/90,b/90,c/90} & $(0,3)$ \\
\bottomrule
\end{tabular*}
\endgroup
\caption{Massive multiplets in $\mathcal M_{k+1,k+2}+\lambda\,\phi_{2,1}$.
Vertices denote vacua, directed edges between distinct vertices denote kinks
and antikinks, and loops denote particles. The counts $(n_K,n_P)$ include
both kink orientations. For $k=5$, $R_1$ is the light multiplet and
$R_0$ is the heavy multiplet.}
\label{tab:spectrum-summary}
\end{table}

The paper is organized as follows: In Section~\ref{sec:phase}, we determine the symmetry preserved by the deformation $\phi_{2,1}$ and describe its action on the massive vacua and quiver multiplets. Section~\ref{sec:circle} constructs the defect circle, derives its transport matrix from $F$-moves, and relates the resulting fractional momenta to the conformal defect Hilbert spaces. Section~\ref{sec:numerics} introduces the TCSA Hamiltonian and the local form factor measurements. In Section~\ref{sec:quiver_reconstruction}, we reconstruct the quivers from the spectra and measured vacuum pair probabilities for $\mathcal M_{4,5}$, both massive phases of $\mathcal M_{5,6}$, and $\mathcal M_{6,7}$. Appendix~\ref{app:recoupling} derives the $F$-symbols from WZW conformal blocks, and Appendix~\ref{app:dictionary} gives the transformation between the momentum and circle bases. Appendix~\ref{app:numerics} gives the numerical conventions and compares the masses of the quiver components. In Appendix~\ref{app:sm}, we compare our TCSA results with the exact spectra from the integrable scattering matrix.

\section{Minimal models, preserved symmetry and massive quivers}
\label{sec:phase}

\subsection{The deformation and the preserved topological lines}

We consider the diagonal unitary minimal models $\mathcal M_{k+1,k+2}$,
with $k\geq3$, and focus on $k=3,4,5$. A primary field
$\phi_{r,s}$ has its conformal weight given by
\begin{align}
 c=1-\frac{6}{(k+1)(k+2)}\,,\qquad
 h_{r,s}=\bar h_{r,s}
 =\frac{((k+2)r-(k+1)s)^2-1}{4(k+1)(k+2)}\,,
 \label{eq:kac}
\end{align}
where the Kac labels obey $1\leq r\leq k$ and $1\leq s\leq k+1$,
with the identification $(r,s)\sim(k+1-r,k+2-s)$.
We normalize the real perturbing field by
$\langle\phi_{2,1}(z,\bar z)\phi_{2,1}(0)\rangle=|z|^{-4h_{2,1}}$
and write the deformed action as
\begin{align}
 S=S_{\mathcal M_{k+1,k+2}}+\lambda\int\dd^2x\,\phi_{2,1}(x)\,,\quad
 {\rm with}\quad
 h_{2,1}=\frac{k+4}{4(k+1)}\,,\ 
 y\equiv2-2h_{2,1}=\frac{3k}{2(k+1)}\,,
 \label{eq:action}
\end{align}
where $y$ is the mass dimension of the coupling $\lambda$, so the masses in a
gapped phase scale as $|\lambda|^{\frac{1}{y}}$. For $k=3,5$, since $\phi_{2,1}$ is odd under the $\mathbb Z_2$ symmetry of $\mathcal M_{k+1,k+2}$, the two signs
of $\lambda$ are related by the $\mathbb Z_2$ symmetry. For $k=4$, on the other hand, $\phi_{2,1}$ is $\mathbb Z_2$ even and the different signs of $\lambda$ thus lead to distinct massive phases, known as ordered phase for $\lambda<0$ and disordered one for $\lambda>0$~\cite{MussardoTakacs2009}.

The symmetry preserved by the $\phi_{2,1}$ perturbation can be determined at the
conformal point. In a diagonal rational CFT, a simple topological line
$\mathcal L$ acts on a bulk conformal family $i$ with charge $q^{\mathcal L}_i=\frac{S_{\mathcal L,i}}{S_{\mathbf v ,i}}$, where $S$ is the modular matrix and $\mathbf v\equiv(1,1)$ denotes the identity family~\cite{Petkova:2000ip}.
The line is preserved when its charge on the perturbing field
equals the vacuum expectation of the line itself, $d_\mathcal L=\frac{S_{\mathcal L,\mathbf v}}{S_{\mathbf v, \mathbf v}}$~\cite{CLSWY}. The modular matrix of the minimal model $\mathcal M_{k+1,k+2}$ is given by,
\begin{align}
 S_{(r,s),(r',s')}
 &=2\sqrt{\frac{2}{(k+1)(k+2)}}(-1)^{1+sr'+rs'}
 \sin\frac{\pi(k+2)rr'}{k+1}
 \sin\frac{\pi(k+1)ss'}{k+2}\,.
 \label{eq:S}
\end{align}
For the perturbation by $\phi_{2,1}$ and a TDL $\mathcal L_{(r,s)}$, the ratio of the corresponding charge and the vacuum expectation is then
\begin{align}
 \frac{q_{(2,1)}^{\mathcal L_{(r,s)}}}{d_{\mathcal L_{(r,s)}}}
 =\frac{S_{(r,s),(2,1)}S_{\mathbf v,\mathbf v}}{S_{\mathbf v,(2,1)}S_{(r,s),\mathbf v}}
 =(-1)^{r+s}\frac{\cos\left(\frac{\pi r}{k+1}\right)}{\cos\left(\frac{\pi}{k+1}\right)}\,.
 \label{eq:preservation}
\end{align}
It equals one for $r=1$ and odd $s$. The preserved simple lines are therefore given by
\begin{align}
 X_i\equiv\mathcal L_{(1,2i+1)}\,,\qquad
 i=0,1,\ldots,J\,,\qquad J=\left\lfloor\frac{k}{2}\right\rfloor\,,
 \label{eq:Clabels}
\end{align}
where $X_0$ is the identity line. We use $i,j,\ell$ for line labels
and $a,b,c,d$ for vacuum labels, reserving $k$ for the model parameter.
Their fusion rules are
\begin{align}
 X_i\otimes X_j
 =\bigoplus_{\ell=|i-j|}^{\min(i+j,k-i-j)}X_\ell\,.
 \label{eq:fusion}
\end{align}
These are the integer spin fusion rules of ${\rm SU}(2)_k$, giving the preserved fusion category $\mathcal C_k\simeq\mathrm{SO}(3)_k$. The quantum dimensions and the corresponding chiral weights and twists are
\begin{align}
 d_j&=\frac{\sin\left(\frac{(2j+1)\pi}{k+2}\right)}{\sin\left(\frac{\pi}{k+2}\right)}\,,
 \qquad
 h_j=j^2-\frac{j(j+1)}{k+2}\,,\qquad
 \theta_j=\ee^{-\frac{2\pi\ii j(j+1)}{k+2}}\,.
 \label{eq:dimensions}
\end{align}
The line $X_1=\mathcal L_{(1,3)}$ has spin label one and generates
$\mathcal C_k$ under fusion. We use $\mathcal C$ for this preserved
category when $k$ is fixed.

For $k=3$, the deformation is the subleading magnetic perturbation of
the tricritical Ising model $\mathcal M_{4,5}$, with
$h_{2,1}=\frac{7}{16}$~\cite{LMC,Colomo:1992nr,Ellem:1997vz,MM,Lencses:2021igo,CGH}.
The preserved category is $\mathcal C_3\simeq\mathrm{Fib}$, with simple
lines $X_0$ and $X_1$. They obey
$X_1\otimes X_1=X_0\oplus X_1$, with $d_1=\frac{1+\sqrt5}{2}$.

For $k=4$, the model is $\mathcal M_{5,6}$ and
$h_{2,1}=\frac{2}{5}$. The preserved category is
$\mathcal C_4\simeq\operatorname{Rep}(S_3)$.
Its simple lines are $X_0,X_1,X_2$, with fusion rules
\begin{align}
 X_1\otimes X_1=X_0\oplus X_1\oplus X_2\,,\qquad
 X_1\otimes X_2=X_1\,,\qquad  X_2\otimes X_2=X_0\,.
 \label{eq:S3fusion}
\end{align}
The quantum dimensions are $d_1=2$ and $d_2=1$.
Both signs of the perturbing coupling preserve this category.
Their different realizations on the massive vacua will be described below.

For $k=5$, the model is $\mathcal M_{6,7}$ and
$h_{2,1}=\frac{3}{8}$. The category $\mathcal C_5\simeq\mathrm{SO}(3)_5$
has three simple lines $X_0$, $X_1$ and $X_2$, with
\begin{align}
 X_1\otimes X_1=X_0\oplus X_1\oplus X_2\,,\qquad
 X_1\otimes X_2=X_1\oplus X_2\,,\qquad
 X_2\otimes X_2=X_0\oplus X_1\,.
 \label{eq:M67fusion}
\end{align}
Their quantum dimensions are $d_0=1$,
$d_1=1+2\cos\left(\frac{2\pi}{7}\right)$ and
$d_2=2\cos\left(\frac{\pi}{7}\right)$.

\subsection{Vacuum modules and kink/particle multiplets}

The realization of the preserved symmetry $\mathcal C_k$ in a massive phase of the deformed theory is specified by its action on the vacua, which naturally furnish a $\mathcal C$-module category of $\mathcal C_k$ denoted by $\mathcal M_{\mathrm{vac}}$~\cite{Ostrik:2001xnt,Huang:2021zvu}. The simple objects $\{v_a,v_b,\dots\}$ in $\mathcal M_{\rm vac}$ are the vacua. A symmetry line $X_i\in\mathcal C$ acts from the left, with multiplicities
\begin{align}
 (n_i)_{a b}
 =\dim\Hom_{\mathcal M_{\mathrm{vac}}}(X_i\mathbin{\triangleright}v_a,v_b)\,,
 \qquad n_i n_j=\sum_\ell N_{ij}^{\ell}\,n_\ell\,,
 \label{eq:nimrep}
\end{align}
where $N_{ij}^{\ell}$ is the multiplicity of $X_\ell$ in $X_i\otimes X_j$, as in Eq.~\eqref{eq:fusion}.
The non-negative integer matrices $n_i$ known as the NIM representations of $X_i$~\cite{Petkova:2001ag,Fuchs:2002cm}, together with the associativity
data for their action, describe how the symmetry lines join different
vacua. Clearly, the same fusion category can act on different vacuum modules.

\begin{figure}[!htbp]
\centering
\begin{tikzpicture}[x=1cm,y=1cm,
  boundary/.style={draw=black,line width=.85pt},
  symmetry/.style={draw=red!75!black,line width=1.10pt},
  line arrow/.style={symmetry,-{Stealth[length=2.0mm,width=1.5mm]}},
  junction/.style={circle,draw=black,fill=white,line width=.65pt,inner sep=1.65pt},
  every node/.style={font=\small}
]
\foreach \x in {0,7.1}{
  \fill[gray!12] (\x-.44,0) rectangle (\x,2.8);
  \fill[gray!12] (\x+2.6,0) rectangle (\x+3.04,2.8);
  \draw[boundary] (\x,0)--(\x,2.8);
  \draw[boundary] (\x+2.6,0)--(\x+2.6,2.8);
}
\draw[gray!70,-{Stealth[length=1.8mm,width=1.3mm]},line width=.7pt]
  (-1.04,.12)--(-1.04,2.67);
\node[gray!70,anchor=east] at (-1.13,2.67) {$t$};

\node at (1.3,3.15) {$\mathsf A_j\circ\mathsf A_i$};
\foreach \y/\index/\element in {.80/i/A,2.00/j/B}{
  \draw[symmetry] (0,\y)--(2.6,\y);
  \draw[line arrow] (.92,\y)--(1.62,\y);
  \node[red!75!black,anchor=south] at (1.3,\y+.08) {$X_\index$};
  \node[junction] at (0,\y) {};
  \node[junction] at (2.6,\y) {};
}
\foreach \y/\leftvac/\rightvac in {.36/a/b,1.40/c/d,2.44/e/f}{
  \node at (-.23,\y) {$\leftvac$};
  \node at (2.83,\y) {$\rightvac$};
}

\node[font=\large] at (5.37,1.47) {$\displaystyle\simeq\;\sum_{\ell}$};
\node[font=\small] at (5.37,.73) {$X_j\otimes X_i\longrightarrow X_\ell$};

\node at (8.4,3.15) {$\mathcal H_{ab}\longrightarrow\mathcal H_{ef}$};
\draw[symmetry] (7.1,1.4)--(9.7,1.4);
\draw[line arrow] (8.02,1.4)--(8.72,1.4);
\node[red!75!black,anchor=south] at (8.4,1.48) {$X_\ell$};
\node[junction] at (7.1,1.4) {};
\node[junction] at (9.7,1.4) {};
\foreach \y/\leftvac/\rightvac in {.68/a/b,2.12/e/f}{
  \node at (6.87,\y) {$\leftvac$};
  \node at (9.93,\y) {$\rightvac$};
}
\end{tikzpicture}
\caption{Multiplication of two strip algebra elements, with time running
upwards, following the diagrammatic construction in~\cite{CHO}.
The lower line $X_i$ maps the vacuum pair $(a,b)$ to $(c,d)$,
and the upper line $X_j$ maps $(c,d)$ to $(e,f)$.
Their product is a sum of strip elements with a fused line $X_\ell$
and compatible end junctions, mapping $(a,b)$ to $(e,f)$.
The symbol $\simeq$ suppresses the coefficients and the junction labels
in this sum. The gray regions lie outside the spatial strip, denoting the vacua.}
\label{fig:strip_product}
\end{figure}
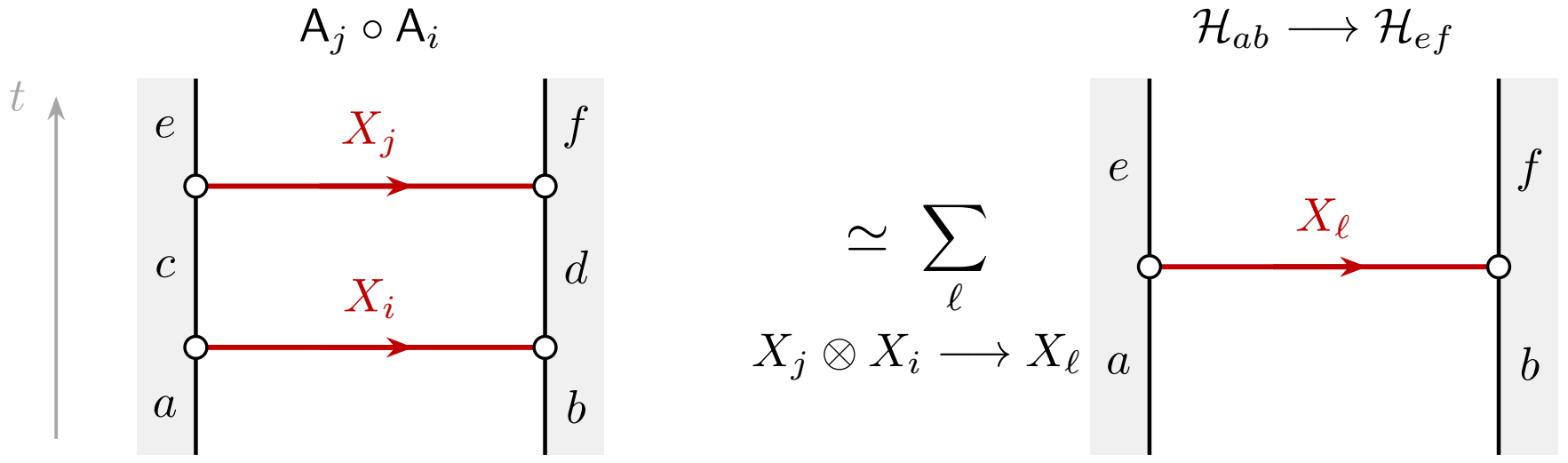

Given by the preserved fusion category $\mathcal C$ and vacua structure $\mathcal M_{\rm vac}$, particles and kinks of a massive theory can be organized by representations of the strip
algebra~\cite{CGH,CHO}, denoted by $\mathbf{Str}_{\mathcal C}(\mathcal M_{\mathrm{vac}})$. To describe its action, let $\mathcal H_{ab}$
be the Hilbert space on the infinite spatial line with vacuum $v_a$ at
$x\to-\infty$ and vacuum $v_b$ at $x\to+\infty$.
The full Hilbert space can be decomposed as
\begin{align}
    \mathcal H=\bigoplus_{a,b}\mathcal H_{ab}\,.
    \label{eq:decompH}
\end{align}
Within the sector of a single stable excitation, $a=b$ describes a
particle above vacuum $a$, while $a\ne b$ describes a kink from vacuum $a$ to $b$. A preserved symmetry line $X_i$ stretched across a spatial slice, with a topological
junction at each end, defines a linear map
$\mathsf A_i:\mathcal H_{ab}\to\mathcal H_{cd}$.
The end junctions specify how the line changes the left and right vacua.
These diagrams span the strip algebra
$\mathbf{Str}_{\mathcal C}(\mathcal M_{\mathrm{vac}})$.
Its multiplication is composition of maps. For
$\mathsf A_i:\mathcal H_{ab}\to\mathcal H_{cd}$ and
$\mathsf A_j:\mathcal H_{cd}\to\mathcal H_{ef}$, the product
$\mathsf A_j\circ\mathsf A_i$ is obtained by placing the diagram for
$\mathsf A_j$ above that for $\mathsf A_i$ in time. Fusing the two
symmetry lines and combining their end junctions gives a linear
combination of strip elements with the same initial and final vacua,
as shown in Figure~\ref{fig:strip_product}. The coefficients are fixed
by the fusion and module associativity data~\cite{CHO}.

Moving a topological line in time direction leaves correlation functions unchanged,
so its map intertwines the Hamiltonians,
\begin{align}
 H_{cd}\circ\mathsf A_i=\mathsf A_i\circ H_{ab}\,,
 \label{eq:strip_intertwiner}
\end{align}
where $H_{ab}$ is the Hamiltonian restricted to Hilbert space $\mathcal H_{ab}$.
As the degenerate vacua have the same bulk energy density, which we choose
as the zero of energy in every sector, the energy of localized
excitations measured relative to their own vacuum background can be compare across different sectors.
Eq.~\eqref{eq:strip_intertwiner} implies that the nonzero image
of an energy eigenstate has the same excitation energy.
These maps also preserve momentum and the space of stable particle
and kink states~\cite{CGH}. At zero momentum the excitation energy
is the rest mass, so all components of one irreducible stable
multiplet have the same mass.

The irreducible representations of a strip algebra are labelled by simple objects
$\rho_i$ of the dual category $\mathcal C^*_{\mathcal M_{\mathrm{vac}}}$.
An object $\rho_i$ assigns to each vacuum $v_a$ a sum of vacua
$v_a\mathbin{\triangleleft}\rho_i$, in a way compatible with the left
action of $\mathcal C$ and its junctions. This defines the right action
of the dual category on $\mathcal M_{\mathrm{vac}}$~\cite{CHO}.
We denote the corresponding representation by $R_i$
and its quiver matrix by $Q_i$, so that
\begin{align}
 R_i=\bigoplus_{a,b}
 \Hom_{\mathcal M_{\mathrm{vac}}}(v_a\mathbin{\triangleleft}\rho_i,v_b)\,,
 \qquad
 (Q_i)_{ab}
 =\dim\Hom_{\mathcal M_{\mathrm{vac}}}
       (v_a\mathbin{\triangleleft}\rho_i,v_b)\,.
 \label{eq:modulequiver}
\end{align}
The quiver has one node for each vacuum and $(Q_i)_{ab}$ directed
arrows from $a$ to $b$. An arrow between different nodes is a kink
component, while a loop at $a$ is a particle component above that vacuum.
The arrows of one such quiver therefore display the kink and particle
states of the multiplet $R_i$.
The allowed quiver representations and their internal mass degeneracies
are universal constraints shared by theories with the same preserved
category $\mathcal C$ and vacuum module $\mathcal M_{\mathrm{vac}}$.
Which of these multiplets occur as stable kink and particle excitations,
and what their masses are, depend on the dynamics of the particular model.
In the absence of a selection rule separating the vacua into disconnected
universe, see in~\cite{Hellerman:2006zs,Seiberg:2010qd,Tanizaki:2019rbk}, kink configurations can interpolate between any pair of vacua, possibly through a sequence of elementary kinks.
Within this massive kink/particle description, the quivers of the stable
multiplets must therefore connect all vacuum nodes when taken together.
For the vacuum modules considered below, the available quivers then
require at least one connected irreducible multiplet.
Determining which such multiplet is realized is a dynamical question
that we address using the approach of defect TCSA.

\subsubsection{The regular vacuum module}

\begin{figure}[!htbp]
\centering
\begin{tikzpicture}[x=1cm,y=1cm,
  particle line/.style={draw=cyan!75!black,line width=.85pt},
  kink arrow/.style={particle line,-{Stealth[length=2.1mm,width=1.55mm]}},
  particle loop/.style={kink arrow},
  vacuum/.style={circle,fill=black,inner sep=1.65pt},
  every node/.style={font=\small}
]
\node[anchor=east,align=right] at (-.85,-0) {$k=3$\\$\mathcal M_{4,5}$};
\begin{scope}[shift={(0,-0)}]
  \node at (1.25,1.1) {$R_0$};
  \quivercircleloop[particle loop]{0,0}{.23}{90}{.12}
  \quivercircleloop[particle loop]{2.5,0}{.23}{90}{.12}
  \node[vacuum] at (0,0) {};
  \node[anchor=north] at (0,-0.22) {$\lvert 0\rangle$};
  \node[vacuum] at (2.5,0) {};
  \node[anchor=north] at (2.5,-0.22) {$\lvert 1\rangle$};
\end{scope}
\begin{scope}[shift={(4.5,-0)}]
  \node at (1.25,1.1) {$R_1$};
  \draw[kink arrow] (0.12,0.035) .. controls (0.7,0.235) and (1.8,0.235) .. (2.38,0.035);
  \draw[kink arrow] (2.38,-0.035) .. controls (1.8,-0.235) and (0.7,-0.235) .. (0.12,-0.035);
  \quivercircleloop[particle loop]{2.5,0}{.23}{0}{.12}
  \node[vacuum] at (0,0) {};
  \node[anchor=north] at (0,-0.22) {$\lvert 0\rangle$};
  \node[vacuum] at (2.5,0) {};
  \node[anchor=north] at (2.5,-0.22) {$\lvert 1\rangle$};
\end{scope}
\node[anchor=east,align=right] at (-.85,-2.4) {$k=4$\\$\mathcal M_{5,6}$};
\begin{scope}[shift={(0,-2.4)}]
  \node at (1.25,1.1) {$R_0$};
  \quivercircleloop[particle loop]{0,0}{.23}{90}{.12}
  \quivercircleloop[particle loop]{1.25,0}{.23}{90}{.12}
  \quivercircleloop[particle loop]{2.5,0}{.23}{90}{.12}
  \node[vacuum] at (0,0) {};
  \node[anchor=north] at (0,-0.22) {$\lvert 0\rangle$};
  \node[vacuum] at (1.25,0) {};
  \node[anchor=north] at (1.25,-0.22) {$\lvert 1\rangle$};
  \node[vacuum] at (2.5,0) {};
  \node[anchor=north] at (2.5,-0.22) {$\lvert 2\rangle$};
\end{scope}
\begin{scope}[shift={(4.5,-2.4)}]
  \node at (1.25,1.1) {$R_1$};
  \draw[kink arrow] (0.12,0.035) .. controls (0.35,0.235) and (0.9,0.235) .. (1.13,0.035);
  \draw[kink arrow] (1.13,-0.035) .. controls (0.9,-0.235) and (0.35,-0.235) .. (0.12,-0.035);
  \quivercircleloop[particle loop]{1.25,0}{.23}{90}{.12}
  \draw[kink arrow] (1.37,0.035) .. controls (1.6,0.235) and (2.15,0.235) .. (2.38,0.035);
  \draw[kink arrow] (2.38,-0.035) .. controls (2.15,-0.235) and (1.6,-0.235) .. (1.37,-0.035);
  \node[vacuum] at (0,0) {};
  \node[anchor=north] at (0,-0.22) {$\lvert 0\rangle$};
  \node[vacuum] at (1.25,0) {};
  \node[anchor=north] at (1.25,-0.22) {$\lvert 1\rangle$};
  \node[vacuum] at (2.5,0) {};
  \node[anchor=north] at (2.5,-0.22) {$\lvert 2\rangle$};
\end{scope}
\begin{scope}[shift={(9,-2.4)}]
  \node at (1.25,1.1) {$R_2$};
  \draw[kink arrow] (0.12,0.2) .. controls (0.7,0.99) and (1.8,0.99) .. (2.38,0.2);
  \quivercircleloop[particle loop]{1.25,0}{.23}{90}{.12}
  \draw[kink arrow] (2.38,-0.2) .. controls (1.8,-0.99) and (0.7,-0.99) .. (0.12,-0.2);
  \node[vacuum] at (0,0) {};
  \node[anchor=north] at (0,-0.3) {$\lvert 0\rangle$};
  \node[vacuum] at (1.25,0) {};
  \node[anchor=north] at (1.25,-0.22) {$\lvert 1\rangle$};
  \node[vacuum] at (2.5,0) {};
  \node[anchor=north] at (2.5,-0.3) {$\lvert 2\rangle$};
\end{scope}
\node[anchor=east,align=right] at (-.85,-4.8) {$k=5$\\$\mathcal M_{6,7}$};
\begin{scope}[shift={(0,-4.8)}]
  \node at (1.25,1.1) {$R_0$};
  \quivercircleloop[particle loop]{0,0}{.23}{90}{.12}
  \quivercircleloop[particle loop]{1.25,0}{.23}{90}{.12}
  \quivercircleloop[particle loop]{2.5,0}{.23}{90}{.12}
  \node[vacuum] at (0,0) {};
  \node[anchor=north] at (0,-0.22) {$\lvert 0\rangle$};
  \node[vacuum] at (1.25,0) {};
  \node[anchor=north] at (1.25,-0.22) {$\lvert 1\rangle$};
  \node[vacuum] at (2.5,0) {};
  \node[anchor=north] at (2.5,-0.22) {$\lvert 2\rangle$};
\end{scope}
\begin{scope}[shift={(4.5,-4.8)}]
  \node at (1.25,1.1) {$R_1$};
  \draw[kink arrow] (0.12,0.035) .. controls (0.35,0.235) and (0.9,0.235) .. (1.13,0.035);
  \draw[kink arrow] (1.13,-0.035) .. controls (0.9,-0.235) and (0.35,-0.235) .. (0.12,-0.035);
  \quivercircleloop[particle loop]{1.25,0}{.23}{90}{.12}
  \draw[kink arrow] (1.37,0.035) .. controls (1.6,0.235) and (2.15,0.235) .. (2.38,0.035);
  \draw[kink arrow] (2.38,-0.035) .. controls (2.15,-0.235) and (1.6,-0.235) .. (1.37,-0.035);
  \quivercircleloop[particle loop]{2.5,0}{.23}{0}{.12}
  \node[vacuum] at (0,0) {};
  \node[anchor=north] at (0,-0.22) {$\lvert 0\rangle$};
  \node[vacuum] at (1.25,0) {};
  \node[anchor=north] at (1.25,-0.22) {$\lvert 1\rangle$};
  \node[vacuum] at (2.5,0) {};
  \node[anchor=north] at (2.5,-0.22) {$\lvert 2\rangle$};
\end{scope}
\begin{scope}[shift={(9,-4.8)}]
  \node at (1.25,1.1) {$R_2$};
  \draw[kink arrow] (0.12,0.2) .. controls (0.7,0.99) and (1.8,0.99) .. (2.38,0.2);
  \quivercircleloop[particle loop]{1.25,0}{.23}{90}{.12}
  \draw[kink arrow] (1.37,0.035) .. controls (1.6,0.235) and (2.15,0.235) .. (2.38,0.035);
  \draw[kink arrow] (2.38,-0.2) .. controls (1.8,-0.99) and (0.7,-0.99) .. (0.12,-0.2);
  \draw[kink arrow] (2.38,-0.035) .. controls (2.15,-0.235) and (1.6,-0.235) .. (1.37,-0.035);
  \node[vacuum] at (0,0) {};
  \node[anchor=north] at (0,-0.3) {$\lvert 0\rangle$};
  \node[vacuum] at (1.25,0) {};
  \node[anchor=north] at (1.25,-0.22) {$\lvert 1\rangle$};
  \node[vacuum] at (2.5,0) {};
  \node[anchor=north] at (2.5,-0.3) {$\lvert 2\rangle$};
\end{scope}
\end{tikzpicture}
\caption{All irreducible quivers of the regular vacuum module for
$\mathcal C_k=\mathrm{SO}(3)_k$, with $k=3,4,5$. Each node is a vacuum, each directed edge between different nodes is a kink, and each circular loop is a particle. The arrows in each panel give the components of one irreducible multiplet.}
\label{fig:quivers}
\end{figure}
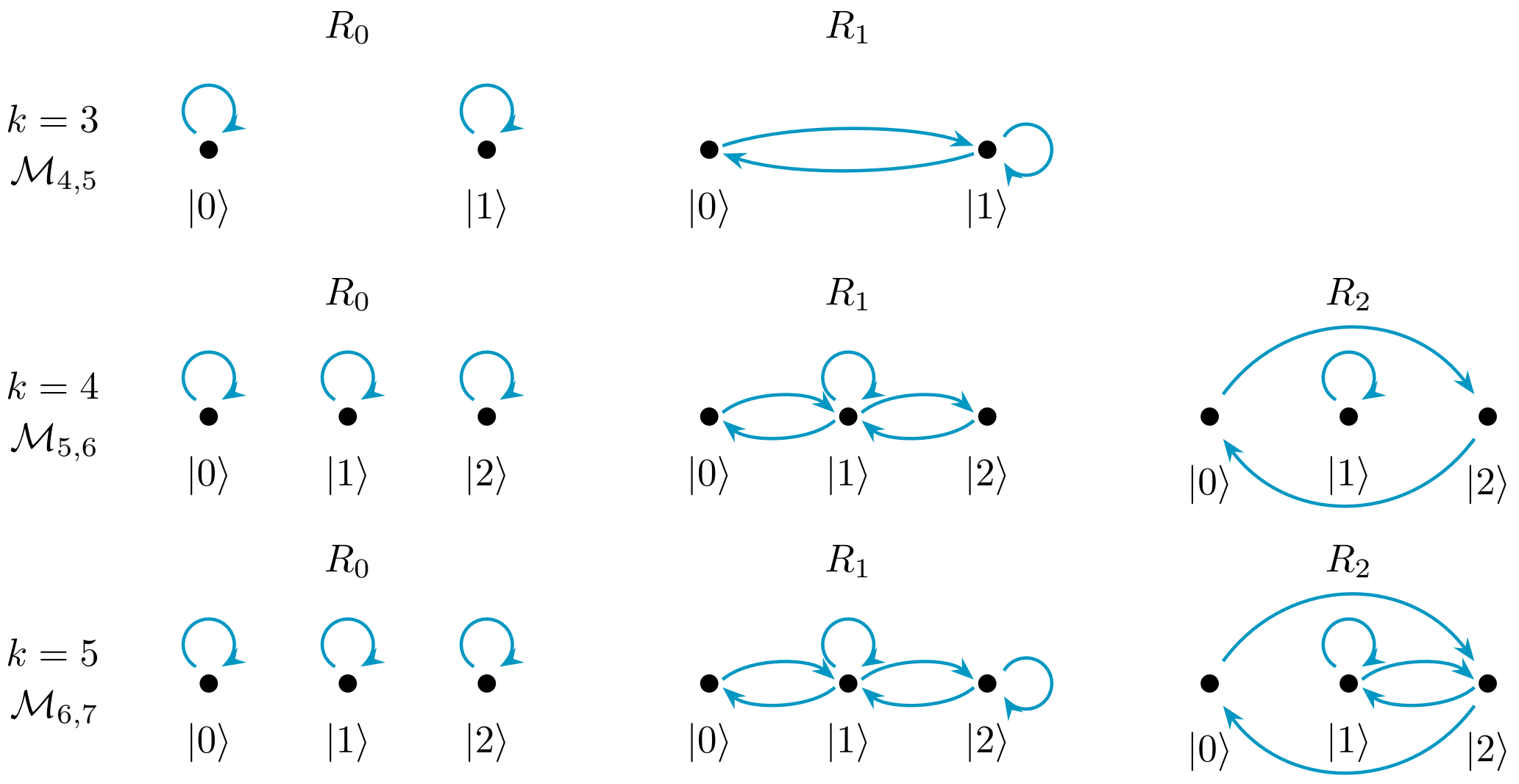

In the regular phase, $\mathcal M_{\mathrm{vac}}=\mathcal C$ and the
$J+1$ vacua are labelled by the simple objects of $\mathcal C$.
We write $\ket a$ for the vacuum $v_a$, i.e. $\ket0,\ket1,\ket2$ corresponding
to the lines $X_0,X_1,X_2$ when they are present. With this identification,
the symmetry acts by fusion, $(n_i)_{ab}=N_{ia}^{b}$.
We also use these numerical labels for the vacuum channels in $F$-symbols.
The dual category is again $\mathcal C$, acting by fusion from the right,
so its simple objects $X_j$ label the irreducible multiplets~\cite{CGH,CHO},
\begin{align}
 R_j=\bigoplus_{a,b}\Hom_{\mathcal C}(v_a\otimes X_j,v_b)\,,
 \qquad (Q_j)_{ab}=N_{aj}^{b}\,.
 \label{eq:Ralpha}
\end{align}
Figure~\ref{fig:quivers} shows all these quivers for $k=3,4,5$.
There are two irreducible representations for $k=3$ and three for
each of $k=4,5$.
The identity representation $R_0$ has one loop at each vacuum
and therefore consists entirely of particles. For the multiplet $R_1$, corresponding to the generator $X_1$, the fusion rule in Eq.~\eqref{eq:fusion}
joins $a$ to $a-1,a$, and $a+1$ when $0<a<J$.
The vacuum $a=0$ is connected only to $a=1$, whereas, for the upper endpoint
$a=J$, an self-connected loop is present for odd $k$ and absent for even $k$.
Thus its numbers of components are
\begin{align}
 \begin{array}{c@{\quad}c@{\quad}c@{\quad}c}
 &\text{vacua}&\text{oriented kinks}&\text{neutral particles}\\\hline
 k=2J+1&J+1&2J&J\\
 k=2J&J+1&2J&J-1
 \end{array}\,,
 \label{eq:counts}
\end{align}
giving $\dim R_1=3J$ for odd $k$ and $3J-1$ for even $k$.
In the regular phase, each simple line $X_i\in\mathcal C_k$ labels
a quiver representation $R_i$, as shown in Figure~\ref{fig:quivers}.
Its particle and kink components are read from the loops and arrows
according to the rules above.

\subsubsection{The disordered vacuum module of \texorpdfstring{$\mathcal M_{5,6}$}{M5,6}}
\label{subsec:draft2_m56_two_vacua}

For $\lambda>0$, the $\phi_{2,1}$ deformation of $\mathcal M_{5,6}$ has two
degenerate vacua that will be verified later by TCSA. The preserved category remains
$\C\simeq\operatorname{Rep}(S_3)$, with simple lines
$X_0$, $X_1$ and $X_2$~\cite{CLSWY}.
The two vacua form the module $\mathcal M_H=\operatorname{Rep}(H)$,
where $H=\{e,t\}\simeq\mathbb Z_2$ is generated by a transposition $t$ in $S_3$~\cite{Ostrik:2001xnt}.
We label its trivial and sign representations by $v_0$ and $v_1$, respectively,
so that $t$ is represented by $+1$ on $v_0$ and $-1$ on $v_1$.
To act on a vacuum, we first restrict the $S_3$ representation $X_i$ to $H$,
retaining its representation space and the matrices assigned to elements of $H$.
We then take its tensor product with the vacuum representation:
\begin{align}
 X_i\mathbin{\triangleright}v_a
 =\bigl(\operatorname{Res}^{S_3}_H X_i\bigr)\otimes v_a\,,
 \qquad a\in\{0,1\}\,,\notag
\end{align}
where $\operatorname{Res}^{S_3}_H$ denotes restriction to $H$, and $\otimes$
is the ordinary tensor product of $H$ representations, with $H$ acting on both factors.
The lines $X_0$ and $X_2$ correspond to the trivial and sign representations of
$S_3$, while $X_1$ is its standard representation of dimension two. In this representation, the matrix
representing $t$ has eigenvalues $+1$ and $-1$. Hence
\begin{align}
 \operatorname{Res}^{S_3}_H X_0=v_0\,,\qquad
 \operatorname{Res}^{S_3}_H X_2=v_1\,,\qquad
 \operatorname{Res}^{S_3}_H X_1=v_0\oplus v_1\,.
 \label{eq:m56_disordered_restriction}
\end{align}
The tensor products satisfy $v_0\otimes v_a=v_a$ and $v_1\otimes v_1=v_0$.
Thus $X_0$ preserves each vacuum, $X_2$ exchanges the two vacua, and $X_1$
sends either vacuum to $v_0\oplus v_1$. For example,
\begin{align}
 X_1\mathbin{\triangleright}v_1
 =(v_0\oplus v_1)\otimes v_1=v_1\oplus v_0\,.\notag
\end{align}
The symmetry action, in the vacuum order $(v_0,v_1)$, is therefore
\begin{align}
 n_0=\begin{pmatrix}1&0\\0&1\end{pmatrix}\,,\qquad
 n_2=\begin{pmatrix}0&1\\1&0\end{pmatrix}\,,\qquad
 n_1=\begin{pmatrix}1&1\\1&1\end{pmatrix}\,,
 \label{eq:m56_disordered_nimrep}
\end{align}
where $(n_i)_{ab}=\dim\operatorname{Hom}_{\mathcal M_H}(X_i\mathbin{\triangleright}v_a,v_b)$
counts the multiplicity of the output vacuum $v_b$ for the input vacuum $v_a$.
The exchange of $v_0$ and $v_1$ by $X_2$ realizes a spontaneous $X_2$ line symmetry breaking in this phase.

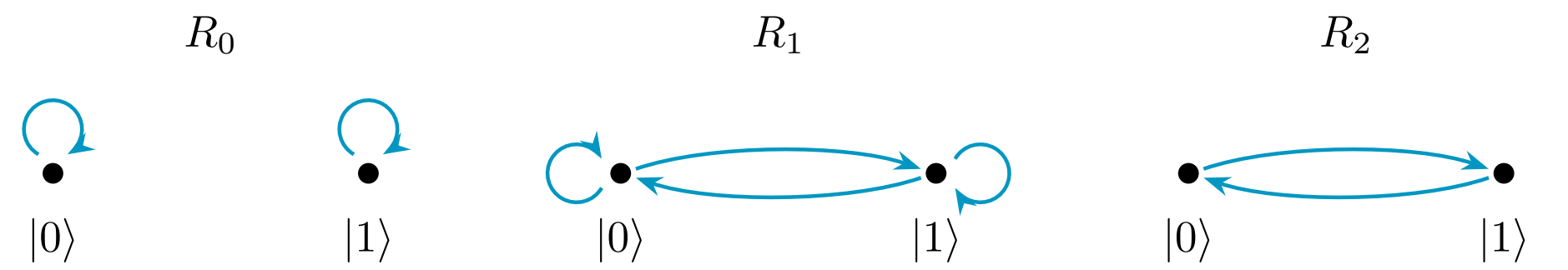
\begin{figure}[tb]
\centering
\begin{tikzpicture}[x=1cm,y=1cm,
  particle line/.style={draw=cyan!75!black,line width=.85pt},
  kink arrow/.style={particle line,-{Stealth[length=2.1mm,width=1.55mm]}},
  particle loop/.style={kink arrow},
  vacuum/.style={circle,fill=black,inner sep=1.65pt},
  every node/.style={font=\small}
]
\begin{scope}[shift={(0,0)}]
  \node at (1.25,1.1) {$R_0$};
  \quivercircleloop[particle loop]{0,0}{.23}{90}{.12}
  \quivercircleloop[particle loop]{2.5,0}{.23}{90}{.12}
  \node[vacuum] at (0,0) {};
  \node[anchor=north] at (0,-0.22) {$\lvert 0\rangle$};
  \node[vacuum] at (2.5,0) {};
  \node[anchor=north] at (2.5,-0.22) {$\lvert 1\rangle$};
\end{scope}
\begin{scope}[shift={(4.5,0)}]
  \node at (1.25,1.1) {$R_1$};
  \quivercircleloop[particle loop]{0,0}{.23}{180}{.12}
  \draw[kink arrow] (0.12,0.035) .. controls (0.7,0.235) and (1.8,0.235) .. (2.38,0.035);
  \draw[kink arrow] (2.38,-0.035) .. controls (1.8,-0.235) and (0.7,-0.235) .. (0.12,-0.035);
  \quivercircleloop[particle loop]{2.5,0}{.23}{0}{.12}
  \node[vacuum] at (0,0) {};
  \node[anchor=north] at (0,-0.22) {$\lvert 0\rangle$};
  \node[vacuum] at (2.5,0) {};
  \node[anchor=north] at (2.5,-0.22) {$\lvert 1\rangle$};
\end{scope}
\begin{scope}[shift={(9,0)}]
  \node at (1.25,1.1) {$R_2$};
  \draw[kink arrow] (0.12,0.035) .. controls (0.7,0.235) and (1.8,0.235) .. (2.38,0.035);
  \draw[kink arrow] (2.38,-0.035) .. controls (1.8,-0.235) and (0.7,-0.235) .. (0.12,-0.035);
  \node[vacuum] at (0,0) {};
  \node[anchor=north] at (0,-0.22) {$\lvert 0\rangle$};
  \node[vacuum] at (2.5,0) {};
  \node[anchor=north] at (2.5,-0.22) {$\lvert 1\rangle$};
\end{scope}
\end{tikzpicture}
\caption{All three irreducible quivers of the module with two vacua in
the $\lambda>0$ phase of $\mathcal M_{5,6}+\lambda\,\phi_{2,1}$.}
\label{fig:m56_disordered_quivers}
\end{figure}

Applying the strip representation construction of~\cite{CHO} to this vacuum
module, one can show that its three irreducible strip representations
$R_0,R_1,R_2$ have quiver matrices
\begin{align}
 Q_i=n_i\,,\qquad i=0,1,2\,.
 \label{eq:m56_disordered_quivers}
\end{align}
The three quivers are shown in Figure~\ref{fig:m56_disordered_quivers}.
The multiplet $R_0$ contains one particle above each vacuum,
$R_2$ contains a kink and its antikink, and $R_1$ contains
both particles and both oriented kinks.
Although their adjacency matrices obey $Q_1=Q_0+Q_2$,
the strip algebra representation $R_1$ is not equivalent to the sum of $R_0$ and $R_2$, 
\begin{align}
 R_1\not\simeq R_0\oplus R_2\,.
 \label{eq:m56_disordered_irreducible}
\end{align}
The strip algebra mixes the particle and kink components of $R_1$
into a single irreducible representation. In contrast, the particle and kink subspaces are separately invariant in $R_0\oplus R_2$.

The defect circle setup that we will discuss in Section~\ref{sec:circle} makes this distinction accessible to TCSA through
fractional momenta and local field form factors. In particular, the
circle with an $X_1$ defect supports light states with
$s=\pm\frac{1}{3}$ modulo integers for $R_1$, whereas these channels
are absent for $R_0\oplus R_2$, where
$s=\frac{pL}{2\pi}$ for momentum $p$ and circumference $L$.
The spectra and vacuum compositions measured in
Section~\ref{subsec:m56_disordered_quiver_reconstruction} identify the two
kinks and two particles as the irreducible multiplet $R_1$.
We now develop the defect circle construction that relates these
momentum eigenstates to the kink and particle components of a quiver.

\FloatBarrier

\section[From quivers to defect circles]{From a vacuum quiver to a defect circle}
\label{sec:circle}

\subsection{Closing a kink and transport through the defect}
\label{sec:circle1}

We first construct the defect circle for the regular vacuum
module. Consider a single kink or particle component $a\to b$ on an interval.
Identifying the ends periodically imposes $a=b$, so only the neutral particle
components of a multiplet can close individually on the periodic circle.
Inserting a line $X_i$ along the time direction supplies a junction that
can turn the vacuum $b$ back into $a$. On the regular vacuum module
$\mathcal M_{\rm vac}=\C$, the symmetry line $X_i$ acts from the left,
$v_b\mapsto X_i\otimes v_b$. The object
$\alpha\in\C^*_{\mathcal M}\simeq\C$, which labels the kink/particle
multiplet $R_\alpha$, acts from the right,
$v_a\mapsto v_a\otimes\alpha$~\cite{CHO}.
The closure line $X_i$ and the multiplet label $\alpha$ are independent.
These two actions give the internal space of $R_\alpha$ on the circle
closed by $X_i$,
\begin{align}
 V_{X_i,\alpha}=\bigoplus_{a,b}
 \Hom(v_a\otimes\alpha,v_b)\otimes\Hom(X_i\otimes v_b,v_a).
 \label{eq:closure}
\end{align}
The first and second factors describe the kink/particle junction and the defect
junction, respectively. For a fixed multiplet $\alpha=X_i$, let's consider to close the multiplet $R_i$ with the same defect line $X_i$ too. In the regular module
of $\mathcal C$, each $X_i$ is its own dual and every allowed fusion
channel occurs once, so $N_{ai}^{b}=N_{ib}^{a}\in\{0,1\}$.
Each quiver edge therefore labels an orthonormal basis state
$\ket{ab}$ in $V_{X_i,X_i}$, and
\begin{align}
 \dim V_{X_i,X_i}=\sum_{a,b}(N_{ai}^{b})^2
 =\sum_{a,b}N_{ai}^{b}=\dim R_i\,.
 \label{eq:closedcount}
\end{align}
This counts all kink and particle components of $R_i$, including both
arrows between distinct vacua and loops at a single vacuum. Specifically, taking $\alpha=X_0$ and $i=0$ as an trivial insertion, it counts the number of the vacua in $\mathcal M_{\rm vac}=\mathcal C$,
\begin{align}
 \dim V_{X_0,X_0}=\sum_{a=0}^{J}\dim\Hom(v_a,v_a)=J+1\,.
\end{align}
In addition, the neutral particle components of $R_i$ multiplet be obtained by choosing
the closing line $X_0$ and the excitation label $\alpha=X_i$,
\begin{align}
 \dim V_{X_0,X_i}=\sum_{a=0}^{J}\dim\Hom(v_a\otimes X_i,v_a)
 =\Tr n_i\,.
\end{align}

\begin{figure}[tb]
\centering
\begin{tikzpicture}[thick,>=Stealth,font=\small,
  seam/.style={red!75!black,line width=1.15pt},
  particle/.style={blue!75!black,line width=1.25pt,line cap=round}]
\node[anchor=east] at (-.6,6.25){(a)};
\draw (2.1,6.25) circle (.85);
\fill[blue!75!black] (2.1,7.10) circle (.065);
\fill[red!75!black] (2.95,6.25) circle (.065);
\node[above] at (2.1,7.45){kink/particle};
\node[right] at (3.25,6.25){$X_1$};
\node at (1.68,6.55){$v_a$};
\node at (2.52,6.55){$v_b$};
\node at (2.1,4.95){one spatial slice};
\node at (8.2,6.75){$\displaystyle
  v_a\xrightarrow{\ \alpha\ }v_b\xrightarrow{\ X_1\ }v_a$};
\node[align=center] at (8.2,5.65){the kink/particle and defect junctions\\[3pt]close the component $\ket{ab}$};
\node[anchor=east] at (-.6,1.5){(b)};
\begin{scope}[shift={(2.1,.35)}]
\node at (0,3.15){before: $\ket{bd}$};
\draw (0,0)--(0,.45)--(.6,1.225);
\draw[seam] (0,.45)--(-1.2,2.0);
\draw (.6,1.225)--(0,2.0);
\draw[particle] (.6,1.225)--(1.2,2.0);
\node[below=6pt] at (0,0){$v_b$};
\node[right=4pt] at (.39,.89){$v_d$};
\node[red!75!black,above=8pt] at (-1.2,2.0){$X_1$};
\node[above=8pt] at (0,2.0){$v_b$};
\node[blue!75!black,above=8pt] at (1.2,2.0){$\alpha=X_j$};
\end{scope}
\node at (6.4,1.5){$\displaystyle
  =\sum_a\bigl(F^{1bj}_b\bigr)_{ad}$};
\begin{scope}[shift={(10.7,.35)}]
\node at (0,3.15){after: $\ket{ab}$};
\draw (0,0)--(0,.45)--(-.6,1.225);
\draw[particle] (0,.45)--(1.2,2.0);
\draw (-.6,1.225)--(0,2.0);
\draw[seam] (-.6,1.225)--(-1.2,2.0);
\node[below=6pt] at (0,0){$v_b$};
\node[left=4pt] at (-.39,.89){$v_a$};
\node[red!75!black,above=8pt] at (-1.2,2.0){$X_1$};
\node[above=8pt] at (0,2.0){$v_b$};
\node[blue!75!black,above=8pt] at (1.2,2.0){$\alpha=X_j$};
\end{scope}
\end{tikzpicture}
\caption{Closure and transport through the defect.
(a) The line $X_1$ converts $v_b$ back to $v_a$, closing the state
$\ket{ab}$ on a spatial slice. The red point is the intersection with
the defect extending in time.
(b) Crossing a state through the defect changes the intermediate vacuum from
$v_d$ to $v_a$. The states before and after the crossing related by the corresponding $F$-move. Taking $j=1$ gives the transport matrix in Eq.~\eqref{eq:T}.}
\label{fig:closure}
\end{figure}
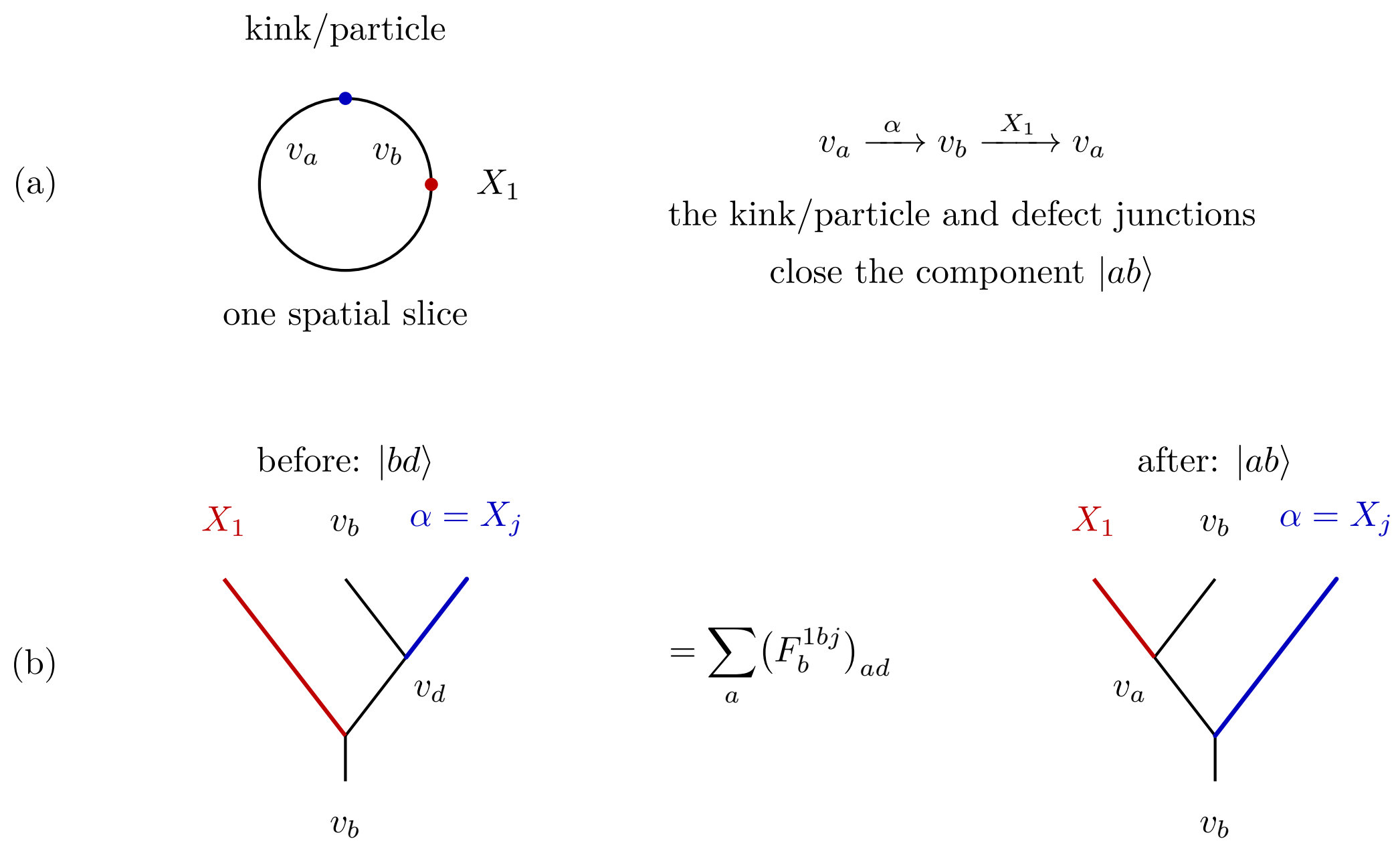

We now specialize to the multiplet of interest, $R_1$, by taking
$i=1$ and $\alpha=X_1$. The circle with the $X_1$ defect accommodates
all $\dim R_1$ kink and particle components. Transporting an excitation
once around this circle carries it through the defect, as shown in
Figure~\ref{fig:recoupling}. To determine the resulting change of basis,
we resolve the crossing procedure into the two junctions in Figure~\ref{fig:closure}.
For the moment, keep the excitation label general, $\alpha=X_j$.
Before the crossing, the state $\ket{bd}$ is closed by
$v_b\otimes\alpha\to v_d$ and $X_1\otimes v_d\to v_b$.
After the crossing, the junctions are $X_1\otimes v_b\to v_a$
and $v_a\otimes\alpha\to v_b$, giving the state $\ket{ab}$.

On the regular vacuum module, the left action of $X_1$ and
the right action of $\alpha$ are related by the associator
of $\C$. The $F$-move in Figure~\ref{fig:closure}(b) represent the
adjoints of the fusion junctions above. The tree before crossing
first groups $v_b\otimes\alpha$ into $v_d$, while the tree after
crossing groups $X_1\otimes v_b$ into $v_a$.
Expanding the left tree in the latter
basis gives the coefficient $(F^{1bj}_b)_{ad}$.
The row $a$ and column $d$ thus label the outgoing and incoming
intermediate vacua, respectively, with external vacuum $v_b$ fixed.
Setting $j=1$ gives the transport operator on $V_{X_1,X_1}$,
\begin{align}
 T\ket{bd}
 =\sum_a (F^{1b1}_b)_{ad}\ket{ab},
 \qquad
 T_{(a,b),(c,d)}
 =\delta_{bc}(F^{1b1}_b)_{ad}\,,
 \label{eq:T}
\end{align}
where the matrix rows $(a,b)$ and columns $(c,d)$ are the outgoing
and incoming vacuum pairs. The incoming left vacuum is the outgoing
right vacuum, so $c=b$, which accounts for $\delta_{bc}$. The operator $T$ is unitary inherited from the unitary property of the $F$-matrix.

For the diagonal minimal model, the $F$-symbol has a finite
quantum-Racah expression. Set
$\zeta=\ee^{\frac{\ii\pi}{k+2}}$, and define
$\qnum{n}=\frac{\sin\left(\frac{n\pi}{k+2}\right)}{\sin\left(\frac{\pi}{k+2}\right)}$. In the real orthonormal
vertex gauge,
\begin{align}
 \F{i}{j}{\ell}{m}{p}{q}
 =(-1)^{i+j+\ell+m}\sqrt{\qnum{2p+1}\qnum{2q+1}}
 \sixj{i}{j}{p}{\ell}{m}{q}.
 \label{eq:Fmain}
\end{align}
Here $i,j,\ell,m$ are simple object labels, and $p,q$ label the
intermediate channels in $(X_i\otimes X_j)\otimes X_\ell$ and
$X_i\otimes(X_j\otimes X_\ell)$, respectively.
The quantum $6j$ symbol on the right is the finite factorial sum
derived in Appendix~\ref{app:recoupling}. 


For a general vacuum module $\mathcal M_{\mathrm{vac}}$, the same crossing
uses the compatibility between the left action of $\C$ and the right
action of its dual category $\C^*_{\mathcal M_{\mathrm{vac}}}$~\cite{CHO},
\begin{align}
 X_i\mathbin{\triangleright}(v_b\mathbin{\triangleleft}\alpha)
 \simeq (X_i\mathbin{\triangleright}v_b)\mathbin{\triangleleft}\alpha\,.
 \notag
\end{align}
We denote the corresponding mixed module $F$-symbol by
$\bigl(\widetilde F^{i b\alpha}_{b}\bigr)_{ad}$, with the same
incoming channel $d$ and outgoing channel $a$ as in Figure~\ref{fig:closure}(b).
For the modules considered here, each allowed junction has multiplicity one.
Replacing the regular associator in~Eq.~\eqref{eq:T} gives
\begin{align}
 T^{\mathcal M_{\mathrm{vac}}}_{X_i,\alpha}\ket{bd}
 &=\sum_a\bigl(\widetilde F^{i b\alpha}_{b}\bigr)_{ad}\ket{ab}\,,
 \qquad
 \bigl(T^{\mathcal M_{\mathrm{vac}}}_{X_i,\alpha}\bigr)_{(a,b),(c,d)}
 =\delta_{bc}\bigl(\widetilde F^{i b\alpha}_{b}\bigr)_{ad}\,.
 \label{eq:Tmodule}
\end{align}
In the disordered phase of $\mathcal M_{5,6}$,\,
$\mathcal M_H=\operatorname{Rep}(\mathbb Z_2)$ is the module with two vacua
described in Section~\ref{subsec:draft2_m56_two_vacua}.
It can be realized by the objects with half-integer spin in ${\rm SU}(2)_4$,
with $v_a$ assigned spin $\widehat a=a+\frac12$ for $a=0,1$.
The left action of $X_1$ and the right action of $\rho_1$, which labels
the multiplet $R_1$, are realized by fusion with the object labelled by
$j=1$ in ${\rm SU}(2)_4$. The mixed module $F$ matrix is therefore inherited
from ${\rm SU}(2)_4$. In the expression of Eq.~\eqref{eq:Fmain}, define
$\widetilde F_b\equiv \widetilde F^{1 b\rho_1}_{b}$ and evaluate at $k=4$:
\begin{align}
 \left(\widetilde F_b\right)_{ad}
 =\left(F^{1\,\widehat b\,1}_{\widehat b}\right)_{\widehat a\widehat d}\,\quad
 {\rm implying}
 \qquad
 \widetilde F_0=\frac12
 \begin{pmatrix}-1&\sqrt3\\\sqrt3&1\end{pmatrix}\,,
 \qquad
 \widetilde F_1=\frac12
 \begin{pmatrix}1&-\sqrt3\\-\sqrt3&-1\end{pmatrix}\,,
 \label{eq:Fmodule56}
\end{align}
where rows $a$ and columns $d$ are ordered as $(0,1)$, and the subscript $b=0,1$ in $\widetilde F_b$ labels the vacuum $v_b$. All four vacuum pairs are allowed since $Q_1=n_1$ has every entry equal
to one in Eq.~\eqref{eq:m56_disordered_nimrep}.
To assemble these blocks into the transport matrix, we spell out
Eq.~\eqref{eq:Tmodule} specific to this module:
\begin{align}
 T_{\mathrm{dis}}\ket{bd}
 =\sum_{a=0}^{1}(\widetilde F_b)_{ad}\ket{ab}\,.
 \notag
\end{align}
Ordering both the rows and columns as
$(\ket{00},\ket{01},\ket{10},\ket{11})$ gives
\begin{align}
 T_{\mathcal M_{5,6}}^{\mathrm{dis.}}\equiv T^{\mathcal M_H}_{X_1,\rho_1}
 =\begin{pmatrix}
 (\widetilde F_0)_{00}&(\widetilde F_0)_{01}&0&0\\
 0&0&(\widetilde F_1)_{00}&(\widetilde F_1)_{01}\\
 (\widetilde F_0)_{10}&(\widetilde F_0)_{11}&0&0\\
 0&0&(\widetilde F_1)_{10}&(\widetilde F_1)_{11}
 \end{pmatrix}
 =\frac12\begin{pmatrix}
 -1&\sqrt3&0&0\\
 0&0&1&-\sqrt3\\
 \sqrt3&1&0&0\\
 0&0&-\sqrt3&-1
 \end{pmatrix}\,.
 \label{eq:Tdisorder}
\end{align}
This matrix is also unitary and has eigenvalues $1,-1,\ee^{\frac{2\pi\ii}3}$
and $\ee^{-\frac{2\pi\ii}3}$. We next diagonalize the transport matrix to determine the fractional momentum shifts and the change of basis between vacuum pairs and momentum channels.

\subsection{Fractional momenta and conformal defect Hilbert spaces}

On an interval, $\ket{ab}$ denotes a physical kink or particle with adjacent
vacua $a$ and $b$. Closing these components through the defect as in
Section~\ref{sec:circle1} gives the circle basis. To obtain the momentum
basis, we diagonalize the transport matrix $T$ and collect its orthonormal
eigenvectors as the columns of a unitary matrix $U$. Writing $A=ab$ for the
allowed vacuum pairs, we have
\begin{align}
 (TU)_{A,\xi}=\ee^{2\pi\ii q_\xi}\,U_{A,\xi},\qquad
 U^\dagger U=\mathbf 1\,,
 \label{eq:diag}
\end{align}
where the rows of $U$ are labelled by $A$ and the columns by the internal
channels $\xi$. The phase $\ee^{2\pi\ii q_\xi}$ determines the fractional
momentum shift $q_\xi\in\mathbb Q$ modulo one. The two bases are related by
\begin{align}
 \ket{\xi}=\sum_{a,b}U_{ab,\xi}\ket{ab},\qquad
 \ket{ab}=\sum_\xi U_{ab,\xi}^*\ket{\xi}\,.
 \label{eq:dictionary}
\end{align}

To relate these internal channels to spatial momentum, let $\ket{\Psi}$
be a state of a single kink or particle on the defect circle, and let
$\ket{x;ab}$ denote its circle component $\ket{ab}$ localized at position
$x$. We write the wavefunction as
\begin{align}
 \psi_{ab}(x)=\langle x;ab|\Psi\rangle,\qquad
 \psi(x)=\bigl(\psi_{ab}(x)\bigr)_{(a,b)}\,,
 \label{eq:psivec}
\end{align}
where $\psi(x)$ is the column vector of amplitudes in the allowed circle
components. Now moving the state around the circle through the defect, the wave function, acted by operator $T$ in Eq.~\eqref{eq:T}, must obey
\begin{align}
 \psi(x+L)=T\psi(x)\,,
 \label{eq:bc}
\end{align}
where $L$ is the circumference of the defect circle. A mode with momentum
$p$ in channel $\xi$ has
$\psi(x)\propto U_{\cdot,\xi}\,\ee^{\ii px}$, where $U_{\cdot,\xi}$ is the
$\xi$th column of $U$. Substituting this wavefunction
into~Eq.~\eqref{eq:bc} and using~Eq.~\eqref{eq:diag} gives
\begin{align}
 \ee^{\ii pL}=\ee^{2\pi\ii q_\xi},\qquad
 p\equiv p_{\xi,n}=\frac{2\pi}{L}(n+q_\xi),\quad n\in\mathbb Z\,.
 \label{eq:pquant}
\end{align}
The spatial phase acquired in one circuit thus proportional to $q_\xi\pmod 1$ the exponent of the eigenvalue of $T$. That is the reason we call the states $\ket{\xi,n}$ momentum basis. The normalized momentum state $\ket{\xi,n}$ has wavefunction
\begin{align}
 \psi^{(\xi,n)}_{ab}(x)
 =\langle x;ab|\xi,n\rangle
 =\frac{U_{ab,\xi}}{\sqrt L}\ee^{\ii p_{\xi,n}x}\,.
 \notag
\end{align}

For the $R_1$ multiplet, the energies of these single excitations at
large $L$ take the form
\begin{align}
 E_{\xi,n}(L)-E_0^{(X_1)}(L)
 =\sqrt{m_{X_1}^2+\left(\frac{2\pi(n+q_\xi)}{L}\right)^2}
 +\delta E_{\xi,n}(L)\,,
 \label{eq:dispersion}
\end{align}
where $m_{X_1}$ is the common rest mass of the multiplet and
$E_0^{(X_1)}(L)$ is a vacuum energy in $\mathcal H_{X_1}$.
The correction $\delta E_{\xi,n}(L)$ includes effects of finite volume and,
in TCSA, truncation. States in the multiplet share this rest mass, while their momenta and $\delta E_{\xi,n}(L)$ determine the energies at finite $L$.

More importantly, We can also determine the allowed momenta from the ultraviolet Hilbert
space. Let $\V_i$ be the irreducible chiral Virasoro module of primary $i$,
with character $\chi_i(\tau)=\Tr_{\V_i}\mathfrak q^{L_0-\frac{c}{24}}$, where
$\mathfrak q=\ee^{2\pi\ii\tau}$ and $\operatorname{Im}\tau>0$. The line $X_1$
inserted along the time direction defines the defect Hilbert space and
its torus partition function~\cite{Petkova:2000ip},
\begin{align}
 \Hc_{X_1}=\bigoplus_{i,j}N_{X_1 i}^{j}\V_i\otimes\overline{\V}_j,
 \qquad Z_{X_1}(\tau,\bar\tau)=\sum_{i,j}N_{X_1 i}^{j}
                         \chi_i(\tau)\bar\chi_j(\bar\tau)\,,
 \label{eq:Zg}
\end{align}
where $N_{X_1 i}^{j}$ is the fusion multiplicity and the barred quantities
refer to the antichiral sector. To compare this conformal description with the moment states in the massive IR phase, consider to translate the UV states by one circumference $L$ via the operator $\ee^{\ii L\hat P}$, where $\hat P$ is the momentum operator. At the conformal point,
\begin{align}
 \hat P=\frac{2\pi}{L}(L_0-\bar L_0)\,.
 \notag
\end{align}
A state in $\V_i\otimes\overline{\V}_j$ at left and right descendant
levels $n_L,n_R\in\mathbb Z_{\ge0}$ has spin
$s=h_i-h_j+n_L-n_R$, where $h_i,h_j$ are the primary conformal weights.
The integer descendant levels therefore give the same eigenvalue of
$\ee^{\ii L\hat P}$ throughout each conformal family:
\begin{align}
 \ee^{2\pi\ii s}
 =\ee^{2\pi\ii(h_i-h_j+n_L-n_R)}
 =\ee^{2\pi\ii(h_i-h_j)}\,.
 \notag
\end{align}
Now, since the $\phi_{2,1}$-deformation preserves $X_1$ and commutes with
$\hat P$, the deformed Hilbert space retains these momentum sectors all through along the RG flow. These spin data are RG invariants, providing non-trivial spin selection rules discussed in~\cite{CLSWY}. In our case, for a single kink or particle moving around one circuit crossing the defect in Eq.~\eqref{eq:bc}, we precisely have the fractional momenta in Eq.~\eqref{eq:pquant} matched with the spin sectors $s$ at the conformal point,
\begin{align}
 \ee^{2\pi\ii q_\xi}=\ee^{2\pi\ii s},
 \qquad s\equiv q_\xi\pmod1\,.
 \notag
\end{align}
The character decomposition gives the allowed residues in the full
defect Hilbert space $\mathcal H_{X_1}$, while the junction construction determines how these residues occur within a given multiplet and with what internal
multiplicities.

We now evaluate the character decomposition for the preserved line
$X_1=(1,3)$. Note that a primary $(r,s)$ fusion with $(1,3)$
preserves $r$ and changes the second label $s$ by $0,\pm2$. We thus have
\begin{align}
 Z_{X_1}=\frac{1}{2}\sum_{r=1}^{k}\left(
 \sum_{s=2}^{k}|\chi_{h_{r,s}}|^2+\sum_{s=1}^{k-1}
 \left(\chi_{h_{r,s+2}}\bar\chi_{h_{r,s}}
       +\chi_{h_{r,s}}\bar\chi_{h_{r,s+2}}\right)\right)\,,
 \label{eq:Zgeneral}
\end{align}
where the character subscripts specify the conformal weights.
Eq.~\eqref{eq:kac} then gives
\begin{align}
 h_{r,s+2}-h_{r,s}=-r+\frac{(k+1)(s+1)}{k+2}
 \equiv-\frac{s+1}{k+2}\pmod1\,.
 \label{eq:spindiff}
\end{align}
The allowed momentum residues in $Z_{X_1}$ are consequently $0$ and
$\frac{j}{k+2}$ for $j=2,\ldots,k$.

For the $R_1$ multiplet in the regular vacuum module, the transport matrix
in~Eq.~\eqref{eq:T} realizes these phases with multiplicities encoded by
the characteristic polynomial
\begin{align}
 \det(\lambda-T)=(\lambda-1)^J
  \prod_{j=2}^{k}\left(\lambda-\ee^{\frac{2\pi\ii j}{k+2}}\right)\,,
 \label{eq:charpoly}
\end{align}
where $J=\left\lfloor\frac{k}{2}\right\rfloor$. In the cases of $\mathcal M_{5,6}$ and $\mathcal M_{6,7}$, there are two degenerate channels with zero fractional momentum. We later distinguish them by a $\mathbb Z_2$ eigenvalues
in $\mathcal M_{5,6}$ and by the eigenvalues of the topological operator
associated with $X_1$ in $\mathcal M_{6,7}$. The general construction uses
the tube algebra~\cite{Muger} and is presented in Appendix~\ref{app:dictionary}, where we derive Eq.~\eqref{eq:charpoly} and construct the squared amplitudes $|U_{ab,\xi}|^2$ between the circle and momentum basis states.

\subsection{Explicit momentum channels}

We now construct the momentum basis for the light multiplet $R_1$ in
the four cases discussed above. In each case, the quiver specifies the
circle basis, and diagonalizing $T$ gives its relation to the momentum basis.
We denote the momentum states by $\ket{s}$, choosing fractional
representatives for $s$ modulo integers, and retain $0_A,0_B$ for
degenerate zero momentum channels. We write
$\omega_n=\ee^{\frac{2\pi\ii}{n}}$ for $n=3,5$.
\providecommand{\quivercircleloop}[5][particle loop]{%
  \begingroup
  \pgfmathsetmacro{\quiverLoopOffset}{#3+#5}%
  \pgfmathsetmacro{\quiverLoopStart}{#4+150}%
  \draw[#1]
    ($(#2)+(#4:\quiverLoopOffset)+(\quiverLoopStart:#3)$)
    arc[start angle=\quiverLoopStart,delta angle=-300,radius=#3];%
  \endgroup
}
%

\providecommand{\momentumExampleQuiver}[3]{%
  \begingroup
  \begin{tikzpicture}[x=1cm,y=1cm,
    baseline=-.5ex,
    particle line/.style={draw=cyan!75!black,line width=1.05pt},
    kink arrow/.style={particle line,
      -{Stealth[length=2.6mm,width=2.0mm]}},
    particle loop/.style={kink arrow},
    vacuum/.style={circle,fill=black,inner sep=1.8pt},
    every node/.style={font=\small}
  ]
    \pgfmathtruncatemacro{\momentumLastVacuum}{#1-1}%
    \pgfmathtruncatemacro{\momentumLastEdge}{#1-2}%
    \foreach \a in {0,...,\momentumLastVacuum}{%
      \node[vacuum] (mq\a) at ({#2*\a},0) {};
      \node[anchor=north] at ({#2*\a},-.24)
        {$\lvert\a\rangle$};
    }%
    \foreach \a in {0,...,\momentumLastEdge}{%
      \pgfmathtruncatemacro{\b}{\a+1}%
      \draw[kink arrow] ($(mq\a)+(.14,.04)$)
        .. controls ($(mq\a)+({.28*#2},.28)$)
                 and ($(mq\b)+({-.28*#2},.28)$)
        .. ($(mq\b)+(-.14,.04)$);
      \draw[kink arrow] ($(mq\b)+(-.14,-.04)$)
        .. controls ($(mq\b)+({-.28*#2},-.28)$)
                 and ($(mq\a)+({.28*#2},-.28)$)
        .. ($(mq\a)+(.14,-.04)$);
    }%
    \foreach \vertex/\direction in {#3}{%
      \quivercircleloop[particle loop]{mq\vertex}{.30}{\direction}{.14}%
    }%
  \end{tikzpicture}%
  \endgroup
}

\providecommand{\momentumQuiverMfourfive}{%
  \momentumExampleQuiver{2}{3.4}{1/0}}

\providecommand{\momentumQuiverMfivesixOrdered}{%
  \momentumExampleQuiver{3}{2.8}{1/90}}

\providecommand{\momentumQuiverMfivesixDisordered}{%
  \momentumExampleQuiver{2}{3.4}{0/180,1/0}}

\providecommand{\momentumQuiverMsixseven}{%
  \momentumExampleQuiver{3}{2.8}{1/90,2/0}}

\subsubsection{\texorpdfstring{$\mathcal M_{4,5}+\lambda\,\phi_{2,1}$}{M4,5 + lambda phi2,1}}
\label{sec:3.3}

The $R_1$ multiplet contains two oriented kinks and one particle,
with the quiver
\begin{center}
\momentumQuiverMfourfive
\end{center}
We order the circle basis $\ket{ab}$ as $(01,10,11)$ and write
$\varphi=\frac{1+\sqrt5}{2}$. Eqs.~\eqref{eq:T} and~\eqref{eq:Fmain}
then give
\begin{align}
 T_{\mathcal M_{4,5}}=\begin{pmatrix}
 0&\varphi^{-1}&-\varphi^{-1/2}\\
 1&0&0\\
 0&-\varphi^{-1/2}&-\varphi^{-1}
 \end{pmatrix}\,.
 \label{eq:T45}
\end{align}
The momentum basis has $s=0,+\frac{2}{5},-\frac{2}{5}$ and is
related to the circle basis by
\begin{align}
 \begin{pmatrix}
 \ket{0}\\[7pt]\ket{+\frac{2}{5}}\\[7pt]\ket{-\frac{2}{5}}
 \end{pmatrix}
 &=\frac{1}{\sqrt{2\varphi-1}}
 \begin{pmatrix}
 1&1&-\varphi^{-3/2}\\[7pt]
 \omega_5^2\,\varphi^{-1/2}&\varphi^{-1/2}&\omega_5\\[7pt]
 \omega_5^{-2}\,\varphi^{-1/2}&\varphi^{-1/2}&\omega_5^{-1}
 \end{pmatrix}
 \begin{pmatrix}\ket{01}\\[7pt]\ket{10}\\[7pt]\ket{11}\end{pmatrix}\,.
 \label{eq:circle45}
\end{align}
Conversely,
\begin{align}
 \begin{pmatrix}\ket{01}\\[7pt]\ket{10}\\[7pt]\ket{11}\end{pmatrix}
 &=\frac{1}{\sqrt{2\varphi-1}}
 \begin{pmatrix}
 1&\omega_5^{-2}\,\varphi^{-1/2}&\omega_5^2\,\varphi^{-1/2}\\[7pt]
 1&\varphi^{-1/2}&\varphi^{-1/2}\\[7pt]
 -\varphi^{-3/2}&\omega_5^{-1}&\omega_5
 \end{pmatrix}
 \begin{pmatrix}
 \ket{0}\\[7pt]\ket{+\frac{2}{5}}\\[7pt]\ket{-\frac{2}{5}}
 \end{pmatrix}\,.
 \label{eq:inverse45}
\end{align}
Table~\ref{tab:mixing45} gives the
squared amplitudes. The two kinks have equal weights in
each momentum channel, while their relative phases distinguish
the two orientations. All three momentum channels contain
both kink and neutral particle components. 

\begin{table}[tb]
\centering
\small
\setlength{\tabcolsep}{10pt}
$\displaystyle\frac{1}{2\varphi-1}\;\vcenter{\hbox{%
\begin{tabular}{cccc}
\toprule
$ab$ & $0$ & $+\frac{2}{5}$ & $-\frac{2}{5}$\\
\midrule
$01$ & $1$ & $\varphi^{-1}$ & $\varphi^{-1}$\\
$10$ & $1$ & $\varphi^{-1}$ & $\varphi^{-1}$\\
$11$ & $\varphi^{-3}$ & $1$ & $1$\\
\bottomrule
\end{tabular}}}$
\caption{Theoretical internal weights $|U_{ab,s}|^2$ for the $R_1$
multiplet of $\mathcal M_{4,5}+\lambda\,\phi_{2,1}$. The factor $\displaystyle\frac{1}{2\varphi-1}$ multiplies every entry in the table. The phases are specified in Eq.~\eqref{eq:circle45}.}
\label{tab:mixing45}
\end{table}

The corresponding UV defect partition function $Z_{X_1}$ decomposed in terms of momentum channels are given by
\begin{align}
 Z_{0}=|\chi_{\frac{1}{10}}|^2+|\chi_{\frac{3}{5}}|^2+|\chi_{\frac{3}{80}}|^2\,,\qquad
 Z_{+\frac{2}{5}}&=\chi_0\bar\chi_{\frac{3}{5}}
       +\chi_{\frac{3}{2}}\bar\chi_{\frac{1}{10}}
       +\chi_{\frac{7}{16}}\bar\chi_{\frac{3}{80}},\qquad
 Z_{-\frac{2}{5}}=\overline{Z_{+\frac{2}{5}}}\,.
 \label{eq:Z45}
\end{align}
Complex conjugation here exchanges the chiral and antichiral
characters. In the massive spectrum, the $q=0$ sector contains a defect vacuum and a light excitation belonging to the multiplet $R_1$. The other two members of the multiplet occur in the sectors with fractional momenta $q=\pm 2/5$. Their rest masses can therefore be compared after accounting for the momentum contribution to their energies.

\subsubsection{\texorpdfstring{$\mathcal M_{5,6}+\lambda_{<0}\,\phi_{2,1}$}{M5,6 + lambda phi2,1 (lambda < 0)}: ordered phase}
\label{subsec:M56}

For $\lambda<0$, the three vacua form the regular module. The
$R_1$ quiver contains four oriented kinks and one particle,
\begin{center}
\momentumQuiverMfivesixOrdered
\end{center}
Ordering the circle basis as $(01,10,12,21,11)$, we obtain
\begin{align}
 T_{\mathcal M_{5,6}}^{\rm ord.}=\begin{pmatrix}
 0&\frac{1}{2}&-\frac{1}{2}&0&\frac{1}{\sqrt2}\\
 1&0&0&0&0\\
 0&0&0&1&0\\
 0&-\frac{1}{2}&\frac{1}{2}&0&\frac{1}{\sqrt2}\\
 0&\frac{1}{\sqrt2}&\frac{1}{\sqrt2}&0&0
 \end{pmatrix}\,.
 \label{eq:T56}
\end{align}
The five momentum states are
\begin{align}
 \begin{pmatrix}
 \ket{0_A}\\[7pt]\ket{0_B}\\[7pt]\ket{\frac{1}{2}}\\[7pt]
 \ket{+\frac{1}{3}}\\[7pt]\ket{-\frac{1}{3}}
 \end{pmatrix}
 &=\frac{1}{2\sqrt3}
 \begin{pmatrix}
 \sqrt2&\sqrt2&\sqrt2&\sqrt2&2\\[7pt]
 \sqrt3&\sqrt3&-\sqrt3&-\sqrt3&0\\[7pt]
 -\sqrt3&\sqrt3&-\sqrt3&\sqrt3&0\\[7pt]
 \sqrt2\omega_3&\sqrt2&\sqrt2&\sqrt2\omega_3&2\omega_3^2\\[7pt]
 \sqrt2\omega_3^2&\sqrt2&\sqrt2&\sqrt2\omega_3^2&2\omega_3
 \end{pmatrix}
 \begin{pmatrix}
 \ket{01}\\[7pt]\ket{10}\\[7pt]\ket{12}\\[7pt]\ket{21}\\[7pt]\ket{11}
 \end{pmatrix}\,.
 \label{eq:U56}
\end{align}
Their transport eigenvalues are $1,1,-1,\omega_3,\omega_3^2$ in this
order. The inverse transformation is
\begin{align}
 \begin{pmatrix}
 \ket{01}\\[7pt]\ket{10}\\[7pt]\ket{12}\\[7pt]\ket{21}\\[7pt]\ket{11}
 \end{pmatrix}
 &=\frac{1}{2\sqrt3}
 \begin{pmatrix}
 \sqrt2&\sqrt3&-\sqrt3&\sqrt2\omega_3^2&\sqrt2\omega_3\\[7pt]
 \sqrt2&\sqrt3&\sqrt3&\sqrt2&\sqrt2\\[7pt]
 \sqrt2&-\sqrt3&-\sqrt3&\sqrt2&\sqrt2\\[7pt]
 \sqrt2&-\sqrt3&\sqrt3&\sqrt2\omega_3^2&\sqrt2\omega_3\\[7pt]
 2&0&0&2\omega_3&2\omega_3^2
 \end{pmatrix}
 \begin{pmatrix}
 \ket{0_A}\\[7pt]\ket{0_B}\\[7pt]\ket{\frac{1}{2}}\\[7pt]
 \ket{+\frac{1}{3}}\\[7pt]\ket{-\frac{1}{3}}
 \end{pmatrix}\,.
 \label{eq:inverse56}
\end{align}
Each oriented kink is a superposition of all five momentum
states. The states $\ket{0_B}$ and $\ket{\frac{1}{2}}$ contain only kink
components, while the particle $\ket{11}$ is a superposition of
$\ket{0_A}$ and $\ket{\pm\frac{1}{3}}$.

\begin{table}[tb]
\centering
\small
\setlength{\tabcolsep}{14pt}
\renewcommand{\arraystretch}{1.2}
\begin{tabular}{cccccc}
\toprule
$ab$ & $0_A$ & $0_B$ & $\frac{1}{2}$ & $+\frac{1}{3}$ & $-\frac{1}{3}$\\
\midrule
$01$ & $\frac{1}{6}$ & $\frac{1}{4}$ & $\frac{1}{4}$ & $\frac{1}{6}$ & $\frac{1}{6}$\\
$10$ & $\frac{1}{6}$ & $\frac{1}{4}$ & $\frac{1}{4}$ & $\frac{1}{6}$ & $\frac{1}{6}$\\
$12$ & $\frac{1}{6}$ & $\frac{1}{4}$ & $\frac{1}{4}$ & $\frac{1}{6}$ & $\frac{1}{6}$\\
$21$ & $\frac{1}{6}$ & $\frac{1}{4}$ & $\frac{1}{4}$ & $\frac{1}{6}$ & $\frac{1}{6}$\\
$11$ & $\frac{1}{3}$ & $0$ & $0$ & $\frac{1}{3}$ & $\frac{1}{3}$\\
\bottomrule
\end{tabular}
\caption{Theoretical internal weights $|U_{ab,s}|^2$ for the $R_1$
multiplet in the ordered phase of $\mathcal M_{5,6}+\lambda\,\phi_{2,1}$.
The phases are specified in Eq.~\eqref{eq:U56}.}
\label{tab:mixing}
\end{table}

The UV defect partition function $Z_{X_1}$ associated with these channels are
\begin{align}
 &Z_{0_A}=|\chi_{\frac{2}{3}}|^2+|\chi_{\frac{1}{15}}|^2\,,\qquad
 Z_{0_B}=|\chi_{\frac{1}{8}}|^2+|\chi_{\frac{13}{8}}|^2
              +|\chi_{\frac{1}{40}}|^2+|\chi_{\frac{21}{40}}|^2\,,\notag\\[5pt]
 &Z_{\frac{1}{2}}=\chi_{\frac{1}{8}}\bar\chi_{\frac{13}{8}}
              +\chi_{\frac{13}{8}}\bar\chi_{\frac{1}{8}}
              +\chi_{\frac{1}{40}}\bar\chi_{\frac{21}{40}}
              +\chi_{\frac{21}{40}}\bar\chi_{\frac{1}{40}}\,,\notag\\[5pt]
 &Z_{+\frac{1}{3}}=\chi_0\bar\chi_{\frac{2}{3}}+\chi_3\bar\chi_{\frac{2}{3}}
              +\chi_{\frac{2}{5}}\bar\chi_{\frac{1}{15}}
              +\chi_{\frac{7}{5}}\bar\chi_{\frac{1}{15}},\qquad
 Z_{-\frac{1}{3}}=\overline{Z_{+\frac{1}{3}}}\,.
 \label{eq:Z56}
\end{align}
To determine the symmetry charges of these channels, we consider the
torus partition function with $X_1$ along time and $X_2$ along space.
The spatial line acts on $\mathcal H_{X_1}$ through
its junction with the temporal defect~\cite{CLSWY,Chen:2026cqz}. With the crossing
convention of Appendix~\ref{app:dictionary}, the ultraviolet partition function is
\begin{align}
 Z_{X_1|X_2}
 =\Tr_{\mathcal H_{X_1}}\left(
 X_2\,\mathfrak q^{L_0-\frac{c}{24}}
 \overline{\mathfrak q}^{\bar L_0-\frac{c}{24}}\right)
 =Z_{0_A}-Z_{0_B}-Z_{\frac{1}{2}}
   +Z_{+\frac{1}{3}}+Z_{-\frac{1}{3}}\,.
 \label{eq:Z56_crossed}
\end{align}
The signs give the $\mathbb Z_2$ charges: $0_A$ and
$\pm\frac{1}{3}$ are even, while $0_B$ and $\frac{1}{2}$ are odd.
These charges are preserved by the $\phi_{2,1}$ deformation.
In the ordered phase, the lowest level in $0_B$ is the defect vacuum.
The lowest level in $0_A$ and the next level in $0_B$ are the two
zero momentum members of the light multiplet $R_1$. We measure all
gaps in different momentum channels from the defect vacuum energy. 

\subsubsection{\texorpdfstring{$\mathcal M_{5,6}+\lambda_{>0}\,\phi_{2,1}$}{M5,6 + lambda phi2,1 (lambda > 0)}: disordered phase}
\label{subsec:draft2_m56_disordered_circle}

For $\lambda>0$, the massive phase contains two vacua forming the module $\mathcal M_H$ described
in Section~\ref{subsec:draft2_m56_two_vacua}. The $R_1$ quiver contains
two oriented kinks and one particle above each vacuum,
\begin{center}
\momentumQuiverMfivesixDisordered
\end{center}
We order the circle basis as $(00,01,10,11)$, so that $00,11$ are
particle components and $01,10$ are kink components. The transport
matrix is given in~Eq.~\eqref{eq:Tdisorder},
\begin{align}
    T_{\mathcal M_{5,6}}^{\mathrm{dis.}} =\frac12\begin{pmatrix}
 -1&\sqrt3&0&0\\
 0&0&1&-\sqrt3\\
 \sqrt3&1&0&0\\
 0&0&-\sqrt3&-1
 \end{pmatrix}\,.
\end{align}
with eigenvalues $1,\,-1,\,\omega_3,\,\omega_3^2$. The corresponding momentum basis is
\begin{align}
 \begin{pmatrix}
 \ket{0_B}\\[7pt]\ket{\frac{1}{2}}\\[7pt]\ket{+\frac{1}{3}}\\[7pt]\ket{-\frac{1}{3}}
 \end{pmatrix}
 &=\frac{1}{2\sqrt2}
 \begin{pmatrix}
 1&\sqrt3&\sqrt3&-1\\[7pt]
 \sqrt3&-1&-1&-\sqrt3\\[7pt]
 \sqrt2&\ii\sqrt2&-\ii\sqrt2&\sqrt2\\[7pt]
 \sqrt2&-\ii\sqrt2&\ii\sqrt2&\sqrt2
 \end{pmatrix}
 \begin{pmatrix}\ket{00}\\[7pt]\ket{01}\\[7pt]\ket{10}\\[7pt]\ket{11}\end{pmatrix}\,,
 \label{eq:U56disorder}
\end{align}
and the inverse transformation is
\begin{align}
 \begin{pmatrix}\ket{00}\\[7pt]\ket{01}\\[7pt]\ket{10}\\[7pt]\ket{11}\end{pmatrix}
 &=\frac{1}{2\sqrt2}
 \begin{pmatrix}
 1&\sqrt3&\sqrt2&\sqrt2\\[7pt]
 \sqrt3&-1&-\ii\sqrt2&\ii\sqrt2\\[7pt]
 \sqrt3&-1&\ii\sqrt2&-\ii\sqrt2\\[7pt]
 -1&-\sqrt3&\sqrt2&\sqrt2
 \end{pmatrix}
 \begin{pmatrix}
 \ket{0_B}\\[7pt]\ket{\frac{1}{2}}\\[7pt]\ket{+\frac{1}{3}}\\[7pt]\ket{-\frac{1}{3}}
 \end{pmatrix}\,.
 \label{eq:inverse56disorder}
\end{align}
All four momentum channels mix kink and particle components. Their
squared amplitudes are listed in Table~\ref{tab:mixing56disorder}.
In particular, the channels with $q=\pm\frac{1}{3}$ have equal weights
on all four circle basis states, with opposite relative phases on
the two kink orientations.

We can now explain why this multiplet contains $0_B$ but no $0_A$.
Using the mixed and left module $F$-symbols, together with the crossing
convention of Appendix~\ref{app:dictionary}, we obtain
the action of the spatial $X_2$ line on the circle basis of the
$R_1$ multiplet,
\begin{align}
 X_2\ket{00}=\ket{11}\,,\qquad
 X_2\ket{11}=\ket{00}\,,\qquad
 X_2\ket{01}=-\ket{10}\,,\qquad
 X_2\ket{10}=-\ket{01}\,.
 \label{eq:X2_56disorder_circle}
\end{align}
Thus $X_2$ exchanges the two particle components and the two kink
components. In the momentum basis of~Eq.~\eqref{eq:U56disorder}, this action
is diagonal,
\begin{align}
 X_2\big|_{R_1}=\operatorname{diag}(-1,-1,+1,+1)\,,
 \label{eq:D2_56disorder}
\end{align}
Notice that the transport matrix has only one eigenvector with eigenvalue one,
and its $X_2$ charge is $-1$. Eq.~\eqref{eq:Z56_crossed} therefore
identifies it with $0_B$. The even components of $R_1$ instead have
fractional momenta $\pm\frac{1}{3}$, leaving no $0_A$ member in this
light multiplet. The defect spectrum also contains two vacuum levels at zero momentum,
with $X_2$ charges $+1,-1$. In TCSA these are the lowest levels in
$0_A$ and $0_B$, respectively. Thus the $0_A$ sector is included in the
numerical calculation, but its lowest level is a defect vacuum and does
not enter the light $R_1$ multiplet. To measure this multiplet, we select
the first excited level in $0_B$ and the lowest light levels at
$s=\frac{1}{2},\pm\frac{1}{3}$, measuring all four gaps from the lowest
level in $0_B$. These are the four momentum channels used in
Section~\ref{subsec:m56_disordered_quiver_reconstruction}.

At last, It's worth to mentioning that, although $R_1$ and $R_0\oplus R_2$ have the same quiver adjacency,
the channels with $q=\pm\frac{1}{3}$ occur only for $R_1$. Their presence as we will see in the TCSA defect spectrum therefore distinguishes the irreducible
multiplet from the sum of a particle multiplet and a kink multiplet.

\begin{table}[H]
\centering
\small
\setlength{\tabcolsep}{14pt}
\renewcommand{\arraystretch}{1.2}
\begin{tabular}{ccccc}
\toprule
$ab$ & $0_B$ & $\frac{1}{2}$ & $+\frac{1}{3}$ & $-\frac{1}{3}$\\
\midrule
$00$ & $\frac{1}{8}$ & $\frac{3}{8}$ & $\frac{1}{4}$ & $\frac{1}{4}$\\
$01$ & $\frac{3}{8}$ & $\frac{1}{8}$ & $\frac{1}{4}$ & $\frac{1}{4}$\\
$10$ & $\frac{3}{8}$ & $\frac{1}{8}$ & $\frac{1}{4}$ & $\frac{1}{4}$\\
$11$ & $\frac{1}{8}$ & $\frac{3}{8}$ & $\frac{1}{4}$ & $\frac{1}{4}$\\
\bottomrule
\end{tabular}
\caption{Theoretical internal weights $|U_{ab,s}|^2$ for the $R_1$
multiplet in the disordered phase of $\mathcal M_{5,6}+\lambda\,\phi_{2,1}$. The phases are specified in~Eq.~\eqref{eq:U56disorder}.}
\label{tab:mixing56disorder}
\end{table}

\subsubsection{\texorpdfstring{$\mathcal M_{6,7}+\lambda\,\phi_{2,1}$}{M6,7 + lambda phi2,1}}

The light $R_1$ multiplet contains four oriented kinks and two
particles on the three vacua. Its quiver is
\begin{center}
\momentumQuiverMsixseven
\end{center}
We order the circle basis $\ket{ab}$ as $(01,10,12,21,11,22)$.
The spatial $X_1$ line distinguishes the two zero momentum channels
$0_A$ and $0_B$ by their negative and positive eigenvalues, respectively.
A systematic treatment of such degeneracies is given in
Appendix~\ref{app:dictionary}.

With $d_1=1+2\cos\left(\frac{2\pi}{7}\right)$ and
$d_2=2\cos\left(\frac{\pi}{7}\right)$ the quantum dimensions, the
change of basis can be written as
\begingroup
\footnotesize
\setlength{\arraycolsep}{1pt}
\begin{align}
 \begin{pmatrix}
 \ket{01}\\[7pt]\ket{10}\\[7pt]\ket{12}\\[7pt]\ket{21}\\[7pt]\ket{11}\\[7pt]\ket{22}
 \end{pmatrix}
 &=\sqrt{\frac{3-d_1}{7}}
 \begin{pmatrix}
 \sqrt{d_1}&\frac{d_2}{\sqrt{d_1}}&\ee^{-\frac{2\pi\ii}{7}}&\ee^{\frac{2\pi\ii}{7}}&\sqrt{d_2}\ee^{\frac{4\pi\ii}{7}}&\sqrt{d_2}\ee^{-\frac{4\pi\ii}{7}}\\[7pt]
 \sqrt{d_1}&\frac{d_2}{\sqrt{d_1}}&\ee^{\frac{2\pi\ii}{7}}&\ee^{-\frac{2\pi\ii}{7}}&\sqrt{d_2}\ee^{-\frac{4\pi\ii}{7}}&\sqrt{d_2}\ee^{\frac{4\pi\ii}{7}}\\[7pt]
 -(d_2-1)^{\frac{3}{2}}&\sqrt{2d_2-1}&\sqrt{d_2}\ee^{-\frac{4\pi\ii}{7}}&\sqrt{d_2}\ee^{\frac{4\pi\ii}{7}}&1-\frac{d_2}{d_1}\ee^{\frac{3\pi\ii}{7}}&1-\frac{d_2}{d_1}\ee^{-\frac{3\pi\ii}{7}}\\[7pt]
 -(d_2-1)^{\frac{3}{2}}&\sqrt{2d_2-1}&\sqrt{d_2}\ee^{\frac{4\pi\ii}{7}}&\sqrt{d_2}\ee^{-\frac{4\pi\ii}{7}}&1-\frac{d_2}{d_1}\ee^{-\frac{3\pi\ii}{7}}&1-\frac{d_2}{d_1}\ee^{\frac{3\pi\ii}{7}}\\[7pt]
 -1-\frac{2}{d_1}&\frac{1}{d_1}&\sqrt{d_1}&\sqrt{d_1}&-(d_2-1)^{\frac{3}{2}}&-(d_2-1)^{\frac{3}{2}}\\[7pt]
 \frac{1}{d_1}&-1&\frac{d_2}{\sqrt{d_1}}&\frac{d_2}{\sqrt{d_1}}&\sqrt{2d_2-1}&\sqrt{2d_2-1}
 \end{pmatrix}
 \begin{pmatrix}
 \ket{0_A}\\[7pt]\ket{0_B}\\[7pt]\ket{+\frac{2}{7}}\\[7pt]\ket{-\frac{2}{7}}\\[7pt]\ket{+\frac{3}{7}}\\[7pt]\ket{-\frac{3}{7}}
 \end{pmatrix}\,.
 \label{eq:inverse67}
\end{align}
\endgroup
Table~\ref{tab:m67_theoretical_weights} gives their squared amplitudes. Both zero momentum channels contain kink and particle components, and each also contains a defect vacuum. More generally, for $R_1$ in the regular vacuum module,
Eq.~\eqref{eq:charpoly} gives $k-1$ channels with nonzero fractional
momentum and $J$ channels with zero fractional momentum.

\begin{table}[tb]
\centering
\footnotesize
\setlength{\tabcolsep}{4pt}
\renewcommand{\arraystretch}{1.5}
$\displaystyle\frac{1}{7}\;\vcenter{\hbox{%
\begin{tabular}{ccccccc}
\toprule
$ab$ & $0_A$ & $0_B$ & $+\frac{2}{7}$ & $-\frac{2}{7}$ & $+\frac{3}{7}$ & $-\frac{3}{7}$\\
\midrule
$01$ & $2d_1-d_2-1$ & $2d_1-3d_2+2$ & $3-d_1$ & $3-d_1$ & $2d_2-d_1$ & $2d_2-d_1$\\
$10$ & $2d_1-d_2-1$ & $2d_1-3d_2+2$ & $3-d_1$ & $3-d_1$ & $2d_2-d_1$ & $2d_2-d_1$\\
$12$ & $12d_2-5d_1-10$ & $4d_2-d_1-3$ & $2d_2-d_1$ & $2d_2-d_1$ & $4d_1-10d_2+10$ & $4d_1-10d_2+10$\\
$21$ & $12d_2-5d_1-10$ & $4d_2-d_1-3$ & $2d_2-d_1$ & $2d_2-d_1$ & $4d_1-10d_2+10$ & $4d_1-10d_2+10$\\
$11$ & $7d_1-20d_2+23$ & $6-d_1-2d_2$ & $2d_1-d_2-1$ & $2d_1-d_2-1$ & $12d_2-5d_1-10$ & $12d_2-5d_1-10$\\
$22$ & $6-d_1-2d_2$ & $3-d_1$ & $2d_1-3d_2+2$ & $2d_1-3d_2+2$ & $4d_2-d_1-3$ & $4d_2-d_1-3$\\
\bottomrule
\end{tabular}}}$
\caption{Analytic internal weights $|U_{ab,s}|^2$ for the
$R_1$ multiplet of $\mathcal M_{6,7}+\lambda\,\phi_{2,1}$. The factor $\frac{1}{7}$ multiplies
every entry in the table. The phases are specified in Eq.~\eqref{eq:inverse67}.}
\label{tab:m67_theoretical_weights}
\end{table}

\par

\section{Defect TCSA and measurements of vacuum composition}
\label{sec:numerics}

In TCSA, the spectrum of the deformed Hamiltonian can be numerically computed, in which a state has a definite momentum. In contrast, a physical kink or particle state has definite vacua on its two sides. To compare the two descriptions, in Section~\ref{sec:circle}, we establish the relations between states in momentum and circle basis. In this section, we will introduce the TCSA measurement on the energy of each momentum state in the presence of a topological defect, known as defect TCSA, and the local form factors of a local field on either side of the defect. The energies give the masses in the momentum basis. The form factor distinguishes the adjacent vacua and thus gives access to the squared magnitudes of the circle basis coefficients in Eq.~\eqref{eq:dictionary}.

\subsection{The Hamiltonian and masses in the momentum basis}

We place the deformed theory on a circle of circumference $L$ with a
preserved line $X_i$ along time. For $i=0$, the identity line $X_0$ gives the
periodic Hilbert space, and thus the ordinary TCSA. For the defect $X_1$, Eq.~\eqref{eq:Zg} gives
the left and right Virasoro modules in $\Hc_{X_1}$. In each module we
retain descendants with left and right levels
$n_L,n_R\leq N_{\mathrm{cut}}$. On this truncated space the Hamiltonian is
\begin{align}
 H_{X_i}(L)
 &=\frac{2\pi}{L}\left(L_0+\bar L_0-\frac{c}{12}\right)
   +\lambda L\left(\frac{2\pi}{L}\right)^\Delta
    \mathcal V_0^{(X_i)}\,,
 \label{eq:H}
\end{align}
where $\Delta=2h_{2,1}$ and $\mathcal V_0^{(X_i)}$ is the zero angular
Fourier mode of $\phi_{2,1}$ on the unit circle. We fix the defect $X_i$
and choose $\ket{\alpha},\ket{\beta}$ in an orthonormal conformal basis
of $\Hc_{X_i}$. Their spins are $s_\alpha$
and $s_\beta$, the eigenvalues of $L_0-\bar L_0$. Under a rotation by
$\theta$, the matrix element of the scalar field acquires the phase
$\ee^{\ii(s_\alpha-s_\beta)\theta}$. Since $\phi_{2,1}$ can pass through
$X_i$, this matrix element is periodic in $\theta$. Its zero mode
therefore selects $s_\alpha=s_\beta$, giving
\begin{align}
 \left(\mathcal V_0^{(X_i)}\right)_{\alpha\beta}
 &=\frac{1}{2\pi}\int_0^{2\pi}\dd\theta\,
 \langle\alpha|\phi_{2,1}(\ee^{\ii\theta},\ee^{-\ii\theta})
 |\beta\rangle=\delta_{s_\alpha,s_\beta}\,
 \langle\alpha|\phi_{2,1}(1,1)|\beta\rangle\notag\\[5pt]
 &=\delta_{s_\alpha, s_\beta}\langle\Psi_\alpha^\dagger(\infty)\,\phi_{2,1}(1,1)\,
\Psi_\beta(0)\rangle\,,
 \label{eq:V0_defect}
\end{align}
where we have use the state-operator correspondence to rewrite the matrix element in terms of the three point correlators. For normalized primary states, it equals the coefficient of $\Psi_\alpha$ in the OPE of the bulk field $\phi_{2,1}$
with $\Psi_\beta$. For $X_i=X_0$ this is the usual bulk OPE coefficient.
The left and right Virasoro Ward identities then give the matrix
elements between descendants~\cite{CFTCSA}. We thus diagonalize the Hamiltonian at fixed spin. For an excitation in channel $\xi$, the physical momentum is $p_\xi=\frac{2\pi s_\xi}{L}$, with $s_\xi=q_\xi\pmod 1$ as in Eq.~\eqref{eq:pquant}.

For a single kink or particle, the dispersion relation in
Eq.~\eqref{eq:dispersion} gives a finite volume estimate of its rest mass,
\begin{align}
 m_\xi(L,N_{\mathrm{cut}})
 &=\sqrt{\left(E^{(X_i)}_\xi(L,N_{\mathrm{cut}})
       -E_0^{(X_i)}(L,N_{\mathrm{cut}})\right)^2
       -\left(\frac{2\pi s_\xi}{L}\right)^2}\,,
 \label{eq:mest}
\end{align}
where $E_0^{(X_i)}$ is a zero momentum vacuum energy in the same defect
Hilbert space. At fixed circumference $L$, we use several values of $N_{\mathrm{cut}}$ to estimate the channel mass in the limit $N_{\mathrm{cut}}\to\infty$, denoted by $m_\xi^{(\infty)}(L)$~\cite{Feverati:2006ni,Giokas:2011ix,Lencses:2014tba}.
Appendix~\ref{app:numerics} gives the cutoff fits for each model.


\subsection{Local form factors and vacuum composition}
\label{sec:projections}

On the $X_1$ defect circle, we can determine the kink and particle
composition of a momentum state $\ket{\xi}$ by measuring the vacua on
both sides of the defect. Within the $R_1$ multiplet, each
allowed pair of vacua $(a,b)$ identifies a unique circle basis state
$\ket{ab}$. It is an oriented kink from $a$ to $b$ for $a\ne b$, or a
particle above vacuum $a$ for $a=b$. If an observable distinguishes
$\ket{ab}$ by an eigenvalue unique to this state within the multiplet,
we can construct a projector $\Pi_{ab}$ for that eigenvalue to select the target state $\ket{ab}$.
Its expectation value in the momentum state $\ket{\xi}$ then gives the squared
magnitude of the coefficient of $\ket{ab}$ in Eq.~\eqref{eq:dictionary},
\begin{align}
 \Pi_{ab}=\ket{ab}\bra{ab}\,,\qquad
 w_{ab,\xi}=\bra{\xi}\Pi_{ab}\ket{\xi}
 =|\langle ab|\xi\rangle|^2=|U_{ab,\xi}|^2\,.
 \label{eq:idealweights}
\end{align}
The local scalar field $\mathcal O=\phi_{2,2}$ has distinct expectation
values in the massive vacua, and thus is a desired probe. We use its finite volume matrix elements to
identify the vacuum on each side and thus measure the probabilities in
Eq.~\eqref{eq:idealweights}.

\paragraph{Local probe on an interval.}
We read the probe operator over a short interval $I$ lying on one side of the
defect. Averaging over $I$ preserves a constant vacuum expectation value
and suppresses contributions that oscillate rapidly across the interval.
At the chiral cutoff $N_{\mathrm{cut}}$ the probe is
\begin{align}
 \mathcal O_I=\frac{1}{|I|}\int_I\dd x\,
 \ee^{\ii \hat Px}\,\mathcal O_{\rm cyl}(0)\,\ee^{-\ii \hat Px}\,,
 \label{eq:averagedprobe}
\end{align}
where $\mathcal O_{\rm cyl}(0)$ is the probe operator at $x=0$ on
the cylinder at fixed time. Write $u=\frac{x}{L}$ and denote the center
and width of $I$ in this coordinate by $c_I$ and $\ell_I$, respectively. We compute the point insertion
$V_{\alpha\beta}=\langle\alpha|\mathcal O(1,1)|\beta\rangle$ on the $z$-plane
in the orthonormal conformal basis used in Eq.~\eqref{eq:V0_defect}.
At $x=0$ the map to the cylinder gives the factor
$\left(\frac{2\pi}{L}\right)^{2h}$, where the chiral weights of
$\phi_{2,2}$ are $h=\bar h=\frac{3}{80},\frac{1}{40},\frac{1}{56}$ in
$\mathcal M_{4,5}$, $\mathcal M_{5,6}$ and $\mathcal M_{6,7}$, respectively.
For basis states of spins $s_\alpha,s_\beta$, translation to $x$ provides an extra phase $\ee^{2\pi\ii\Delta s\,u}$ with $\Delta s=s_\alpha-s_\beta$.
The interval integral in Eq.~\eqref{eq:averagedprobe} then gives
\begin{align}
 (\mathcal O_I)_{\alpha\beta}
 &=\left(\frac{2\pi}{L}\right)^{2h}V_{\alpha\beta}
   \frac{1}{\ell_I}\int_{c_I-\frac{\ell_I}{2}}^{c_I+\frac{\ell_I}{2}}
       \dd u\,\ee^{2\pi\ii\Delta s\,u}\notag\\
 &=\left(\frac{2\pi}{L}\right)^{2h}V_{\alpha\beta}\,
   \underbrace{\ee^{2\pi\ii\Delta s\,c_I}}_{\text{probe position}}\,
   \underbrace{\operatorname{sinc}(\pi\Delta s\,\ell_I)}_{\text{probe width}}\,,
 \label{eq:probematrix}
\end{align}
where $\operatorname{sinc}(z)=\frac{\sin z}{z}$. Eq.~\eqref{eq:probematrix} gives
the full probe matrix by multiplying each matrix element by its
interval factor. For two given normalized TCSA eigenvectors $v_m$ and $v_n$, the finite volume form factor of the averaged
field is $v_m^\dagger\mathcal O_I\,v_n$~\cite{CFTCSA,PozsgayTakacs2008FF,Pozsgay:2007gx}.

\paragraph{Reference vacua and vacuum projectors.}
We first calibrate the operator values on the periodic circle at the same
$L$ and $N_{\mathrm{cut}}$. Let $\ket{e_j}$ be the low lying energy
eigenstates that span the vacuum subspace. At finite volume these states
can be superpositions of the massive vacua. In the infinite volume limit
a local field has vanishing matrix elements between distinct vacua, and
its diagonal entries are the corresponding vacuum expectation values.
We therefore choose the reference vacua by diagonalizing the probe
within this subspace,
\begin{align}
 (O_{\rm vac})_{jk}=\bra{e_j}\mathcal O_I\ket{e_k}\,,\qquad
 O_{\rm vac} u^{(\nu)}=\mu_\nu u^{(\nu)}\,,\qquad
 \ket{\Omega_\nu}=\sum_j u_j^{(\nu)}\ket{e_j}\,,
 \label{eq:Ovac}
\end{align}
where $\nu=1,\ldots,n$ orders the $n$ reference vacua by increasing
field value $\mu_\nu$. All the $\ket{e_j}$
have zero momentum, so the phase and sinc in Eq.~\eqref{eq:probematrix}
are both one in $O_{\rm vac}$.

To obtain the distribution of probe readings in each reference vacuum, we
diagonalize $\mathcal O_I$ in the full truncated periodic Hilbert space.
Let $\ket{r}$ be its orthonormal eigenstates with eigenvalues $g_r$.
The probability of reading $g$ in $\ket{\Omega_\nu}$ is
\begin{align}
 P(g\mid\Omega_\nu)=\sum_{r:\,g_r=g}
       \bigl|\langle r\vert\Omega_\nu\rangle\bigr|^2\,.
\label{eq:Prob_g}
\end{align}
For two vacua, denote the reference states in order of increasing field value
by $\ket{\Omega_-}$ and $\ket{\Omega_+}$. A threshold $t$ assigns readings
$g<t$ to the first vacuum and $g\geq t$ to the second. Giving the two vacua
equal weight, the probability of a correct assignment is
\begin{align}
 A(t)=\frac{1}{2}\left(P(g<t\mid\Omega_-)+P(g\geq t\mid\Omega_+)\right)\,.
\label{eq:At}
\end{align}
The first term sums the probabilities below $t$ in $\ket{\Omega_-}$,
and the second sums those at or above $t$ in $\ket{\Omega_+}$.
Between successive distinct probe eigenvalues these assignments stay fixed.
We therefore evaluate $A(t)$ at the midpoint of each gap and select a
threshold that maximizes it. For $n$ vacua the $n-1$ ordered thresholds
divide the readings into intervals $\mathcal B_{(\nu)}$. We maximize
$\frac{1}{n}\sum_\nu P(g\in\mathcal B_{(\nu)}\mid\Omega_\nu)$ by the same rule,
assigning a reading at a threshold to the upper interval.
Among equally accurate choices we take the thresholds closest in squared
distance to the midpoints of the adjacent reference values. If a tie remains,
we take the lower threshold at the first boundary where the choices differ.
For example, in the disordered phase
of $\mathcal M_{5,6}$ at $L=60$ and $N_{\mathrm{cut}}=5$, the two
reference values are approximately $-0.898$ and $+0.898$. The rule above
selects the threshold $-0.227775$ and correctly identifies the reference
vacuum with mean probability $98.36\%$.

To identify the quiver vertex $a$ corresponding to each calibrated
interval, we evaluate a spatial $X_1$ line between the
same periodic reference vacua. We compare the magnitudes
$\bigl|\bra{\Omega_\nu}X_1\ket{\Omega_{\nu'}}\bigr|$ with the entries
of the vacuum action matrix $n_1$ defined in Eq.~\eqref{eq:nimrep} to find the
corresponding ordering of the vertices. If exchanging two vertices
leaves $n_1$ unchanged, we assign the smaller
vertex label to the lower reference operator value. In the disordered
$\mathcal M_{5,6}$ example, $X_1$ sends either vacuum to $v_0\oplus v_1$,
so the lower interval is labelled $0$ and the upper interval $1$.
We write $\mathcal B_a$ for the calibrated interval carrying vertex
label $a$. Section~\ref{sec:quiver_reconstruction} gives the assignments for all four cases. We then use these intervals on the defect circle. With the defect at $x=0$,
we take two intervals $I_L,I_R$ of equal width $\ell L$, with centers at
$c_L=\epsilon+\frac{\ell}{2}$ and $c_R=1-c_L$ in units of $L$. The
nearer end of each interval is a distance $\epsilon L$ from the defect.
For $j=L,R$, let $\ket{r,j}$ be an eigenstate of $\mathcal O_{I_j}$
with eigenvalue $g_r^j$. The projector onto a reading in $\mathcal B_a$
on side $j$ is
\begin{align}
 Q_a^j=\mathbf 1_{\mathcal B_a}(\mathcal O_{I_j})
     =\sum_{r:\,g_r^j\in\mathcal B_a}\ket{r,j}\bra{r,j}\,.
 \label{eq:localprojectors}
\end{align}
We construct these spectral projectors in the full truncated defect
space before evaluating them on the states of the $R_1$ multiplet.

\paragraph{Joint measurement and squared amplitudes.}
Now let $v_\xi$ be the normalized TCSA eigenvector of a state in
momentum channel $\xi$. Reading $a$ on the left and then $b$ on the
right gives probability $\|Q_b^R Q_a^L v_\xi\|^2$. Keeping the matrix
elements between all selected states gives
\begin{align}
 (E_{ab}^{L\to R})_{\xi\zeta}
 &=(Q_b^R Q_a^L v_\xi)^\dagger(Q_b^R Q_a^L v_\zeta)
   =v_\xi^\dagger Q_a^L Q_b^R Q_a^L v_\zeta\,.
 \label{eq:jointmeasurement}
\end{align}
Since the left and right projectors may fail to commute at finite cutoff,
we also evaluate the squared amplitude in a reverse order, for which the product in
Eq.~\eqref{eq:jointmeasurement} is $Q_b^R Q_a^L Q_b^R$, and take the mean value
\begin{align}
 E_{ab}=\frac{1}{2}\left(E_{ab}^{L\to R}+E_{ab}^{R\to L}\right)\,.
 \label{eq:coherentmeasurement}
\end{align}
The diagonal entry $(E_{ab})_{\xi\xi}$ gives the probability of
reading $(a,b)$ in the momentum state $\ket{\xi}$. To recover the quiver, we sum these measured entries over a complete momentum basis of the multiplet before normalizing any row.
In an ideal measurement, $\sum_\xi(E_{ab})_{\xi\xi}$ is the trace of the projector onto the components with vacuum pair $(a,b)$ and hence counts them. For a pair with a single closing junction, this gives the multiplicity of the arrow $a\to b$.
The entries with
$\xi\ne\zeta$ measure the interference between two momentum states
for that same outcome. For the ideal projector in Eq.~\eqref{eq:idealweights},
$\bra{\xi}\Pi_{ab}\ket{\zeta}=U_{ab,\xi}^*U_{ab,\zeta}$, so each
diagonal entry gives the weight $|U_{ab,\xi}|^2$ directly.

The probes resolve the two vacua but leave the position of the
excitation undetermined. We model its position as uniformly distributed
between the probe centers. Taking the midpoint $u=\frac{1}{2}$ as
the phase origin, we average the relative translation phase of two
channels with spins $s_\xi$ and $s_\zeta$ over this interval,
\begin{align}
 K_{\xi\zeta}
 &=\frac{1}{d}\int_{c_L}^{c_R}\dd u\,
 \ee^{2\pi\ii(s_\zeta-s_\xi)\left(u-\frac{1}{2}\right)}
 =\operatorname{sinc}\!\left(\pi d(s_\zeta-s_\xi)\right)\,,
 \qquad d=c_R-c_L\,.
 \label{eq:positionkernel}
\end{align}
This average suppresses interference only between channels of unequal
spin. Since the probes distinguish the vacua with finite accuracy,
each diagonal probability only approximates the corresponding weight.
Interference between two channels probes the product of their
amplitudes and thus further constrains their magnitudes. For each vacuum
pair $(a,b)$ we therefore fit one complex amplitude $C_{ab,\xi}$ per
momentum channel to the full measured matrix,
\begin{align}
 (E_{ab})_{\xi\zeta}\simeq K_{\xi\zeta}\,C_{ab,\xi}^*C_{ab,\zeta}\,,
 \label{eq:amplitudemodel}
\end{align}
by minimizing
\begin{align}
 \sum_{\xi,\zeta}\left|(E_{ab})_{\xi\zeta}
              -K_{\xi\zeta}\,C_{ab,\xi}^*C_{ab,\zeta}\right|^2\,.
 \label{eq:amplitudefit}
\end{align}
Normalizing this row over the momentum channels removes a common
response factor for the given vacuum pair. The resulting estimate for
$|U_{ab,\xi}|^2$ is
\begin{align}
 \widehat w_{ab,\xi}
 =\frac{|C_{ab,\xi}|^2}{\displaystyle\sum_\zeta |C_{ab,\zeta}|^2}\,.
 \label{eq:reconstructedweights}
\end{align}
We use $N_{\mathrm{cut}}=5$, $\epsilon=0.005$ and $\ell=0.02$, giving
$d=0.97$. The circumference is $L=100$ for $\mathcal M_{4,5}$ and
$L=60$ for $\mathcal M_{5,6}$ and $\mathcal M_{6,7}$. Each column $\xi$
of the resulting table gives the composition of one momentum state. 


\subsection{Kink and particle masses and their identifications}

To estimate the mass of each kink or particle component, we combine the
weights $\widehat w_{ab,\xi}$ measured in Section~\ref{sec:projections}
with the masses extracted from the spectrum in each momentum channel.
In the momentum basis the squared masses are diagonal.
Changing to the circle basis with Eq.~\eqref{eq:dictionary} gives the
diagonal matrix element for each kink or particle component. However, we here will only rely on numerical measurements that are independent of any theoretical predictions. So we replace the exact weights by the fitted values, and estimate the mass of a kink or a particle by
\begin{align}
 M_{ab}
 &=\left(\sum_\xi\widehat w_{ab,\xi}
                \left(m_\xi^{(\infty)}\right)^2\right)^{1/2}\,.
 \label{eq:vacuumweightmass}
\end{align}
Here $m_\xi^{(\infty)}$ is evaluated at the volume or averaged over the
volumes specified in Appendix~\ref{app:numerics}.
Ideally, for a multiplet with a common rest mass, every row gives the same
$M_{ab}$. At finite volume the spread of these estimates records the
residual splitting of the momentum channels as seen by each component.

To identify neutral particles in the spectrum with periodic boundary
conditions, we measure how strongly a local scalar operator $\mathcal O$
connects a normalized excitation $\ket{n,L}$ to the vacuum subspace.
Its energy gap is $\Delta E_n(L)=E_n(L)-E_{\rm vac}(L)$, where $E_n(L)$
is the energy of the excitation and $E_{\rm vac}(L)$ is the energy of
the reference vacuum at the same circumference $L$.
On the periodic circle, the vacuum subspace is spanned
by the low energy states that become degenerate vacua as $L$ grows.
For an orthonormal basis $\ket{\Omega_a(L)}$ of this subspace, we sum
the squared form factors,
\begin{align}
 |F_{n,L}|^2
 &=\sum_a\bigl|\bra{\Omega_a(L)}\mathcal O(0)\ket{n,L}\bigr|^2\,.
 \notag
\end{align}
This is the squared norm of the projection of $\mathcal O(0)\ket{n,L}$
onto the vacuum subspace and is therefore independent of the choice
of basis. In $\mathcal M_{4,5}$ the probe $\phi_{2,1}$ preserves the
two symmetry sectors, each of which contains one vacuum. In a basis of
states with definite symmetry charge, only the vacuum in the same
sector as $\ket{n,L}$ can contribute. In $\mathcal M_{5,6}$ and
$\mathcal M_{6,7}$ we use $\phi_{2,2}$. When the vacuum levels are split, we use vacuum energy eigenstates and
weight each squared matrix element by $L\bigl(E_n(L)-E_{\Omega_a}(L)\bigr)$
before summing, where $E_{\Omega_a}(L)$ is the energy of $\ket{\Omega_a(L)}$.
The weighted sum is
\begin{align}
 \mathcal S_n(L)=L\sum_a\bigl(E_n(L)-E_{\Omega_a}(L)\bigr)
 \bigl|\bra{\Omega_a(L)}\mathcal O(0)\ket{n,L}\bigr|^2\,.
 \label{eq:periodic_spectral_weight}
\end{align}
When the splitting is negligible, this reduces to $L\Delta E_n|F_{n,L}|^2$. If the probe couples to the particle, finite volume normalization gives $|F_{n,L}|^2\propto\frac{1}{L\Delta E_n}$,
so $L\Delta E_n|F_{n,L}|^2$ approaches a nonzero constant~\cite{PozsgayTakacs2008FF}.
For a two particle scattering level with fixed quantum numbers, the
particles carry opposite momenta of order $\frac{1}{L}$. Its gap
therefore approaches the sum of their rest masses with a correction
of order $\frac{1}{L^2}$~\cite{Klassen:1992qy}. Each particle contributes a factor of
$\frac{1}{\sqrt{L}}$ to the finite volume matrix element~\cite{PozsgayTakacs2008FF}.
If the probe has a nonzero two particle form factor at threshold,
$|F_{n,L}|^2$ falls as $\frac{1}{L^2}$ and the same quantity
$L\Delta E_n|F_{n,L}|^2$ falls as $\frac{1}{L}$. These asymptotic behaviors distinguish a one particle state from a two particle scattering level.

\FloatBarrier
\section{Quiver reconstruction from defect TCSA}
\label{sec:quiver_reconstruction}
In this section, we will use the defect TCSA developed in previous section to reconstruct the quivers of massive multiplets in the $\phi_{2,1}$-deformed theories. The reconstruction can be performed in the following three steps: First, in the ordinary periodic spectrum, we identify the nearly degenerate vacua. We distinguish them using the local operator $\phi_{2,2}$ and assign their vertex labels by Eq.~\eqref{eq:Ovac} and \eqref{eq:At}. Second, in the defect spectrum, we select the momentum states $\ket{\xi}$
with the degenerate rest mass. In each state we use the calibrated probes to measure the
probability $(E_{ab})_{\xi\xi}$ of finding vacuum $a$ on the left and $b$
on the right. The full matrix $E_{ab}$ also gives an estimate of the squared
overlaps $|U_{ab,\xi}|^2$ between the momentum and circle bases, as described
in Section~\ref{sec:projections}. Thirdly, for each ordered pair $(a,b)$, we sum
$(E_{ab})_{\xi\xi}$ over the complete set of momentum states of the
multiplet. For a pair vacua that the defect can close
in one way, a sum $\sum_\xi (E_{ab})_{\xi\xi}$ near one indicates an arrow $a\to b$, and a sum near zero
gives no arrow in that direction. A loop represents a particle in vacuum $a$,
and an arrow between distinct vacua represents an oriented kink.
Finally, to check the reconstruction, we count one state for each way the defect can
close a kink or particle on the quiver. This count should match the total number
of branches at the same mass in the defect spectrum.
The periodic spectrum also independently checks the total number of particle loops.

\subsection{\texorpdfstring{$\mathcal M_{4,5}+\lambda\,\phi_{2,1}$}{M4,5 + lambda phi2,1}}
\label{subsec:m45_quiver_reconstruction}

\begin{figure}[!htbp]
\centering
\includegraphics[width=0.65\linewidth]{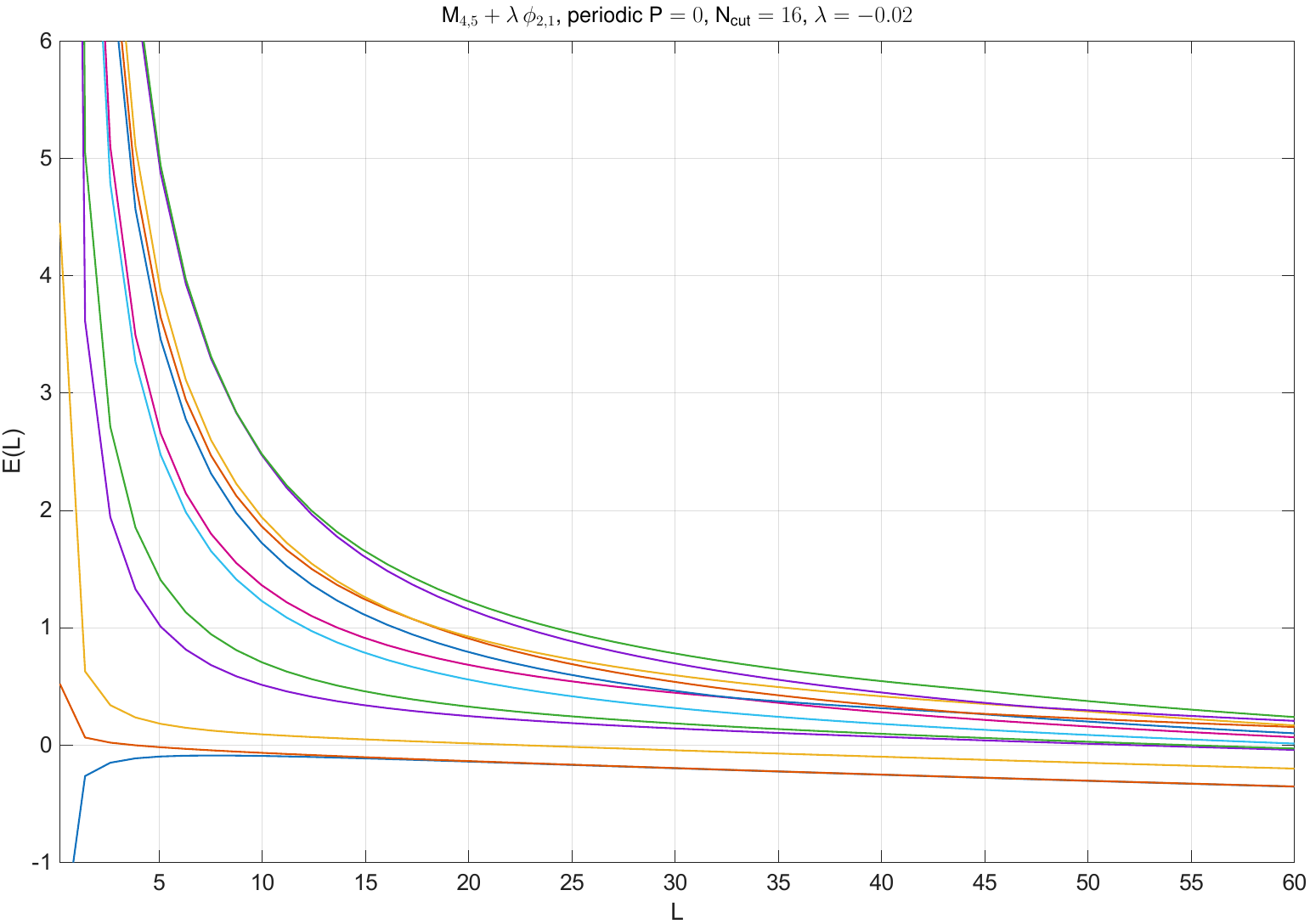}
\par\medskip
\begingroup\small
\setlength{\tabcolsep}{6pt}
\begin{tabular}{@{}rrrrr@{}}
\toprule
$L$ & $E_1-E_0$ & $E_2-E_0$ & $E_3-E_0$ & $E_4-E_0$\\
\midrule
20 & 0.00427 & 0.15546 & 0.38766 & 0.46817\\
40 & 0.00084 & 0.15308 & 0.32282 & 0.34839\\
60 & 0.00133 & 0.15388 & 0.31317 & 0.32500\\
\bottomrule
\end{tabular}
\endgroup
\caption{Periodic spectrum of $\mathcal M_{4,5}+\lambda\,\phi_{2,1}$ at zero momentum, $\lambda=-0.02$ and $N_{\mathrm{cut}}=16$. The curves show the periodic energies $E_n(L)$. The table gives $E_n(L)-E_0(L)$ from a separate TCSA diagonalization at $L=20,40,60$. The first gap measures the splitting of the two vacua.}
\label{fig:m45_recon_periodic}
\end{figure}

\paragraph{Periodic spectrum and vacuum labels.}
We take $\lambda=-0.02$ and order the energies at zero momentum as $E_0,E_1,\ldots$.
Figure~\ref{fig:m45_recon_periodic} shows the periodic energies $E_n(L)$ and tabulates their gaps above the ground state $E_0(L)$.
At the two largest circumferences in the table, the splitting $E_1-E_0$ is less than one percent of $E_2-E_0$. We identify the lowest pair as the two massive vacua.
The gap $E_2-E_0$ stays close to $0.154$, while $E_3-E_0$ and $E_4-E_0$ decrease towards twice this value. Thus the periodic spectrum indicates one neutral particle at this mass. Its location on the quiver remains to be determined.

We label the vacua at $L=100$ and $N_{\mathrm{cut}}=5$, where 1020 states are collected to measure the form factors below. In the basis of the two lowest periodic energy eigenstates, ordered by $E_0,E_1$, the interval average of $\phi_{2,2}$ gives
\begin{align}
 O_{\rm vac}\simeq
 \begin{pmatrix}0&0.81582\\0.81582&0.81629\end{pmatrix}
 \,,\qquad \operatorname{spec}(O_{\rm vac})\simeq\{-0.50407,\,1.32036\}\,.
 \notag
\end{align}
The two corresponding eigenvectors $\ket{\Omega_\mp}$ are our reference vacua. To assign their quiver labels, we measure the VEVs of $X_1$ respect to them. With proper relative phases chosen, we find
\begin{align}
 X_1\big|_{\rm vac}\simeq
 \begin{pmatrix}1.00023&0.99988\\0.99988&-0.00023\end{pmatrix}
 \,,\qquad \operatorname{spec}\!\left(X_1\big|_{\rm vac}\right)
 \simeq\left\{\varphi,\,-\varphi^{-1}\right\}\,.
 \notag
\end{align}
Each column gives the amplitudes for the indicated initial vacuum to end in the two reference vacua. The first column gives both vacua with coefficients close to one, while the second gives only the first vacuum to this accuracy. Comparing with $X_1\otimes X_1=X_0\oplus X_1$ and $X_1\otimes X_0=X_1$, we assign the lower operator value to vacuum $1$ and the higher value to vacuum $0$. We then use the approach discussed in Section~\ref{sec:projections} to identify a threshold at $0.33002$ distinguishing the two vacua with a mean accuracy of $99.04\%$.

\begin{figure}[!htbp]
\centering
\includegraphics[width=0.96\linewidth]{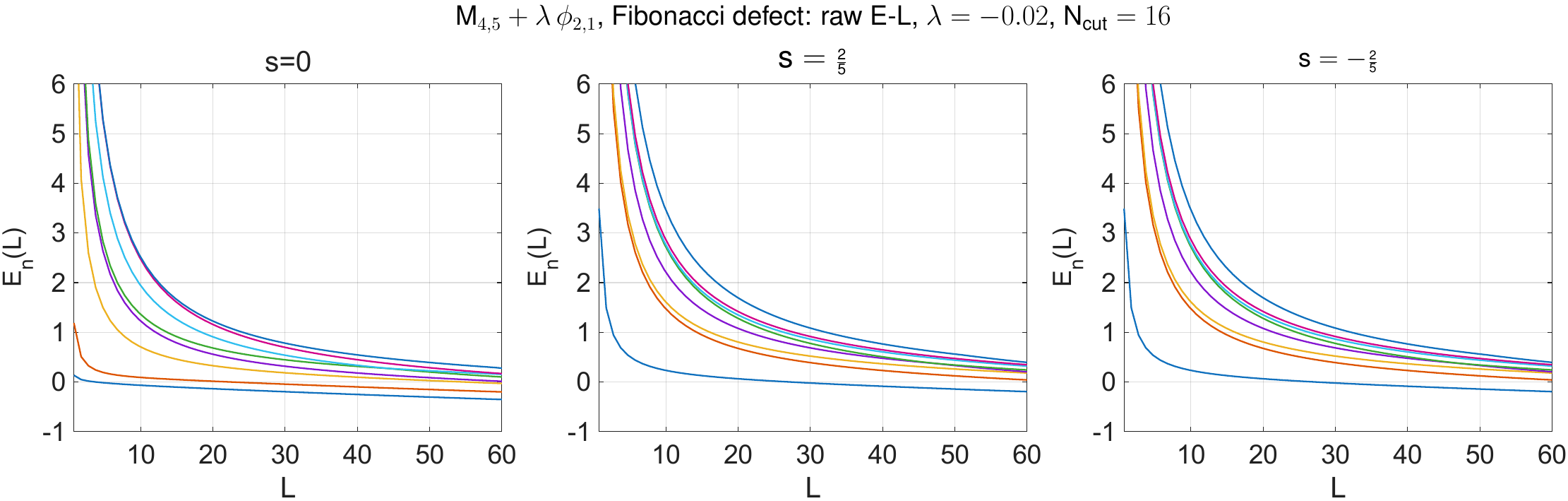}
\caption{Defect spectrum of $\mathcal M_{4,5}+\lambda\,\phi_{2,1}$ on the $X_1$ circle at $\lambda=-0.02$ and $N_{\mathrm{cut}}=16$. From left to right the panels show $s=0,+\frac{2}{5},-\frac{2}{5}$. We select the first excited level in the $s=0$ panel and the lowest level in each of the two fractional momentum panels.}
\label{fig:m45_recon_defect}
\end{figure}

\paragraph{The three states on the defect circle.}
On inserting $X_1=\mathcal L_{(1,3)}$, there are three fractional momentum channels, $s=0,+\frac{2}{5},-\frac{2}{5}$ from defect TCSA as in Eq.~\eqref{eq:Z45}.
The lowest state at $s=0$ is the defect vacuum. We select the first excited state at $s=0$ and the lowest state in each of the two fractional channels in Figure~\ref{fig:m45_recon_defect}.
We measure their gaps from the same defect vacuum and remove the momentum $p=\frac{2\pi s}{L}$ using Eq.~\eqref{eq:mest}. At $L=60$ and $N_{\mathrm{cut}}=16$, this gives $m_0=0.15255$ and $m_{+\frac{2}{5}}=m_{-\frac{2}{5}}=0.15302$, where the subscripts denote the fractional channels.
Using $N_{\mathrm{cut}}=10,12,14,16$ to extrapolate the gaps at fixed $L$ and averaging the resulting masses over the crossover region gives $0.15214$, $0.15213$ and $0.15213$, respectively. The three branches thus have a common rest mass within the numerical resolution.

\paragraph{From vacuum pairs to a quiver.}
We now measure the three degenerate states at $L=100$ and $N_{\mathrm{cut}}=5$, using the periodic calibration above for both $\phi_{2,2}$ probes.
For each momentum state $\ket{s}$, Table~\ref{tab:m45_recon_raw} gives the probability $(E_{ab})_{ss}$ of reading vacuum $a$ on the left and $b$ on the right, averaged over the two measurement orders as in Eq.~\eqref{eq:coherentmeasurement}.
The four vacuum pairs exhaust the possible probe readings in each momentum state, so their measured probabilities already satisfy $\sum_{a,b}(E_{ab})_{ss}=1$. To determine the arrows, we sum across the three states for each fixed pair along rows.

\begin{table}[!htbp]
\centering\small
\setlength{\tabcolsep}{9pt}
\begin{tabular}{crrrr}
\toprule
$ab$ & $s=0$ & $s=+\frac{2}{5}$ & $s=-\frac{2}{5}$ & $\sum_s(E_{ab})_{ss}$\\
\midrule
$00$ & 0.02610 & 0.02283 & 0.02283 & 0.07176\\
$01$ & 0.42867 & 0.26784 & 0.26784 & 0.96435\\
$10$ & 0.42867 & 0.26784 & 0.26784 & 0.96435\\
$11$ & 0.11656 & 0.44149 & 0.44149 & 0.99954\\
\bottomrule
\end{tabular}
\caption{Measured vacuum pair probabilities on the $X_1$ circle in $\mathcal M_{4,5}+\lambda\,\phi_{2,1}$ at $\lambda=-0.02$, $L=100$ and $N_{\mathrm{cut}}=5$. The last column uses the unrounded probabilities.}
\label{tab:m45_recon_raw}
\end{table}

The sums for $01,10,11$ are each close to one. Each pair has one closing junction on $X_1$, so we draw the kink pair $0\rightleftarrows1$ and a particle loop at $1$.
The summed reading for $00$ around $0.07$ measures leakage of the probes, indicating that there is no particle states above the vacuum $0$. It is also consistent with the defect circle construction that the defect $X_1$ admits no $00$ junction. Finally, from the periodic spectra, one can only read off one particle state, also implying that the only particle state must be supported above the vacuum $1$. Overall, we reconstruct the well-known Fibonacci quiver below,
\begin{align}
 Q=\begin{pmatrix}0&1\\1&1\end{pmatrix}
 \qquad &\Longrightarrow \qquad \text{\momentumQuiverMfourfive}
 \notag
\end{align}

\needspace{5\baselineskip}
\paragraph{Comparison with the theoretical squared amplitudes.}
Table~\ref{tab:m45_recon_weights} compares the normalized TCSA weights
$\widehat w_{ab,s}$ from Eq.~\eqref{eq:reconstructedweights}
with the theoretical squared overlaps $|U_{ab,s}|^2$ in Table~\ref{tab:mixing45}.

\begin{table}[!htbp]
\centering\small
\setlength{\tabcolsep}{6pt}
\begin{tabular}{crrr@{\hspace{1.6em}}rrr}
\toprule
& \multicolumn{3}{c}{TCSA $\widehat w_{ab,s}$} & \multicolumn{3}{c}{Theory $|U_{ab,s}|^2$}\\
\cmidrule(lr){2-4}\cmidrule(lr){5-7}
$ab$ & $s=0$ & $s=+\frac{2}{5}$ & $s=-\frac{2}{5}$
     & $s=0$ & $s=+\frac{2}{5}$ & $s=-\frac{2}{5}$\\
\midrule
$01$ & 0.44431 & 0.27785 & 0.27785 & 0.44721 & 0.27639 & 0.27639\\
$10$ & 0.44431 & 0.27785 & 0.27785 & 0.44721 & 0.27639 & 0.27639\\
$11$ & 0.10563 & 0.44718 & 0.44718 & 0.10557 & 0.44721 & 0.44721\\
\bottomrule
\end{tabular}
\caption{Squared amplitudes $|U_{ab,s}|^2$ on the $X_1$ circle in
$\mathcal M_{4,5}+\lambda\,\phi_{2,1}$.
The TCSA data are taken at $\lambda=-0.02$, $L=100$ and $N_{\mathrm{cut}}=5$.}
\label{tab:m45_recon_weights}
\end{table}

The maximum and root mean square absolute deviations from the theoretical weights are 0.291 and 0.168 percentage points, respectively. We test numerical stability by repeating the full vacuum calibration, measurement and fit at $N_{\mathrm{cut}}=4$, by fitting the two probe orders separately, and by moving the intervals from $\epsilon=0.005$ to $0.0025$ or $0.01$ with their width fixed at $\ell=0.02$. The position kernel in Eq.~\eqref{eq:positionkernel} is updated for each geometry. The largest changes from the tabulated weights are 0.168, 0.050 and 0.101 percentage points, respectively. These deterministic variations measure cutoff, probe order and probe position sensitivity at the stated volume and interval width.
\FloatBarrier

\needspace{8\baselineskip}
\subsection{\texorpdfstring{$\mathcal M_{5,6}+\lambda_{<0}\,\phi_{2,1}$}{M5,6 + lambda phi2,1}: ordered phase}
\label{subsec:m56_ordered_quiver_reconstruction}

\begin{figure}[!htbp]
\centering
\includegraphics[width=0.65\linewidth]{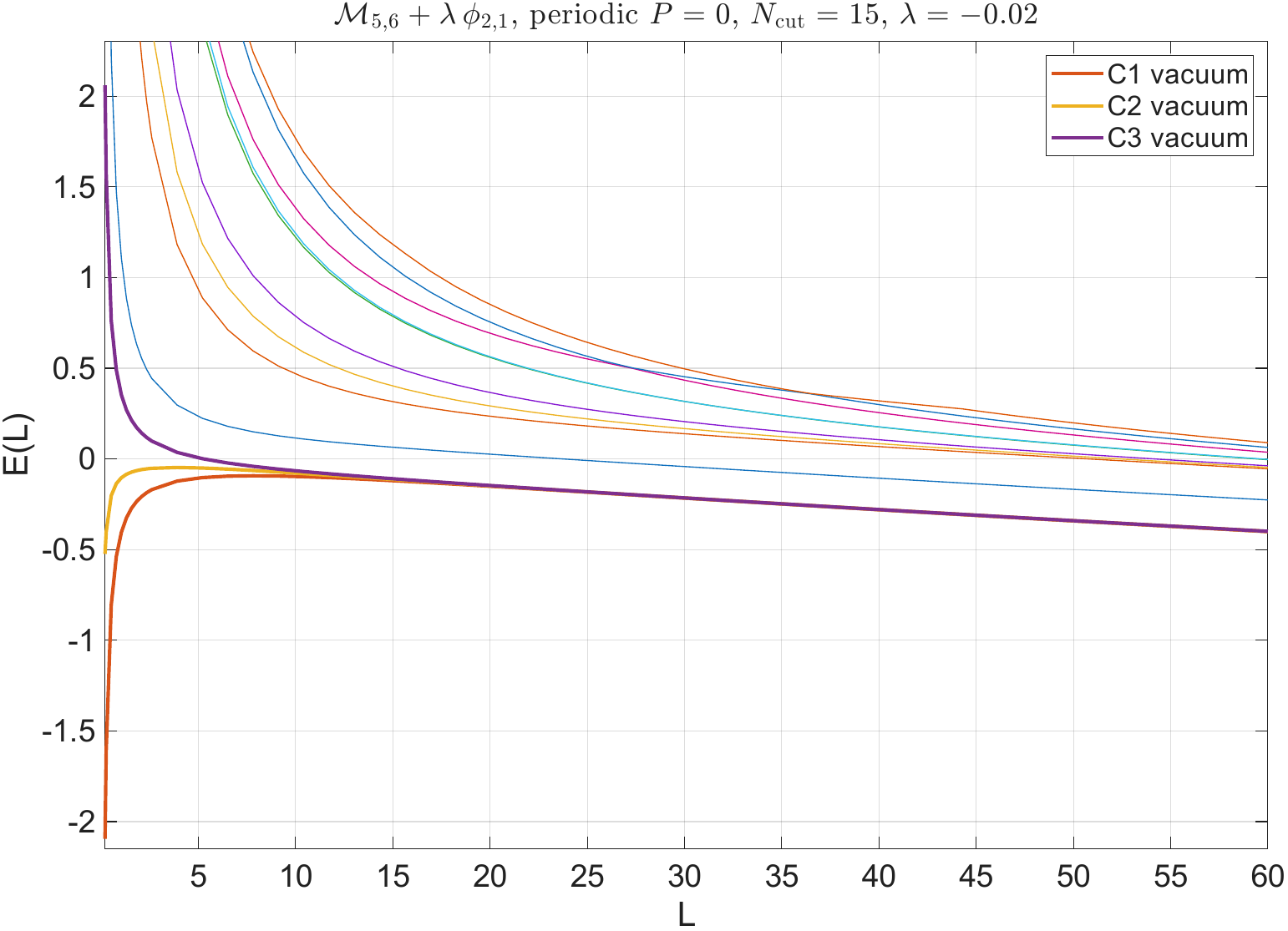}
\par\medskip
\begingroup\small
\setlength{\tabcolsep}{6pt}
\begin{tabular}{@{}rrrrr@{}}
\toprule
$L$ & $E_1-E_0$ & $E_2-E_0$ & $E_3-E_0$ & $E_4-E_0$\\
\midrule
20 & 0.00262 & 0.00396 & 0.17673 & 0.38607\\
40 & 0.00061 & 0.00077 & 0.17380 & 0.34800\\
60 & 0.00127 & 0.00164 & 0.17468 & 0.34629\\
\bottomrule
\end{tabular}
\endgroup
\caption{Periodic spectrum of $\mathcal M_{5,6}+\lambda\,\phi_{2,1}$ at zero momentum, $\lambda=-0.02$ and $N_{\mathrm{cut}}=15$. The curves show the periodic energies $E_n(L)$. The table gives $E_n(L)-E_0(L)$ from a separate TCSA
diagonalization at $L = 20, 40, 60$. The first two gaps measure the splitting of the three vacua.}
\label{fig:m56_ordered_recon_periodic}
\end{figure}

\paragraph{Periodic spectrum and vacuum labels.}
We take $\lambda=-0.02$ and order the energies at zero momentum as $E_0,E_1,\ldots$.
Figure~\ref{fig:m56_ordered_recon_periodic} shows the periodic energies $E_n(L)$ and tabulates their gaps above the ground state $E_0(L)$.
At the two largest circumferences in the table, both vacuum splittings $E_1-E_0$ and $E_2-E_0$ are less than one percent of $E_3-E_0$. We identify the lowest three states as the massive vacua.
The gap $E_3-E_0$ stays close to $0.175$, while $E_4-E_0$ decreases towards twice this value, indicating a two-particle scattering state. Thus the periodic spectrum indicates one neutral particle at this mass. Its location on the quiver remains to be determined.

We label the vacua at $L=60$ and $N_{\mathrm{cut}}=5$, where 2343 states are collected to measure the form factors below. In the basis of the three lowest periodic energy eigenstates, ordered by $E_0,E_1,E_2$, the interval average of $\phi_{2,2}$ gives
\begingroup
\setlength{\arraycolsep}{4pt}
\begin{align}
 O_{\rm vac}\simeq
 \begin{pmatrix}0&0.89822&0\\0.89822&0&1.26671\\0&1.26671&0\end{pmatrix}
 \,,\qquad \operatorname{spec}(O_{\rm vac})\simeq\{-1.55286,\,0,\,1.55286\}\,.
 \notag
\end{align}
\endgroup
The three corresponding eigenvectors are our reference vacua. To assign their quiver labels, we measure $X_1$ in this reference basis, ordered by increasing field value. Choosing relative phases so that the adjacent off diagonal entries are positive, we find
\begingroup
\setlength{\arraycolsep}{4pt}
\begin{align}
 X_1\big|_{\rm vac}\simeq
 \begin{pmatrix}0.00188&1.00093&0.00188\\1.00093&0.99625&1.00093\\0.00188&1.00093&0.00188\end{pmatrix}
 \,,\qquad \operatorname{spec}\!\left(X_1\big|_{\rm vac}\right)
 \simeq\{-1,\,0,\,2\}\,.
 \notag
\end{align}
\endgroup
Each outer column gives the middle vacuum, while the middle column gives all three vacua with coefficients close to one. Comparing with $X_1\otimes X_0=X_1$, $X_1\otimes X_2=X_1$ and $X_1\otimes X_1=X_0\oplus X_1\oplus X_2$, we label the middle reference vacuum by $1$. The other two are exchanged by $X_2$, and we label them $0,2$ in order of increasing field value. The approach in Section~\ref{sec:projections} gives thresholds $-0.83481$ and $+0.83481$, assigning the three intervals to vacua $0,1,2$ with a mean accuracy of $98.12\%$.

\begin{figure}[!htbp]
\centering
\includegraphics[width=0.96\linewidth]{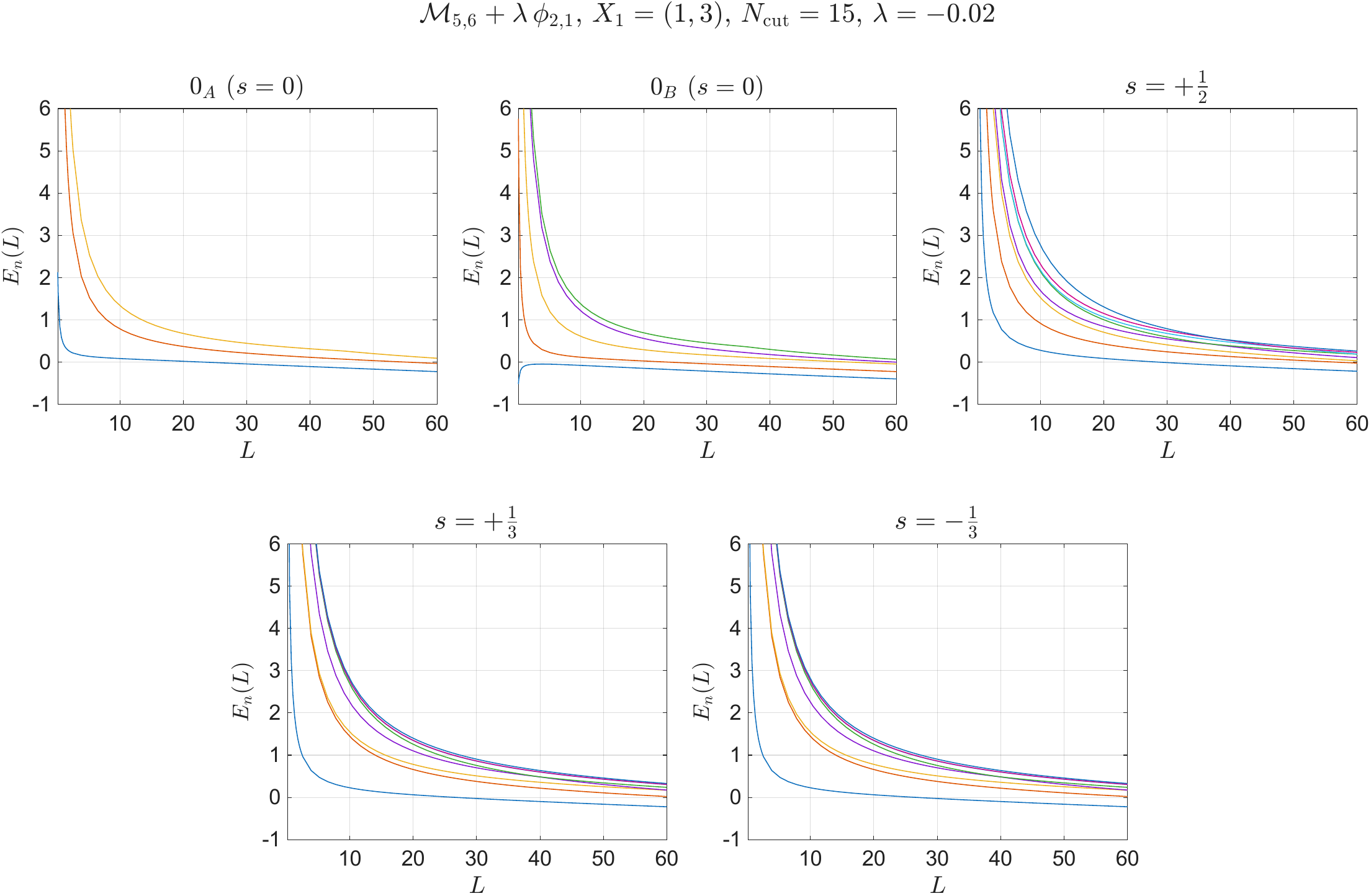}
\caption{Defect spectrum of $\mathcal M_{5,6}+\lambda\,\phi_{2,1}$ on the $X_1$ circle at $\lambda=-0.02$ and $N_{\mathrm{cut}}=15$. From left to right, the top row shows $0_A,0_B,\frac{1}{2}$ and the bottom row shows $+\frac{1}{3},-\frac{1}{3}$. We select the first excited level in the $0_B$ panel and the lowest level in each of the other four panels.}
\label{fig:m56_ordered_recon_defect}
\end{figure}

\paragraph{The five states on the defect circle.}
On inserting $X_1=\mathcal L_{(1,3)}$, the defect TCSA channels in Eq.~\eqref{eq:Z56} are $0_A,0_B,\frac{1}{2},+\frac{1}{3},-\frac{1}{3}$. The spatial $X_2$ line distinguishes the two zero momentum families, with charges $+1$ and $-1$ on $0_A$ and $0_B$, respectively.
The lowest state in $0_B$ is the defect vacuum. We select the lowest state in $0_A$, the first excited state in $0_B$ and the lowest state in each of the three fractional channels in Figure~\ref{fig:m56_ordered_recon_defect}.
We measure their gaps from the same defect vacuum and remove the momentum $p=\frac{2\pi s}{L}$ using Eq.~\eqref{eq:mest}.
At $L=52.100073$ and $N_{\mathrm{cut}}=15$, this gives masses $0.17346$, $0.17331$, $0.17422$, $0.17377$ and $0.17377$ in the channel order above.
Using $N_{\mathrm{cut}}=8,10,12,14,15$ to extrapolate the gaps at this fixed $L$ gives masses $0.17264$, $0.17280$, $0.17254$, $0.17277$ and $0.17277$, respectively. The five branches thus approach a common rest mass within the numerical resolution.

\paragraph{From vacuum pairs to a quiver.}
We now measure the five selected states at $L=60$ and $N_{\mathrm{cut}}=5$, using the periodic calibration above for both $\phi_{2,2}$ probes.
For each momentum state $\ket{s}$, Table~\ref{tab:m56_ordered_recon_raw} gives the probability $(E_{ab})_{ss}$ of reading vacuum $a$ on the left and $b$ on the right, averaged over the two measurement orders as in Eq.~\eqref{eq:coherentmeasurement}.
The nine vacuum pairs exhaust the possible probe readings in each momentum state, so their measured probabilities already satisfy $\sum_{a,b}(E_{ab})_{ss}=1$. To determine the arrows, we sum across the five states for each fixed pair along rows.

\begin{table}[!htbp]
\centering\small
\setlength{\tabcolsep}{7pt}
\begin{tabular}{crrrrrr}
\toprule
$ab$ & $0_A$ & $0_B$ & $s=\frac{1}{2}$ & $s=+\frac{1}{3}$ & $s=-\frac{1}{3}$ & $\sum_s(E_{ab})_{ss}$\\
\midrule
$00$ & 0.00850 & 0.01251 & 0.01954 & 0.01281 & 0.01281 & 0.06617\\
$01$ & 0.16227 & 0.23555 & 0.23190 & 0.16231 & 0.16231 & 0.95434\\
$02$ & 0.00131 & 0.00273 & 0.00262 & 0.00274 & 0.00274 & 0.01213\\
$10$ & 0.16227 & 0.23555 & 0.23190 & 0.16231 & 0.16231 & 0.95434\\
$11$ & 0.33132 & 0.02734 & 0.02808 & 0.31966 & 0.31966 & 1.02605\\
$12$ & 0.16227 & 0.23555 & 0.23190 & 0.16231 & 0.16231 & 0.95434\\
$20$ & 0.00131 & 0.00273 & 0.00262 & 0.00274 & 0.00274 & 0.01213\\
$21$ & 0.16227 & 0.23555 & 0.23190 & 0.16231 & 0.16231 & 0.95434\\
$22$ & 0.00850 & 0.01251 & 0.01954 & 0.01281 & 0.01281 & 0.06617\\
\bottomrule
\end{tabular}
\caption{Measured vacuum pair probabilities on the $X_1$ circle in the ordered phase of $\mathcal M_{5,6}+\lambda\,\phi_{2,1}$ at $\lambda=-0.02$, $L=60$ and $N_{\mathrm{cut}}=5$. The last column uses the unrounded probabilities.}
\label{tab:m56_ordered_recon_raw}
\end{table}
\FloatBarrier

The sums for $01,10,12,21,11$ are each close to one. Each pair has one closing junction on $X_1$, so we draw the two kink pairs $0\rightleftarrows1$ and $1\rightleftarrows2$, together with a particle loop at $1$.
The summed readings for $00,22$ are about $0.066$, and those for $02,20$ are about $0.012$. These small probabilities measure leakage of the probes, with no additional component resolved in the five selected states. The periodic spectrum contains one particle at this mass, consistent with the single loop at vacuum $1$. The four kink components and this particle also account for the five branches in the defect spectrum. We therefore reconstruct the quiver of the measured multiplet as
\begin{align}
 Q=\begin{pmatrix}0&1&0\\1&1&1\\0&1&0\end{pmatrix}
 \qquad &\Longrightarrow \qquad \text{\momentumQuiverMfivesixOrdered}
 \notag
\end{align}

\paragraph{Comparison with the theoretical squared amplitudes.}
Table~\ref{tab:m56_ordered_recon_weights} compares the normalized TCSA weights
$\widehat w_{ab,s}$ from Eq.~\eqref{eq:reconstructedweights}
with the theoretical squared overlaps $|U_{ab,s}|^2$ in Table~\ref{tab:mixing}.

\begin{table}[H]
\centering\small
\setlength{\tabcolsep}{3pt}
\begin{tabular}{crrrrr@{\hspace{1em}}rrrrr}
\toprule
& \multicolumn{5}{c}{TCSA $\widehat w_{ab,s}$} & \multicolumn{5}{c}{Theory $|U_{ab,s}|^2$}\\
\cmidrule(lr){2-6}\cmidrule(lr){7-11}
$ab$ & $0_A$ & $0_B$ & $\frac{1}{2}$ & $+\frac{1}{3}$ & $-\frac{1}{3}$
     & $0_A$ & $0_B$ & $\frac{1}{2}$ & $+\frac{1}{3}$ & $-\frac{1}{3}$\\
\midrule
$01$ & 0.16464 & 0.25134 & 0.24812 & 0.16725 & 0.16865 & 0.16667 & 0.25000 & 0.25000 & 0.16667 & 0.16667\\
$10$ & 0.16464 & 0.25134 & 0.24812 & 0.16725 & 0.16865 & 0.16667 & 0.25000 & 0.25000 & 0.16667 & 0.16667\\
$11$ & 0.33563 & 0.00000 & 0.00000 & 0.33219 & 0.33219 & 0.33333 & 0.00000 & 0.00000 & 0.33333 & 0.33333\\
$12$ & 0.16464 & 0.25134 & 0.24812 & 0.16725 & 0.16865 & 0.16667 & 0.25000 & 0.25000 & 0.16667 & 0.16667\\
$21$ & 0.16464 & 0.25134 & 0.24812 & 0.16725 & 0.16865 & 0.16667 & 0.25000 & 0.25000 & 0.16667 & 0.16667\\
\bottomrule
\end{tabular}
\caption{Squared amplitudes $|U_{ab,s}|^2$ on the $X_1$ circle in the ordered phase of
$\mathcal M_{5,6}+\lambda\,\phi_{2,1}$.
The TCSA data are taken at $\lambda=-0.02$, $L=60$ and $N_{\mathrm{cut}}=5$.}
\label{tab:m56_ordered_recon_weights}
\end{table}

The maximum and root mean square absolute deviations from the theoretical weights are 0.230 and 0.158 percentage points, respectively. Applying the same numerical stability tests as in Section~\ref{subsec:m45_quiver_reconstruction}, the largest changes under reducing the cutoff, fitting the probe orders separately and moving the probes are 0.088, 0.184 and 0.227 percentage points, respectively.
\FloatBarrier

\needspace{8\baselineskip}
\subsection{\texorpdfstring{$\mathcal M_{5,6}+\lambda_{>0}\,\phi_{2,1}$}{M5,6 + lambda phi2,1}: disordered phase}
\label{subsec:m56_disordered_quiver_reconstruction}

\begin{figure}[!htbp]
\centering
\includegraphics[width=0.65\linewidth]{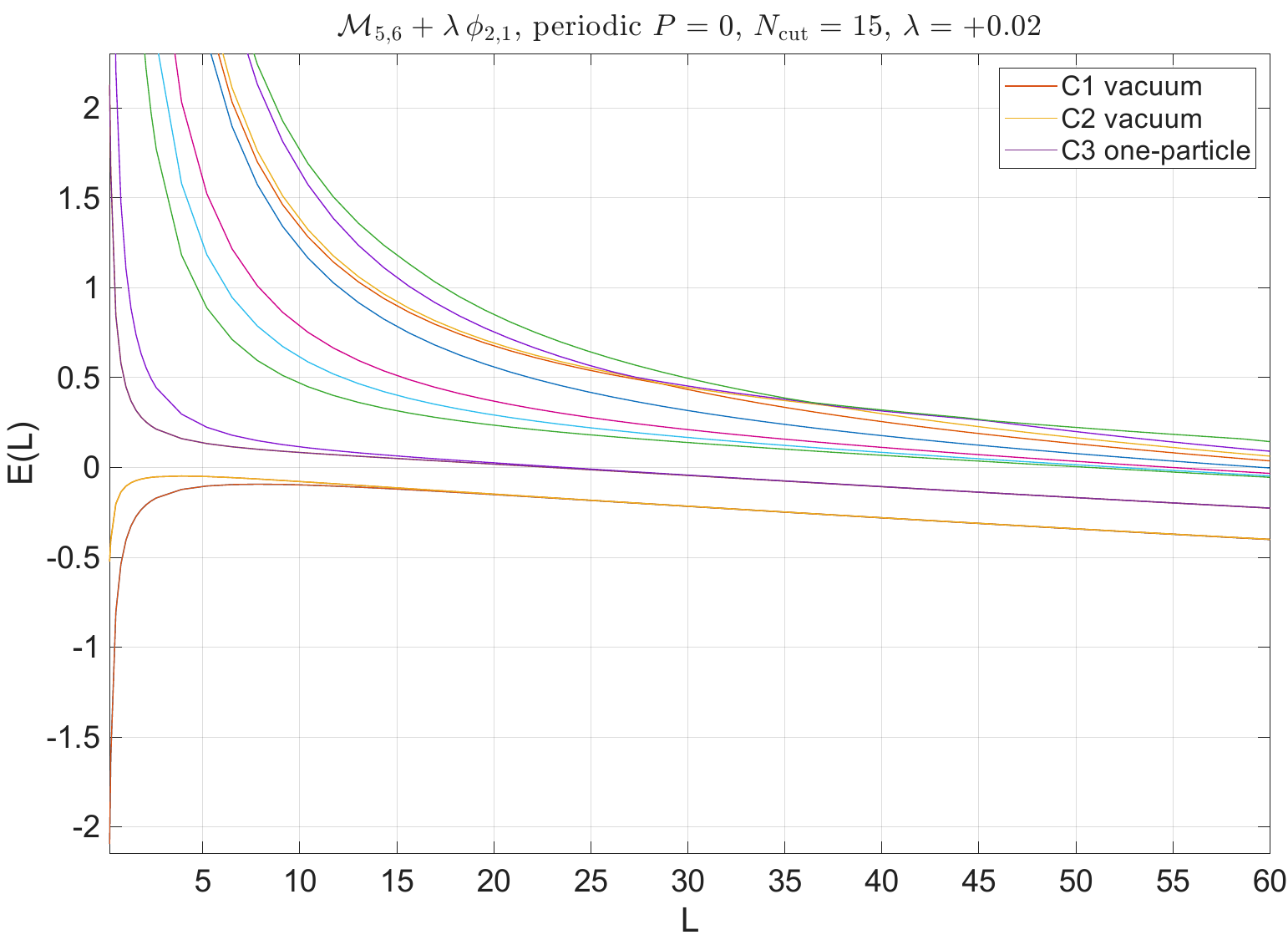}
\par\medskip
\begingroup\small
\setlength{\tabcolsep}{6pt}
\begin{tabular}{@{}rrrrr@{}}
\toprule
$L$ & $E_1-E_0$ & $E_2-E_0$ & $E_3-E_0$ & $E_4-E_0$\\
\midrule
20 & 0.00262 & 0.16945 & 0.17673 & 0.38607\\
40 & 0.00061 & 0.17335 & 0.17380 & 0.34800\\
60 & 0.00127 & 0.17468 & 0.17504 & 0.34629\\
\bottomrule
\end{tabular}
\endgroup
\caption{Periodic spectrum of $\mathcal M_{5,6}+\lambda\,\phi_{2,1}$ at zero momentum, $\lambda=+0.02$ and $N_{\mathrm{cut}}=15$. The curves show the periodic energies $E_n(L)$. The table gives $E_n(L)-E_0(L)$ from a separate TCSA diagonalization at $L=20,40,60$. The first gap measures the splitting of the two vacua.}
\label{fig:m56_disordered_recon_periodic}
\end{figure}

\paragraph{Periodic spectrum and vacuum labels.}
We take $\lambda=+0.02$ and order the energies at zero momentum as $E_0,E_1,\ldots$.
Figure~\ref{fig:m56_disordered_recon_periodic} shows the periodic energies $E_n(L)$ and tabulates their gaps above the ground state $E_0(L)$.
At the two largest circumferences in the table, the splitting $E_1-E_0$ is less than one percent of $E_2-E_0$. We identify the lowest pair as the two massive vacua.
The gaps $E_2-E_0$ and $E_3-E_0$ approach a common value near $0.175$, while $E_4-E_0$ decreases towards twice this value, indicating a two particle scattering state. Thus the periodic spectrum indicates two neutral particles at this mass. Their locations on the quiver remain to be determined.

We label the vacua at $L=60$ and $N_{\mathrm{cut}}=5$, where 2343 states are collected to measure the form factors below. In the basis of the two lowest periodic energy eigenstates, ordered by $E_0,E_1$, the interval average of $\phi_{2,2}$ gives
\begingroup
\setlength{\arraycolsep}{4pt}
\begin{align}
 O_{\rm vac}\simeq
 \begin{pmatrix}0&0.89822\\0.89822&0\end{pmatrix}
 \,,\qquad \operatorname{spec}(O_{\rm vac})\simeq\{-0.89822,\,0.89822\}\,.
 \notag
\end{align}
\endgroup
The two corresponding eigenvectors are our reference vacua. In this reference basis, ordered by increasing field value and with relative phases chosen to make the off diagonal entry positive, we measure
\begingroup
\setlength{\arraycolsep}{4pt}
\begin{align}
 X_1\big|_{\rm vac}\simeq
 \begin{pmatrix}1&1\\1&1\end{pmatrix}
 \,,\qquad \operatorname{spec}\!\left(X_1\big|_{\rm vac}\right)
 \simeq\{0,\,2\}\,.
 \notag
\end{align}
\endgroup
Each column gives both reference vacua with coefficients close to one. Exchanging the two vacua leaves this measured matrix unchanged, and we label the lower field value by $0$ and the higher one by $1$. The approach in Section~\ref{sec:projections} gives a threshold at $-0.22777$, assigning readings below it to vacuum $0$ and readings at or above it to vacuum $1$, with a mean accuracy of $98.36\%$.

\begin{figure}[!htbp]
\centering
\includegraphics[width=0.96\linewidth]{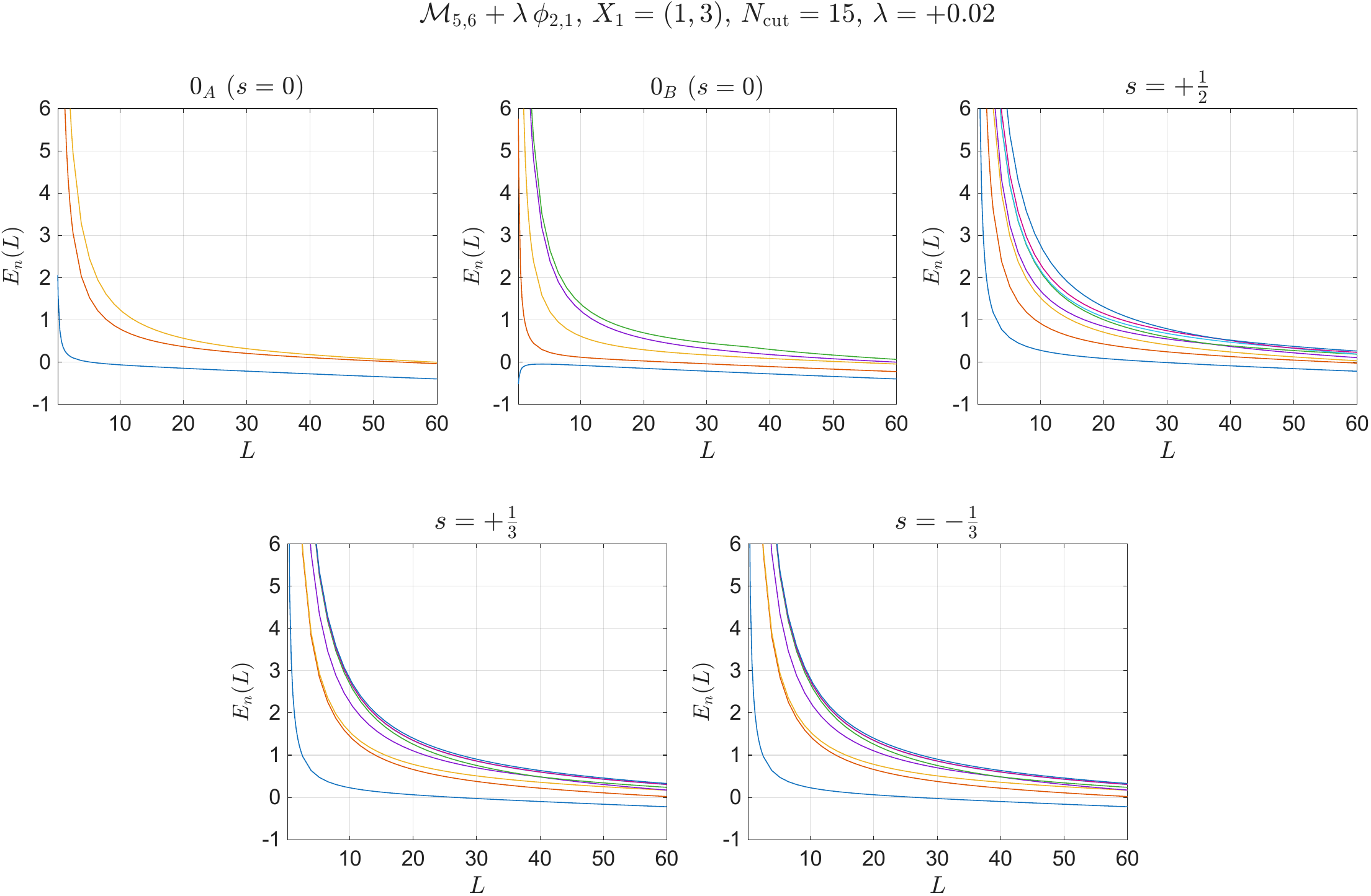}
\caption{Defect spectrum of $\mathcal M_{5,6}+\lambda\,\phi_{2,1}$ on the $X_1$ circle at $\lambda=+0.02$ and $N_{\mathrm{cut}}=15$. From left to right, the top row shows $0_A,0_B,\frac{1}{2}$ and the bottom row shows $+\frac{1}{3},-\frac{1}{3}$. The lowest levels in the first two panels are defect vacua. We select the first excited level in the $0_B$ panel and the lowest level in each of the three fractional momentum panels.}
\label{fig:m56_disordered_recon_defect}
\end{figure}

\paragraph{The four states on the defect circle.}
On inserting $X_1=\mathcal L_{(1,3)}$, the defect TCSA channels in Eq.~\eqref{eq:Z56} are $0_A,0_B,\frac{1}{2},+\frac{1}{3},-\frac{1}{3}$. The spatial $X_2$ line distinguishes the two zero momentum families, with charges $+1$ and $-1$ on $0_A$ and $0_B$, respectively.
The lowest states in $0_A$ and $0_B$ are the two defect vacua. We select the first excited state in $0_B$ and the lowest state in each of the three fractional channels in Figure~\ref{fig:m56_disordered_recon_defect}. The $0_A$ channel contains no excitation in this band.
We measure the four gaps from the same $0_B$ defect vacuum and remove the momentum $p=\frac{2\pi s}{L}$ using Eq.~\eqref{eq:mest}. At $L=52.100073$ and $N_{\mathrm{cut}}=15$, this gives masses $0.17331$, $0.17422$, $0.17377$ and $0.17377$ in the order $0_B,\frac{1}{2},+\frac{1}{3},-\frac{1}{3}$.
Using $N_{\mathrm{cut}}=8,10,12,14,15$ to extrapolate these masses at fixed $L$ gives $0.17280$, $0.17254$, $0.17277$ and $0.17277$, respectively. The four branches thus approach a common rest mass within the numerical resolution.

\paragraph{From vacuum pairs to a quiver.}
We now measure the four selected states at $L=60$ and $N_{\mathrm{cut}}=5$, using the periodic calibration above for both $\phi_{2,2}$ probes.
For each momentum state $\ket{s}$, Table~\ref{tab:m56_disordered_recon_raw} gives the probability $(E_{ab})_{ss}$ of reading vacuum $a$ on the left and $b$ on the right, averaged over the two measurement orders as in Eq.~\eqref{eq:coherentmeasurement}.
The four vacuum pairs exhaust the possible probe readings in each momentum state, so their measured probabilities already satisfy $\sum_{a,b}(E_{ab})_{ss}=1$. To determine the arrows, we sum across the four states for each fixed pair along rows.

\begin{table}[!htbp]
\centering\small
\setlength{\tabcolsep}{8pt}
\begin{tabular}{crrrrr}
\toprule
$ab$ & $0_B$ & $s=\frac{1}{2}$ & $s=+\frac{1}{3}$ & $s=-\frac{1}{3}$ & $\sum_s(E_{ab})_{ss}$\\
\midrule
$00$ & 0.13334 & 0.35352 & 0.24069 & 0.24069 & 0.96823\\
$01$ & 0.35041 & 0.13074 & 0.24087 & 0.24087 & 0.96289\\
$10$ & 0.35041 & 0.13074 & 0.24087 & 0.24087 & 0.96289\\
$11$ & 0.16585 & 0.38499 & 0.27757 & 0.27757 & 1.10598\\
\bottomrule
\end{tabular}
\caption{Measured vacuum pair probabilities on the $X_1$ circle in the disordered phase of $\mathcal M_{5,6}+\lambda\,\phi_{2,1}$ at $\lambda=+0.02$, $L=60$ and $N_{\mathrm{cut}}=5$. The last column uses the unrounded probabilities.}
\label{tab:m56_disordered_recon_raw}
\end{table}
\FloatBarrier

The sums for $00,01,10,11$ are each close to one. We therefore draw a particle loop at each vacuum and the kink pair $0\rightleftarrows1$. The two loops agree with the two particles observed in the periodic spectrum. Each of the four components closes once on the defect circle, accounting for the four measured momentum branches. We reconstruct the quiver as
\begin{align}
 Q=\begin{pmatrix}1&1\\1&1\end{pmatrix}
 \qquad &\Longrightarrow \qquad \text{\momentumQuiverMfivesixDisordered}
 \notag
\end{align}

\paragraph{Comparison with the theoretical squared amplitudes.}
Table~\ref{tab:m56_disordered_recon_weights} compares the normalized TCSA weights
$\widehat w_{ab,s}$ from Eq.~\eqref{eq:reconstructedweights}
with the theoretical squared overlaps $|U_{ab,s}|^2$ in Table~\ref{tab:mixing56disorder}.

\begin{table}[H]
\centering\small
\setlength{\tabcolsep}{4pt}
\begin{tabular}{crrrr@{\hspace{1.3em}}rrrr}
\toprule
& \multicolumn{4}{c}{TCSA $\widehat w_{ab,s}$} & \multicolumn{4}{c}{Theory $|U_{ab,s}|^2$}\\
\cmidrule(lr){2-5}\cmidrule(lr){6-9}
$ab$ & $0_B$ & $\frac{1}{2}$ & $+\frac{1}{3}$ & $-\frac{1}{3}$
     & $0_B$ & $\frac{1}{2}$ & $+\frac{1}{3}$ & $-\frac{1}{3}$\\
\midrule
$00$ & 0.12770 & 0.37047 & 0.25092 & 0.25091 & 0.12500 & 0.37500 & 0.25000 & 0.25000\\
$01$ & 0.36788 & 0.12710 & 0.25118 & 0.25384 & 0.37500 & 0.12500 & 0.25000 & 0.25000\\
$10$ & 0.36788 & 0.12710 & 0.25118 & 0.25384 & 0.37500 & 0.12500 & 0.25000 & 0.25000\\
$11$ & 0.12969 & 0.35782 & 0.25504 & 0.25745 & 0.12500 & 0.37500 & 0.25000 & 0.25000\\
\bottomrule
\end{tabular}
\caption{Squared amplitudes $|U_{ab,s}|^2$ on the $X_1$ circle in the disordered phase of
$\mathcal M_{5,6}+\lambda\,\phi_{2,1}$.
The TCSA data are taken at $\lambda=+0.02$, $L=60$ and $N_{\mathrm{cut}}=5$.}
\label{tab:m56_disordered_recon_weights}
\end{table}

The maximum and root mean square absolute deviations from the theoretical weights are 1.718 and 0.597 percentage points, respectively. Applying the same numerical stability tests as in Section~\ref{subsec:m45_quiver_reconstruction}, the largest changes under reducing the cutoff, fitting the probe orders separately and moving the probes are 0.114, 0.163 and 0.255 percentage points, respectively.
\FloatBarrier

\needspace{8\baselineskip}
\subsection{\texorpdfstring{$\mathcal M_{6,7}+\lambda\,\phi_{2,1}$}{M6,7 + lambda phi2,1}}
\label{subsec:m67_quiver_reconstruction}

\begin{figure}[!htbp]
\centering
\includegraphics[width=0.65\linewidth]{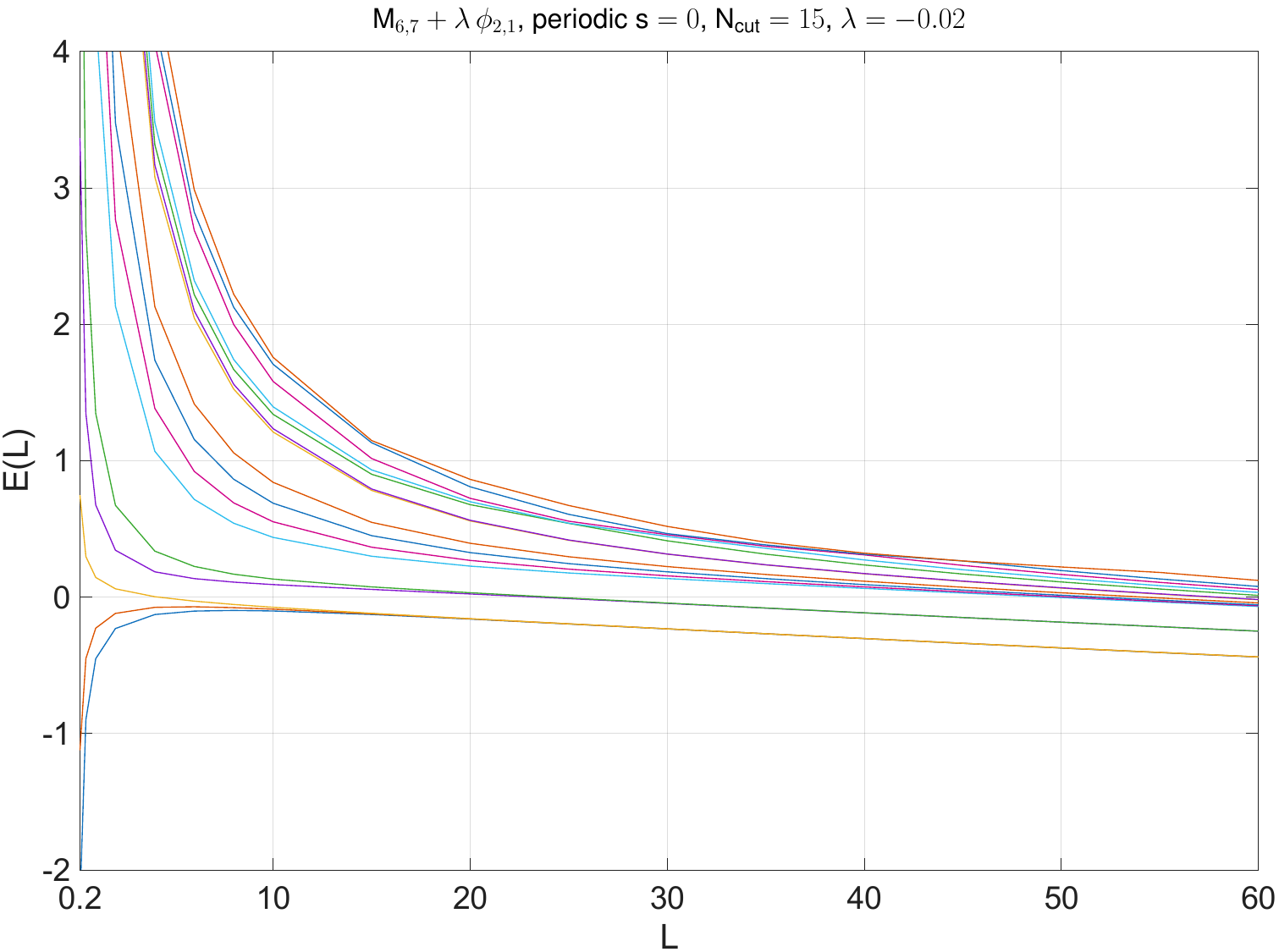}
\par\medskip
\begingroup\small
\setlength{\tabcolsep}{4pt}
\begin{tabular}{@{}rrrrrrrr@{}}
\toprule
$L$ & $E_1-E_0$ & $E_2-E_0$ & $E_3-E_0$ & $E_4-E_0$ & $E_5-E_0$ & $E_6-E_0$ & $E_7-E_0$\\
\midrule
20 & 0.00169 & 0.00304 & 0.18411 & 0.19325 & 0.38812 & 0.42995 & 0.48735\\
40 & 0.00045 & 0.00058 & 0.18866 & 0.18917 & 0.36820 & 0.37926 & 0.39583\\
60 & 0.00107 & 0.00146 & 0.18993 & 0.19024 & 0.37097 & 0.37639 & 0.38333\\
\bottomrule
\end{tabular}
\endgroup
\caption{Periodic spectrum of $\mathcal M_{6,7}+\lambda\,\phi_{2,1}$ at zero momentum, $\lambda=-0.02$ and $N_{\mathrm{cut}}=15$. The curves show the periodic energies $E_n(L)$. The table gives $E_n(L)-E_0(L)$ from a separate TCSA diagonalization at $L=20,40,60$. The first two gaps measure the splitting of the three vacua.}
\label{fig:m67_recon_periodic}
\end{figure}

\paragraph{Periodic spectrum and vacuum labels.}
We take $\lambda=-0.02$ and order the energies at zero momentum as $E_0,E_1,\ldots$.
Figure~\ref{fig:m67_recon_periodic} shows the periodic energies $E_n(L)$ and tabulates their gaps above the ground state $E_0(L)$.
At the two largest circumferences in the table, both $E_1-E_0$ and $E_2-E_0$ are less than one percent of $E_3-E_0$. We identify the lowest three states as the massive vacua.
The gaps $E_3-E_0$ and $E_4-E_0$ approach a common value near $0.19$, indicating two neutral particles at this mass. At $L=60$, the next three gaps lie between $0.371$ and $0.383$, close to twice this value. We examine these higher branches at the end of this subsection and first reconstruct the quiver at the lower mass.

We label the vacua at $L=60$ and $N_{\mathrm{cut}}=5$, where 4107 states are collected to measure the form factors below. In the basis of the three lowest periodic energy eigenstates, ordered by $E_0,E_1,E_2$, the interval average of $\phi_{2,2}$ gives
\begingroup
\setlength{\arraycolsep}{4pt}
\begin{align}
 O_{\rm vac}&\simeq
 \begin{pmatrix}0&0.93088&0\\0.93088&0&0.92863\\0&0.92863&-0.92743\end{pmatrix}
 \,,\notag\\
 \operatorname{spec}(O_{\rm vac})&\simeq\{-1.67356,\,-0.41395,\,1.16007\}\,.
 \notag
\end{align}
\endgroup
The three corresponding eigenvectors are our reference vacua. In this reference basis, ordered by increasing field value and with relative phases chosen to make the adjacent off diagonal entries positive, we measure
\begingroup
\setlength{\arraycolsep}{4pt}
\begin{align}
 X_1\big|_{\rm vac}&\simeq
 \begin{pmatrix}0.00277&1.00136&0.00172\\1.00136&0.99653&1.00001\\0.00172&1.00001&1.00070\end{pmatrix}
 \,,\notag\\
 \operatorname{spec}\!\left(X_1\big|_{\rm vac}\right)&\simeq\{-0.80194,\,0.55496,\,2.24698\}\,.
 \notag
\end{align}
\endgroup
The first reference vacuum is mapped only to the second to this accuracy. Comparing with $X_1\otimes X_0=X_1$, we label them $0$ and $1$, respectively, and label the remaining vacuum $2$. The overall sign of $\phi_{2,2}$ has been chosen so that these labels follow increasing field value. The approach in Section~\ref{sec:projections} gives thresholds $-1.14574$ and $0.30065$. Readings below the first threshold, between the thresholds, and above the second are assigned to vacua $0,1,2$, respectively, with a mean accuracy of $98.37\%$.

\begin{figure}[!t]
\centering
\includegraphics[width=0.96\linewidth]{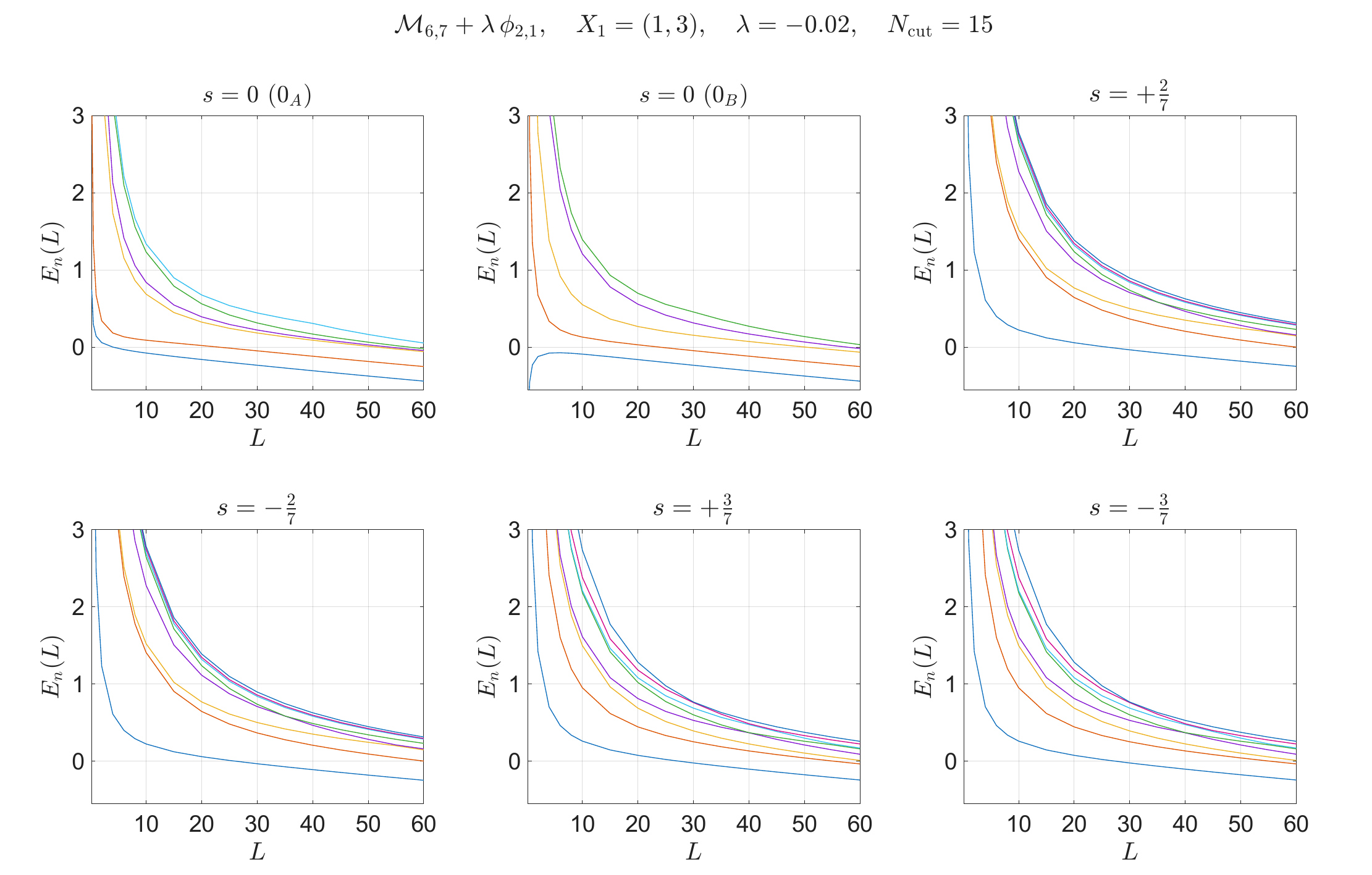}
\caption{Defect spectrum of $\mathcal M_{6,7}+\lambda\,\phi_{2,1}$ on the $X_1$ circle at $\lambda=-0.02$ and $N_{\mathrm{cut}}=15$. From left to right, the top row shows $0_A,0_B,+\frac{2}{7}$ and the bottom row shows $-\frac{2}{7},+\frac{3}{7},-\frac{3}{7}$. The lowest levels in the first two panels are defect vacua. We select the first excited level in each zero momentum panel and the lowest level in each of the four fractional momentum panels.}
\label{fig:m67_recon_defect}
\end{figure}

\paragraph{The six states on the defect circle.}
On inserting $X_1=\mathcal L_{(1,3)}$, the defect TCSA channels in Eq.~\eqref{eq:inverse67} are $0_A,0_B,+\frac{2}{7},-\frac{2}{7},+\frac{3}{7},-\frac{3}{7}$. The spatial $X_1$ line distinguishes the two zero momentum families, with negative and positive eigenvalues on $0_A$ and $0_B$, respectively.
The lowest states in $0_A$ and $0_B$ are the two defect vacua. We select the first excited state in each zero momentum family and the lowest state in each of the four fractional channels in Figure~\ref{fig:m67_recon_defect}.
We measure the six gaps from the same $0_B$ defect vacuum and remove the momentum $p=\frac{2\pi s}{L}$ using Eq.~\eqref{eq:mest}. At $L=50$ and $N_{\mathrm{cut}}=15$, this gives masses $0.18875$, $0.18874$, $0.18918$, $0.18918$, $0.18940$ and $0.18940$ in the channel order above. Using $N_{\mathrm{cut}}=8,10,12,14,15$ to extrapolate these masses at fixed $L$ gives $0.18808$, $0.18821$, $0.18816$, $0.18816$, $0.18802$ and $0.18802$, respectively. The six branches thus approach a common rest mass within the numerical resolution.

\paragraph{From vacuum pairs to a quiver.}
We now measure the six selected states at $L=60$ and $N_{\mathrm{cut}}=5$, using the periodic calibration above for both $\phi_{2,2}$ probes.
For each momentum state $\ket{s}$, Table~\ref{tab:m67_recon_raw} gives the probability $(E_{ab})_{ss}$ of reading vacuum $a$ on the left and $b$ on the right, averaged over the two measurement orders as in Eq.~\eqref{eq:coherentmeasurement}.
The nine vacuum pairs exhaust the possible probe readings in each momentum state, so their measured probabilities already satisfy $\sum_{a,b}(E_{ab})_{ss}=1$. To determine the arrows, we sum across the six states for each fixed pair along rows.

\begin{table}[!t]
\centering\small
\setlength{\tabcolsep}{6pt}
\begin{tabular}{crrrrrrr}
\toprule
$ab$ & $0_A$ & $0_B$ & $+\frac{2}{7}$ & $-\frac{2}{7}$ & $+\frac{3}{7}$ & $-\frac{3}{7}$ & $\sum_s(E_{ab})_{ss}$\\
\midrule
$00$ & 0.01057 & 0.00708 & 0.00741 & 0.00741 & 0.01408 & 0.01408 & 0.06064\\
$01$ & 0.23052 & 0.14750 & 0.10535 & 0.10535 & 0.17960 & 0.17960 & 0.94792\\
$02$ & 0.00143 & 0.00264 & 0.00203 & 0.00203 & 0.00207 & 0.00207 & 0.01227\\
$10$ & 0.23052 & 0.14750 & 0.10535 & 0.10535 & 0.17960 & 0.17960 & 0.94792\\
$11$ & 0.37683 & 0.03919 & 0.23212 & 0.23212 & 0.06741 & 0.06741 & 1.01510\\
$12$ & 0.05866 & 0.26180 & 0.18732 & 0.18732 & 0.13586 & 0.13586 & 0.96682\\
$20$ & 0.00143 & 0.00264 & 0.00203 & 0.00203 & 0.00207 & 0.00207 & 0.01227\\
$21$ & 0.05866 & 0.26180 & 0.18732 & 0.18732 & 0.13586 & 0.13586 & 0.96682\\
$22$ & 0.03139 & 0.12981 & 0.17104 & 0.17104 & 0.28347 & 0.28347 & 1.07023\\
\bottomrule
\end{tabular}
\caption{Measured vacuum pair probabilities on the $X_1$ circle in $\mathcal M_{6,7}+\lambda\,\phi_{2,1}$ at $\lambda=-0.02$, $L=60$ and $N_{\mathrm{cut}}=5$. The last column uses the unrounded probabilities.}
\label{tab:m67_recon_raw}
\end{table}

The sums for $01,10,12,21,11,22$ are each close to one. We therefore draw the two kink pairs $0\rightleftarrows1$ and $1\rightleftarrows2$, together with particle loops at vacua $1$ and $2$.
The two loops account for both particles in the periodic spectrum, leaving no further particle at vacuum $0$ at this mass. The readings for $00,02,20$ are small. These pairs have no closing junction on $X_1$, so we also check the spectrum on $X_2=\mathcal L_{(1,5)}$. It contains three branches at the same light mass, in $s=0,\pm\frac{3}{7}$. For example, at $N_{\mathrm{cut}}=12$ and $L=45.73051$, their masses are $0.18847,0.18935,0.18935$, whereas the lowest states at $s=\pm\frac{1}{7}$ have mass $0.52097$. The measured $11,12,21$ components each close once on $X_2$ and account for all three light branches. Thus no additional $02$ or $20$ kink is needed at this mass. The six components also account for the six light branches on $X_1$.
We reconstruct the quiver as
\begin{align}
 Q=\begin{pmatrix}0&1&0\\1&1&1\\0&1&1\end{pmatrix}
 \qquad &\Longrightarrow \qquad \text{\momentumQuiverMsixseven}
 \notag
\end{align}

\paragraph{Comparison with the theoretical squared amplitudes.}
Table~\ref{tab:m67_recon_weights} compares the normalized TCSA weights
$\widehat w_{ab,s}$ from Eq.~\eqref{eq:reconstructedweights}
with the theoretical squared overlaps $|U_{ab,s}|^2$ in Table~\ref{tab:m67_theoretical_weights}.

\begin{table}[!htbp]
\centering\small
\setlength{\tabcolsep}{5pt}
\begin{tabular}{clrrrrrr}
\toprule
$ab$ & & $0_A$ & $0_B$ & $+\frac{2}{7}$ & $-\frac{2}{7}$ & $+\frac{3}{7}$ & $-\frac{3}{7}$\\
\midrule
$01,10$ & TCSA $\widehat w$ & 0.23787 & 0.15719 & 0.10926 & 0.10926 & 0.19321 & 0.19321\\
 & Theory $|U|^2$ & 0.24172 & 0.15545 & 0.10757 & 0.10757 & 0.19384 & 0.19384\\
\addlinespace[3pt]
$12,21$ & TCSA $\widehat w$ & 0.05339 & 0.27656 & 0.19559 & 0.19559 & 0.13944 & 0.13944\\
 & Theory $|U|^2$ & 0.05548 & 0.28011 & 0.19384 & 0.19384 & 0.13836 & 0.13836\\
\addlinespace[3pt]
$11$ & TCSA $\widehat w$ & 0.38703 & 0.02038 & 0.23939 & 0.23939 & 0.05690 & 0.05690\\
 & Theory $|U|^2$ & 0.38430 & 0.02131 & 0.24172 & 0.24172 & 0.05548 & 0.05548\\
\addlinespace[3pt]
$22$ & TCSA $\widehat w$ & 0.02198 & 0.11303 & 0.16188 & 0.16188 & 0.27062 & 0.27062\\
 & Theory $|U|^2$ & 0.02131 & 0.10757 & 0.15545 & 0.15545 & 0.28011 & 0.28011\\
\bottomrule
\end{tabular}
\caption{Squared amplitudes $|U_{ab,s}|^2$ on the $X_1$ circle in $\mathcal M_{6,7}+\lambda\,\phi_{2,1}$. The TCSA data are taken at $\lambda=-0.02$, $L=60$ and $N_{\mathrm{cut}}=5$. Each pair of rows compares the measured $\widehat w_{ab,s}$ with the theoretical $|U_{ab,s}|^2$.}
\label{tab:m67_recon_weights}
\end{table}
The maximum and root mean square absolute deviations from the theoretical weights are 0.949 and 0.340 percentage points, respectively. Applying the same numerical stability tests as in Section~\ref{subsec:m45_quiver_reconstruction}, the largest changes under reducing the cutoff, fitting the probe orders separately and moving the probes are 0.108, 0.196 and 0.191 percentage points, respectively.

\begin{figure}[!htbp]
\centering
\includegraphics[width=\linewidth]{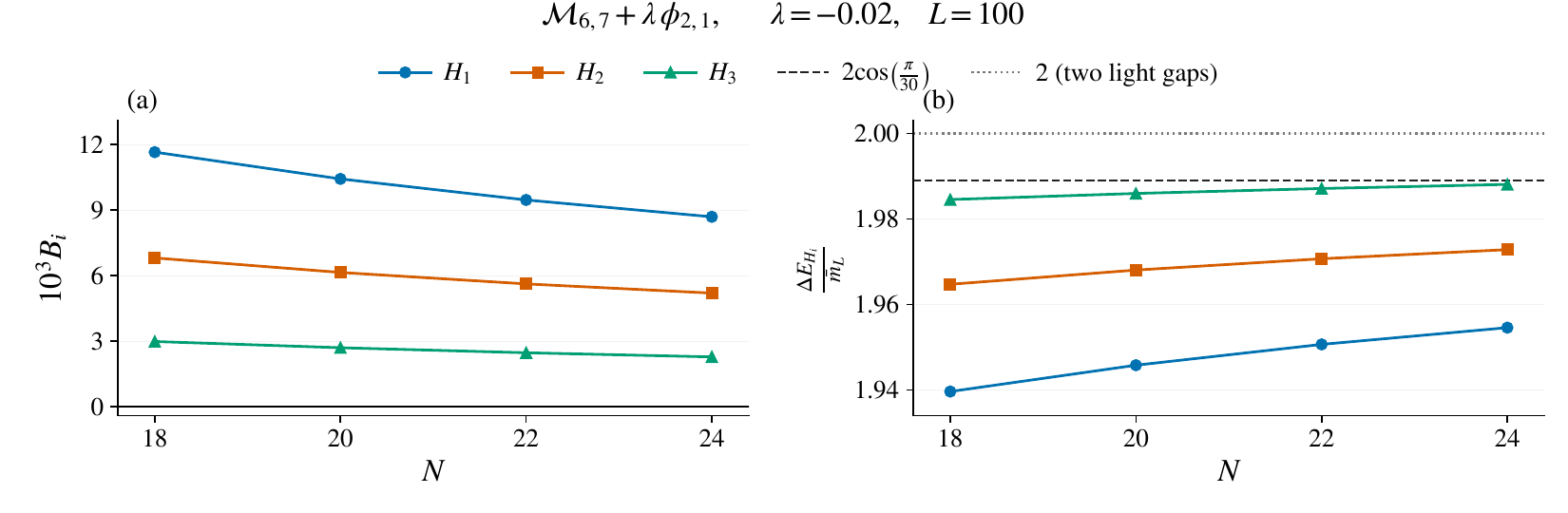}
\par\medskip
\begingroup\small
\setlength{\tabcolsep}{8pt}
\begin{tabular}{@{}rrrrr@{}}
\toprule
$N_{\mathrm{cut}}$ & $B_1$ & $B_2$ & $B_3$ & $\frac{\overline m_H}{\overline m_L}$\\
\midrule
18 & 0.01165466 & 0.00681335 & 0.00298743 & 1.96294\\
20 & 0.01042907 & 0.00614866 & 0.00270277 & 1.96658\\
22 & 0.00946620 & 0.00562454 & 0.00247277 & 1.96947\\
24 & 0.00869515 & 0.00520308 & 0.00228367 & 1.97181\\
\bottomrule
\end{tabular}
\endgroup
\caption{Heavy periodic branches of $\mathcal M_{6,7}+\lambda\,\phi_{2,1}$
at zero momentum, $\lambda=-0.02$ and $L=100$.
The labels $H_i$, $i=1,2,3$, denote the levels $E_{4+i}$.
The left panel shows the differences $B_i$ defined in
Eq.~\eqref{eq:m67_heavy_binding}, and the right panel shows the
individual ratios $\frac{\Delta E_{4+i}}{\overline m_L}$.
The dashed line gives the exact mass ratio
$2\cos\left(\frac{\pi}{30}\right)$, while the dotted line marks two-particle scattering state threshold $2$. The table gives the three $B_i$ and
the ratio of the mean heavy gap to the mean light gap.}
\label{fig:m67_heavy_binding}
\end{figure}

\needspace{8\baselineskip}
\paragraph{The heavy multiplet $R_0$.}
Above the two light particles, the periodic levels $E_5,E_6,E_7$
form a heavier triplet near the two particle threshold. To examine
their binding, we fix $L=100$ and increase the cutoff through
$N_{\mathrm{cut}}=18,20,22,24$. At each cutoff we subtract the same
ground state energy, $\Delta E_n=E_n-E_0$, and define
\begin{align}
 \overline m_L&=\frac{\Delta E_3+\Delta E_4}{2}\,, &
 B_i&=2\overline m_L-\Delta E_{4+i}\,,\qquad i=1,2,3\,,
 \label{eq:m67_heavy_binding}
\end{align}
where $\overline m_L$ is the mean light gap and $i$ labels the three
heavy branches in increasing energy order. Figure~\ref{fig:m67_heavy_binding}
shows that all three $B_i$ remain positive as the cutoff increases.
This supports the shallow bound state interpretation of the three branches.
The differences $B_i$ decrease by about $8\%$ between the last two
cutoffs, so they have not yet reached a stable plateau.

For the mass comparison, we average the three heavy gaps at the same
volume and cutoff, $\overline m_H=\frac{1}{3}\sum_{n=5}^{7}\Delta E_n$.
At $L=100$ and $N_{\mathrm{cut}}=24$, we obtain
\begin{align}
 \overline m_L&\simeq0.191335\,,\qquad
 \overline m_H\simeq0.377276\,,\qquad
 \frac{\overline m_H}{\overline m_L}\simeq1.97181\,.
 \label{eq:m67_measured_heavy_ratio}
\end{align}
The mean ratio moves towards the exact value
$2\cos\left(\frac{\pi}{30}\right)\simeq1.98904$ as the cutoff increases,
with a difference of $0.87\%$ at $N_{\mathrm{cut}}=24$.
The scattering matrix assigns this bound state to the unit fusion
channel, giving the $R_0$ multiplet with one neutral particle above
each vacuum and quiver $Q_H=\mathbf 1_3$, as derived in
Appendix~\ref{app:sm}. The periodic triplet and its mass ratio are
consistent with this assignment.

\FloatBarrier

\section*{Acknowledgments}
We thank Giuseppe Mussardo for helpful discussions. J.~Chen is supported by the Fujian Provincial Natural Science Foundation of China (No.2025J01004), and the National Natural Science Foundation of China (Grants No.12247103). T.C. is supported by the Xiamen University "Undergraduate Innovation Training Program" – Innovation Training (No.2025X1372). We acknowledge substantial assistance from generative AI in refining our form factor method in Section~\ref{sec:projections} for extracting the squared amplitude between the circle and momentum bases, checking the claims, and improving the manuscript. The authors take full responsibility for the contents of this paper.

\section*{Data availability}
The ancillary package provides the complex measurement matrices and
analysis scripts for the four cases in
Section~\ref{sec:quiver_reconstruction}. It also includes the CFT inputs
and code for reproducing the vacuum calibration and measurement with two
probes in $\mathcal M_{4,5}$. The accompanying README describes
the data formats, software requirements, and commands for reproducing
the corresponding measurement tables and checking the results.

\appendix
\section{Fusion matrices from WZW conformal blocks}
\label{app:recoupling}

We obtain the $F$ matrices for the regular vacuum module in Section~\ref{sec:circle} from the four-point conformal blocks of the ${\rm SU}(2)_k$ WZW theory. Following the WZW description in Appendix B of Ref.~\cite{FZ1985}, its primary fields have spins $j=0,\frac{1}{2},\ldots,\frac{k}{2}$ and conformal weights $h_j^{\mathrm{WZW}}=\frac{j(j+1)}{k+2}$. For four fields of spins $i,j,\ell,m$ at $0,z,1,\infty$, respectively, the chiral blocks satisfy the Knizhnik--Zamolodchikov equation~\cite{KZ1984},
\begin{align}
(k+2)\frac{\dd\boldsymbol{\mathcal F}}{\dd z}=\left(\frac{\Omega_{12}}{z}+\frac{\Omega_{23}}{z-1}\right)\boldsymbol{\mathcal F},\qquad \Omega_{rs}=2\,\mathbf S_r\cdot\mathbf S_s.
\label{eq:KZappendix}
\end{align}
Here $\mathbf S_r$ denotes the ${\rm SU}(2)$ generators acting on the $r$th primary field. The blocks in the $i\otimes j\to p$ and $j\otimes \ell\to q$ channels have leading powers $z^{h_p^{\mathrm{WZW}}-h_i^{\mathrm{WZW}}-h_j^{\mathrm{WZW}}}$ near $z=0$ and $(1-z)^{h_q^{\mathrm{WZW}}-h_j^{\mathrm{WZW}}-h_\ell^{\mathrm{WZW}}}$ near $z=1$, respectively. With normalized fusion vertices, the two channel bases are related by
\begin{align}
\mathcal F^{(t)}_q(z)=\sum_p\mathcal F^{(s)}_p(z)\,\bigl(F^{i j \ell}_m\bigr)_{pq}.
\label{eq:blockF}
\end{align}
Thus $F$ transforms between the two bases. The junction move in
Figure~\ref{fig:closure}(b) specializes this relation to $F^{1bj}_b$,
with intermediate channels $(p,q)=(a,d)$. In the real orthonormal
basis used in the main text, the WZW fusion matrix is expressed
through a quantum $6j$ symbol~\cite{ARS1999,DG},
\begin{align}
\F{i}{j}{\ell}{m}{p}{q}=(-1)^{i+j+\ell+m}\sqrt{\qnum{2p+1}\qnum{2q+1}}\sixj{i}{j}{p}{\ell}{m}{q}.
\label{eq:Fappendix}
\end{align}

For minimal model, the coset construction~\cite{GKO}
\begin{equation}
\mathcal M_{k+1,k+2}=\frac{{\rm SU}(2)_{k-1}\times {\rm SU}(2)_1}{{\rm SU}(2)_{k}}
\label{eq:coset}
\end{equation}
identifies the preserved line $X_i=(1,2i+1)$ with spin $i$ in the denominator and the identity representations in both numerator factors. The numerator fusion matrices are therefore one, and the $F$ matrices in this sector are those of ${\rm SU}(2)_{k}$ restricted to integer spins. We set $\zeta=\ee^{\frac{\ii\pi}{k+2}}$ and
\begin{align}
\qnum{n}=\frac{\sin\left(\frac{n\pi}{k+2}\right)}{\sin\left(\frac{\pi}{k+2}\right)},\qquad \qnum{n}!=\prod_{r=1}^{n}\qnum{r},\qquad \qnum{0}!=1.
\label{eq:qints}
\end{align}
The fusion $i\otimes j\to \ell$ is allowed when $i+j+\ell$ is an integer, $|i-j|\leq \ell\leq i+j$ and $i+j+\ell\leq k$. A spin-$i$ line has quantum dimension $d_i=\qnum{2i+1}$. To evaluate~Eq.~\eqref{eq:Fappendix} explicitly, the quantum $6j$ symbol has the finite expression~\cite{DG}
\begin{align}
\begin{aligned}
\sixj{i}{j}{p}{\ell}{m}{q}&=\Delta(i,j,p)\Delta(p,\ell,m)\Delta(j,\ell,q)\Delta(i,q,m)\\
&\quad\times\sum_{z=\max_r A_r}^{\min(\min_r B_r,k)}\frac{(-1)^z\qnum{z+1}!}{\prod_{r=1}^{4}\qnum{z-A_r}!\prod_{r=1}^{3}\qnum{B_r-z}!},
\end{aligned}
\label{eq:6j}
\end{align}
where
\begin{align}
\begin{aligned}
\Delta(i,j,\ell)&=\sqrt{\frac{\qnum{i+j-\ell}!\qnum{i-j+\ell}!\qnum{-i+j+\ell}!}{\qnum{i+j+\ell+1}!}},\\
(A_1,A_2,A_3,A_4)&=(i+j+p,p+\ell+m,j+\ell+q,i+q+m),\\
(B_1,B_2,B_3)&=(i+j+\ell+m,i+\ell+p+q,j+m+p+q).
\end{aligned}\notag
\end{align}
The summation bounds keep every factorial argument nonnegative and implement the level truncation. These formulas also apply to the half-integer labels needed in Appendix~\ref{app:dictionary} for even $k$.

For $\mathcal M_{5,6}$, take $k=4$ and $i=j=\ell=m=1$, so that $\qnum{1}=\qnum{5}=1$, $\qnum{2}=\qnum{4}=\sqrt3$ and $\qnum{3}=2$. Substituting these into Eqs.~\eqref{eq:Fappendix} and~\eqref{eq:6j}, with intermediate spins ordered as $(0,1,2)$, gives
\begin{equation}
F^{111}_1=\begin{pmatrix}
\frac{1}{2}&-\frac{1}{\sqrt2}&\frac{1}{2}\\
-\frac{1}{\sqrt2}&0&\frac{1}{\sqrt2}\\
\frac{1}{2}&\frac{1}{\sqrt2}&\frac{1}{2}
\end{pmatrix}.
\label{eq:F56}
\end{equation}

\section{The generalized charges from a spatial symmetry line}
\label{app:dictionary}
\begingroup
\definecolor{ic}{RGB}{30,93,187}\definecolor{xc}{RGB}{193,50,40}\definecolor{bc}{RGB}{26,132,95}
We consider the regular vacuum module with the particle line $\alpha=X_i$ and the closing line $X_1$. A momentum channel $\xi$ of Eq.~\eqref{eq:diag} is labelled by a pair of ${\rm SU}(2)_k$ spins $(p,q)$ satisfying $N_{pq}^{X_1}N_{pq}^{X_i}=1$. For odd $k$, both spins are integers. For even $k$, we include integer and half-integer pairs with the identification $(p,q)\sim(k/2-p,k/2-q)$~\cite{Bischoff} and choose the representative with $p<k/4$, or with $p=k/4$ and $q\leq k/4$. At the fixed pair $p=q=k/4$, the component containing $X_1$ has odd integer spin and contributes when $i$ is odd. In both cases we write $\xi=(p,q)$ and $\ket{\xi}=\ket{p,q}$ in the chosen labels.

We insert $X_x$ ($x=0,1,\ldots,J$, with $J=\lfloor \tfrac{k}{2}\rfloor$ as in Eq.~(2.5)) along the spatial circle and use the same notation for its action on the defect Hilbert space. In the momentum basis its action is~\cite{DG}
\begin{equation}
\displaystyle X_x\ket{p,q}=\frac{S_{xp}}{S_{0p}}\ket{p,q}
\label{eq:B_Daction}
\end{equation}
Thus the momentum basis matrix element is
\begin{equation}
\begin{aligned}
(X_x)_{(p',q'),(p,q)}
&=\delta_{p'p}\delta_{q'q}\frac{S_{xp}}{S_{0p}}
\end{aligned}
\label{eq:B_Dmomentum}
\end{equation}
Here $S_{rs}=\sqrt{\frac{2}{k+2}}\sin[\frac{(2r+1)(2s+1)\pi}{k+2}]$ is the ${\rm SU}(2)_k$ modular matrix.

We evaluate the same $X_x$ in the vacuum pair basis $\ket{uv}$. The cylinder in Figure~\ref{fig:B_uvcyl} fixes the junctions below and above the insertion: below they are $v_u\otimes X_i\to v_v$ and $X_1\otimes v_v\to v_u$, while above they are $v_a\otimes X_i\to v_b$ and $X_1\otimes v_b\to v_a$. The two blocks on the right show the input $\ket{uv}$ and output $\ket{ab}$ determined by the $X_x$ action.
\begin{figure}[!htbp]
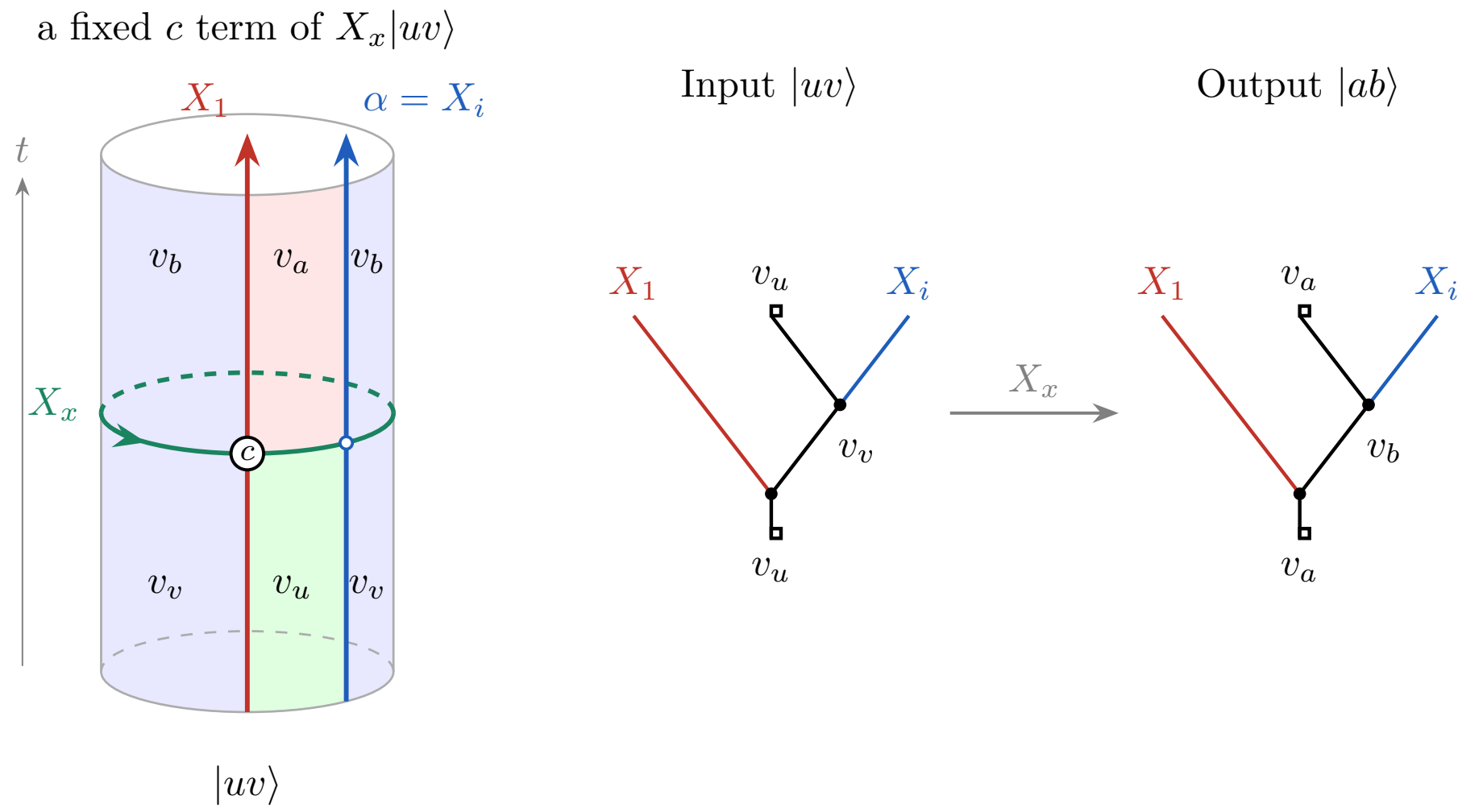

\centering
\resizebox{.98\linewidth}{!}{\begingroup
\input{figures/appendixB/appendix_b_cylinder_helpers.tex}
\input{figures/appendixB/appendix_b_d_helpers.tex}
\begin{tikzpicture}[>=Stealth,font=\small]
\draw[->,gray](-2,.05)--(-2,4.4)node[above]{$t$};
\begin{scope}
\Bcylbase{blue!9}
\begin{scope}
\clip(-1.3,0)--(-1.3,4.6)arc(180:0:1.3 and .36)--(1.3,0)arc(0:-180:1.3 and .36)--cycle;
\fill[green!12](0,-.4)rectangle(.88,5);
\fill[blue!9](-1.30000,2.30000) -- (-1.26750,2.22001) -- (-1.23500,2.18759) -- (-1.20250,2.16321) -- (-1.17000,2.14308) -- (-1.13750,2.12572) -- (-1.10500,2.11036) -- (-1.07250,2.09655) -- (-1.04000,2.08400) -- (-1.00750,2.07249) -- (-0.97500,2.06188) -- (-0.94250,2.05205) -- (-0.91000,2.04291) -- (-0.87750,2.03439) -- (-0.84500,2.02642) -- (-0.81250,2.01898) -- (-0.78000,2.01200) -- (-0.74750,2.00546) -- (-0.71500,1.99934) -- (-0.68250,1.99360) -- (-0.65000,1.98823) -- (-0.61750,1.98321) -- (-0.58500,1.97851) -- (-0.55250,1.97413) -- (-0.52000,1.97005) -- (-0.48750,1.96627) -- (-0.45500,1.96277) -- (-0.42250,1.95954) -- (-0.39000,1.95658) -- (-0.35750,1.95388) -- (-0.32500,1.95143) -- (-0.29250,1.94923) -- (-0.26000,1.94727) -- (-0.22750,1.94556) -- (-0.19500,1.94407) -- (-0.16250,1.94282) -- (-0.13000,1.94180) -- (-0.09750,1.94101) -- (-0.06500,1.94045) -- (-0.03250,1.94011) -- (0.00000,1.94000) -- (0.03250,1.94011) -- (0.06500,1.94045) -- (0.09750,1.94101) -- (0.13000,1.94180) -- (0.16250,1.94282) -- (0.19500,1.94407) -- (0.22750,1.94556) -- (0.26000,1.94727) -- (0.29250,1.94923) -- (0.32500,1.95143) -- (0.35750,1.95388) -- (0.39000,1.95658) -- (0.42250,1.95954) -- (0.45500,1.96277) -- (0.48750,1.96627) -- (0.52000,1.97005) -- (0.55250,1.97413) -- (0.58500,1.97851) -- (0.61750,1.98321) -- (0.65000,1.98823) -- (0.68250,1.99360) -- (0.71500,1.99934) -- (0.74750,2.00546) -- (0.78000,2.01200) -- (0.81250,2.01898) -- (0.84500,2.02642) -- (0.87750,2.03439) -- (0.91000,2.04291) -- (0.94250,2.05205) -- (0.97500,2.06188) -- (1.00750,2.07249) -- (1.04000,2.08400) -- (1.07250,2.09655) -- (1.10500,2.11036) -- (1.13750,2.12572) -- (1.17000,2.14308) -- (1.20250,2.16321) -- (1.23500,2.18759) -- (1.26750,2.22001) -- (1.30000,2.30000) -- (1.3,5) -- (-1.3,5) -- cycle;
\fill[red!10](0.00000,1.94000) -- (0.01100,1.94001) -- (0.02200,1.94005) -- (0.03300,1.94012) -- (0.04400,1.94021) -- (0.05500,1.94032) -- (0.06600,1.94046) -- (0.07700,1.94063) -- (0.08800,1.94083) -- (0.09900,1.94105) -- (0.11000,1.94129) -- (0.12100,1.94156) -- (0.13200,1.94186) -- (0.14300,1.94218) -- (0.15400,1.94253) -- (0.16500,1.94291) -- (0.17600,1.94331) -- (0.18700,1.94374) -- (0.19800,1.94420) -- (0.20900,1.94468) -- (0.22000,1.94519) -- (0.23100,1.94573) -- (0.24200,1.94629) -- (0.25300,1.94688) -- (0.26400,1.94750) -- (0.27500,1.94815) -- (0.28600,1.94882) -- (0.29700,1.94952) -- (0.30800,1.95025) -- (0.31900,1.95101) -- (0.33000,1.95179) -- (0.34100,1.95261) -- (0.35200,1.95345) -- (0.36300,1.95432) -- (0.37400,1.95522) -- (0.38500,1.95615) -- (0.39600,1.95711) -- (0.40700,1.95810) -- (0.41800,1.95912) -- (0.42900,1.96017) -- (0.44000,1.96125) -- (0.45100,1.96236) -- (0.46200,1.96350) -- (0.47300,1.96467) -- (0.48400,1.96588) -- (0.49500,1.96712) -- (0.50600,1.96839) -- (0.51700,1.96969) -- (0.52800,1.97103) -- (0.53900,1.97240) -- (0.55000,1.97381) -- (0.56100,1.97525) -- (0.57200,1.97672) -- (0.58300,1.97823) -- (0.59400,1.97978) -- (0.60500,1.98136) -- (0.61600,1.98298) -- (0.62700,1.98464) -- (0.63800,1.98634) -- (0.64900,1.98807) -- (0.66000,1.98985) -- (0.67100,1.99166) -- (0.68200,1.99352) -- (0.69300,1.99542) -- (0.70400,1.99736) -- (0.71500,1.99934) -- (0.72600,2.00137) -- (0.73700,2.00344) -- (0.74800,2.00556) -- (0.75900,2.00773) -- (0.77000,2.00994) -- (0.78100,2.01221) -- (0.79200,2.01452) -- (0.80300,2.01689) -- (0.81400,2.01931) -- (0.82500,2.02178) -- (0.83600,2.02431) -- (0.84700,2.02690) -- (0.85800,2.02954) -- (0.86900,2.03225) -- (0.88000,2.03502) -- (0.88,5) -- (0,5) -- cycle;
\end{scope}
\Bcylshell\Bxloop\Bdefect\Bparticle
\node at(.40,.75){$v_u$};\node at(-.72,.75){$v_v$};\node at(1.07,.75){$v_v$};
\node at(.40,3.65){$v_a$};\node at(-.72,3.65){$v_b$};\node at(1.07,3.65){$v_b$};
\node[below]at(0,-.70){$|uv\rangle$};\node[above]at(0,5.40){a fixed $c$ term of $X_x|uv\rangle$};
\end{scope}
\node at(4.65,5.2){Input $|uv\rangle$};
\node at(9.35,5.2){Output $|ab\rangle$};
\begin{scope}[shift={(4.3,.25)},x=.72cm,y=.72cm,line width=.9pt]
\rawinputstate
\end{scope}
\draw[->,gray,thick](6.25,2.3)--(7.75,2.3)node[midway,above]{$X_x$};
\begin{scope}[shift={(9,.25)},x=.72cm,y=.72cm,line width=.9pt]
\standardoutputstate
\end{scope}
\end{tikzpicture}
\endgroup}
\caption{Spatial insertion $X_x$ acting on $\ket{uv}$, with the input and output junctions shown to the right. Red denotes the defect $X_1$ and blue the excitation $X_i$.}
\label{fig:B_uvcyl}
\end{figure}
Figure~\ref{fig:B_D} inserts the junction $X_x\otimes X_1\to X_c\to X_1\otimes X_x$ and relates the input and output junctions by recoupling.
\begin{figure}[!htbp]
\centering
\resizebox{.93\linewidth}{!}{\begingroup
\definecolor{ic}{RGB}{30,93,187}\definecolor{xc}{RGB}{193,50,40}\definecolor{bc}{RGB}{26,132,95}
\input{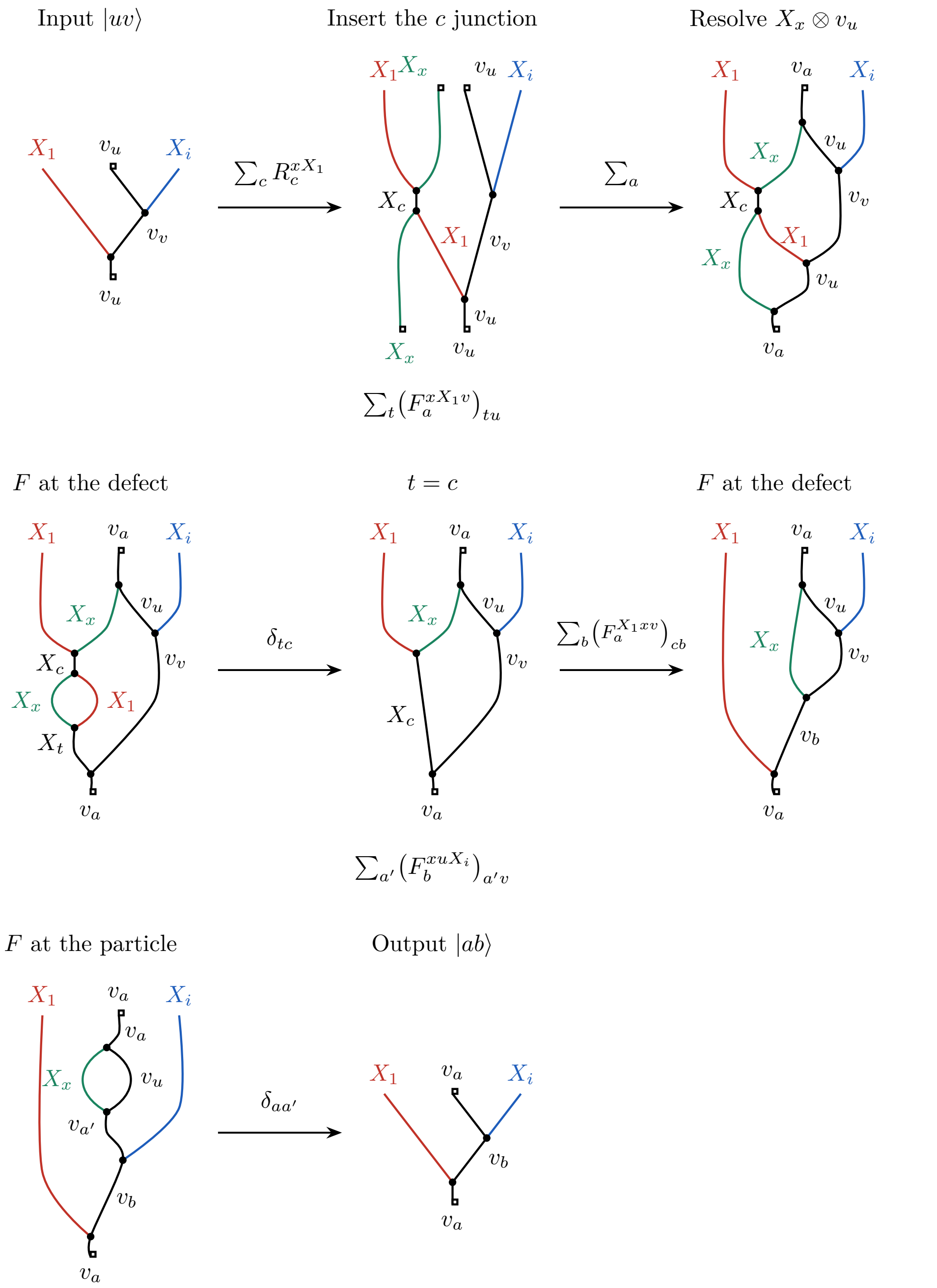}
\begin{tikzpicture}[>=Stealth,font=\small]
\foreach \pos/\txt in {0/{Input $|uv\rangle$},5.1/{Insert the $c$ junction},10.2/{Resolve $X_x\otimes v_u$}}{\node at(\pos,4.65){\txt};}
\foreach \pos/\txt in {0/{$F$ at the defect},5.1/{$t=c$},10.2/{$F$ at the defect}}{\node at(\pos,-2.25){\txt};}
\foreach \pos/\txt in {0/{$F$ at the particle},5.1/{Output $|ab\rangle$}}{\node at(\pos,-9.15){\txt};}
\begin{scope}[shift={(0,0)},x=.6cm,y=.6cm,line width=.9pt]\rawinputstate\end{scope}
\begin{scope}[shift={(5.1,0)},x=.6cm,y=.6cm,line width=.9pt]\unresolvedaction\end{scope}
\begin{scope}[shift={(10.2,0)},x=.6cm,y=.6cm,line width=.9pt]\wholeinitial\cutports\end{scope}
\begin{scope}[shift={(0,-6.9)},x=.6cm,y=.6cm,line width=.9pt]\wholefirstF\cutports\end{scope}
\begin{scope}[shift={(5.1,-6.9)},x=.6cm,y=.6cm,line width=.9pt]\wholeselected\cutports\end{scope}
\begin{scope}[shift={(10.2,-6.9)},x=.6cm,y=.6cm,line width=.9pt]\wholesecondF\cutports\end{scope}
\begin{scope}[shift={(0,-13.8)},x=.6cm,y=.6cm,line width=.9pt]\wholethirdF\cutports\end{scope}
\begin{scope}[shift={(5.1,-13.8)},x=.6cm,y=.6cm,line width=.9pt]\standardoutputstate\end{scope}
\draw[->,thick](1.9,1.85)--(3.75,1.85)node[midway,above=5pt]{$\sum_c R^{x X_1}_c$};
\draw[->,thick](7,1.85)--(8.85,1.85)node[midway,above=5pt]{$\sum_a$};
\node[inner sep=2pt]at(5.1,-1.1){$\sum_t\F{x}{X_1}{v}{a}{t}{u}$};
\draw[->,thick](1.9,-5.05)--(3.75,-5.05)node[midway,above=5pt]{$\delta_{tc}$};
\draw[->,thick](7,-5.05)--(8.85,-5.05)node[midway,above=5pt,font=\scriptsize]{$\sum_b\F{X_1}{x}{v}{a}{c}{b}$};
\node[inner sep=2pt]at(5.1,-8.0){$\sum_{a'}\F{x}{u}{X_i}{b}{a'}{v}$};
\draw[->,thick](1.9,-11.95)--(3.75,-11.95)node[midway,above=5pt]{$\delta_{aa'}$};
\end{tikzpicture}
\endgroup}
\caption{Recoupling the spatial insertion. Read each row from left to right. After attaching the $c$ junction, three $F$-moves and two vertex contractions give the output junctions of Figure~\ref{fig:B_uvcyl}.}
\label{fig:B_D}
\end{figure}
Reading the three $F$ coefficients and summing the braiding gives the vacuum basis matrix element
\begin{equation}
\displaystyle
(X_x)_{ab,uv}
=\sum_{c:\,N_{x X_1}^{c}=1}R^{x X_1}_c
\F{x}{u}{X_i}{b}{a}{v}\F{x}{X_1}{v}{a}{c}{u}\F{X_1}{x}{v}{a}{c}{b}.
\label{eq:B_Dvac}
\end{equation}
The row $ab$ is the output and the column $uv$ is the input. In the conventions of Appendix~\ref{app:recoupling}, the braiding coefficient is
\begin{equation}
R^{rs}_t=(-1)^{t-r-s}\exp\!\left[-\frac{\pi i}{k+2}\{t(t+1)-r(r+1)-s(s+1)\}\right].
\label{eq:B_R}
\end{equation}
The quantum dimensions and twists are $d_\ell=\frac{S_{0\ell}}{S_{00}}$ and $\theta_\ell=\exp\big(-\frac{2\pi i\ell(\ell+1)}{(k+2)} \big)$. The transport matrix and its momentum eigenvalue are
\begin{equation}
\begin{aligned}
T_{ab,uv}&=\delta_{bu}\F{X_1}{b}{X_i}{b}{a}{v},\qquad
T\ket{\xi}=\ee^{2\pi\ii q_\xi}\ket{\xi},\\
q_\xi&=-\frac{p(p+1)-q(q+1)}{k+2}\pmod1.
\end{aligned}
\label{eq:B_Tvac}
\end{equation}
Here $q_\xi$ is the fractional momentum shift of channel $\xi$. The physical momentum is $p_{\xi,n}$ in Eq.~\eqref{eq:pquant}. For even $k$, the reflection changes the expression for $q_\xi$ before reduction modulo one by the integer $p-q$, so the transport eigenvalue is independent of the representative. Together, $T$ and $X_1$ distinguish the allowed channels. The eigenvalue $\frac{S_{X_1 p}}{S_{0p}}=1+2\cos\left(\frac{2(2p+1)\pi}{k+2}\right)$ is the same for the two labels $p$ and $\frac k2-p$. For odd $k$ the reflected label is absent from the integer-spin channel list. For even $k$ the simultaneous identification lets us compare channels with the same $p$. Fusion with the fixed defect $X_1$ permits only $q=p-1,p,p+1$; the particle condition $N_{pq}^{X_i}=1$ selects a subset. The phases $\theta_p/\theta_q$ for $q=p,p-1,p+1$ are $1$, $\ee^{-4\pi\ii p/(k+2)}$ and $\ee^{4\pi\ii(p+1)/(k+2)}$, respectively. An allowed neighbour phase differs from one, and the two neighbour phases coincide only at $p=k/4$, where their labels are in the same identified orbit. Thus the transport phase and the spatial-line eigenvalue determine the channel.

For the $R_1$ multiplet, $i=1$ and the two fusion conditions reduce to $N_{pq}^{X_1}=1$. The diagonal pairs $q=p$ have $q_\xi=0$ and satisfy $1\leq 2p\leq k-1$. For odd $k$ the integer labels give $p=1,\ldots,J$. For even $k$ the reflection identifies $2p$ with $k-2p$, and the fixed pair $p=q=k/4$ contributes for $i=1$. In either case there are $J$ channels with eigenvalue one.

For odd $k$, the pairs $(p,p+1)$ have $p=0,\ldots,J-1$, so the shifts $(2p+2)/(k+2)$ have numerators $2,4,\ldots,k-1$. The reversed pairs give the shifts with numerators $3,5,\ldots,k$ modulo $k+2$. For even $k$, we can count the off-diagonal orbits using the unique representative of the form $(p,p+1)$ in each orbit, with $2p=0,1,\ldots,k-2$. Its shift has numerator $2p+2$, so the same set of nonzero shifts $j/(k+2)$ occurs once in either case. Taking the product over the momentum channels gives
\begin{equation*}
\det(\lambda-T)=\prod_\xi\left(\lambda-\ee^{2\pi\ii q_\xi}\right)
=(\lambda-1)^J\prod_{j=2}^{k}\left(\lambda-\ee^{\frac{2\pi\ii j}{k+2}}\right),
\end{equation*}
which is Eq.~\eqref{eq:charpoly}.

We now determine the internal weights from the two matrix elements in Eqs.~\eqref{eq:B_Dmomentum} and~\eqref{eq:B_Dvac}. As in Eq.~\eqref{eq:dictionary}, $U_{ab,\xi}=\langle ab|\xi\rangle$. In the pair labels this entry is $U_{ab,(p,q)}$. The allowed vacuum pairs satisfy $N_{a X_i}^{b}N_{X_1 b}^{a}=1$. Acting with $X_x$ on the momentum state gives
\begin{equation}
\langle ab|X_x|p,q\rangle
=\frac{S_{xp}}{S_{0p}}\langle ab|p,q\rangle
=\frac{S_{xp}}{S_{0p}}U_{ab,(p,q)}.
\label{eq:B_mixedmomentum}
\end{equation}

For a fixed vacuum pair $ab$, expand the ket in the diagonal matrix element $\langle ab|X_x|ab\rangle$ in the momentum basis using Eq.~\eqref{eq:dictionary}. This gives $\sum_{p,q} U_{ab,(p,q)}^*\langle ab|X_x|p,q\rangle$. Substituting Eq.~\eqref{eq:B_mixedmomentum} then yields
\begin{equation}
\langle ab|X_x|ab\rangle
=\sum_{p,q}\frac{S_{xp}}{S_{0p}}|U_{ab,(p,q)}|^2.
\label{eq:B_diagonalspectral}
\end{equation}
The same calculation with $T^mX_x$ retains the transport eigenvalue $\tau_{pq}=\ee^{2\pi\ii q_\xi}=\theta_p/\theta_q$ and gives, for each nonnegative integer $m$,
\begin{equation}
\begin{aligned}
\langle ab|T^mX_x|ab\rangle
&=\sum_{p,q}\langle ab|T^mX_x|p,q\rangle U_{ab,(p,q)}^*\\
&=\sum_{p,q}\frac{S_{xp}}{S_{0p}}\tau_{pq}^{m}|U_{ab,(p,q)}|^2.
\end{aligned}
\label{eq:B_spectralmoments}
\end{equation}

Let $\mathcal J_p$ be the set of allowed second spin labels for a given first spin label $p$, using the representatives specified above for even $k$. We first separate the label $p$ using modular orthogonality. The full $S$ matrix is unitary and obeys $S_{x,k/2-r}=(-1)^{2x}S_{xr}$. Summing only over the preserved integer-spin lines therefore gives
\begin{equation}
\sum_{x=0}^{J}S_{xp}^*S_{xr}
=\frac12\left(\delta_{pr}+\delta_{p,k/2-r}\right).
\label{eq:B_Sorthogonal}
\end{equation}
Rename the summed spin labels in~Eq.~\eqref{eq:B_spectralmoments} as $(r,s)$, multiply by $S_{xp}^*$, sum over $x$, and use~Eq.~\eqref{eq:B_Sorthogonal}. Among the chosen representatives, the reflected first label coincides with $p$ only at $p=k/4$. Restoring the factor $S_{0p}$ from the eigenvalue denominator then gives
\begin{equation}
\frac{2S_{0p}}{1+\delta_{p,k/4}}
\sum_{x=0}^{J}S_{xp}^*\langle ab|T^mX_x|ab\rangle
=\sum_{s\in\mathcal J_p}\tau_{ps}^{m}|U_{ab,(p,s)}|^2.
\label{eq:B_resolvedmoments}
\end{equation}
The remaining transport phases are distinct, as established above. For a selected $q\in\mathcal J_p$, the polynomial
\begin{equation}
\left.\prod_{r\in\mathcal J_p\setminus\{q\}}
\frac{z-\tau_{pr}}{\tau_{pq}-\tau_{pr}}\right|_{z=\tau_{ps}}
=\delta_{sq}
\label{eq:B_phaseinterpolation}
\end{equation}
selects this one term in~Eq.~\eqref{eq:B_resolvedmoments}: at $s=q$ each factor is one, while for $s\ne q$ the factor $r=s$ vanishes. Taking the corresponding linear combination of the equations for $m=0,\ldots,|\mathcal J_p|-1$ gives
\begin{equation}
|U_{ab,(p,q)}|^2
=\frac{2S_{0p}}{1+\delta_{p,k/4}}
\sum_{x=0}^{J}S_{xp}^*
\left\langle ab\left|
\left[\prod_{r\in\mathcal J_p\setminus\{q\}}
\frac{T-\tau_{pr}\mathrm I}{\tau_{pq}-\tau_{pr}}\right]X_x
\right|ab\right\rangle.
\label{eq:B_weightprojection}
\end{equation}
The modular sum selects the first label $p$, and the polynomial selects the second label $q$. Their product is the orthogonal projector onto the momentum state $\ket{\xi}=\ket{p,q}$; its trace is one, which fixes the normalization of the weights.

To express~Eq.~\eqref{eq:B_weightprojection} entirely in fusion data, expand the polynomial and evaluate the vacuum-basis products. Choosing $T$ from the factors labelled by a subset $E\subseteq\mathcal J_p\setminus\{q\}$ gives $T^{|E|}$; every unchosen factor contributes $-\tau_{pr}$. The numerator consequently expands as
\begin{equation}
\prod_{r\in\mathcal J_p\setminus\{q\}}(T-\tau_{pr}\mathrm I)
=\sum_{E\subseteq\mathcal J_p\setminus\{q\}}
(-1)^{|\mathcal J_p|-1-|E|}
\left[\prod_{r\in\mathcal J_p\setminus(\{q\}\cup E)}\tau_{pr}\right]T^{|E|}.
\label{eq:B_polynomialexpanded}
\end{equation}
Since fusion with $X_1$ allows at most three second labels, only the powers $T^0,T,T^2$ occur. Their matrix elements follow directly from the Kronecker delta in~Eq.~\eqref{eq:B_Tvac}. In particular,
\begin{equation}
\begin{aligned}
\langle ab|TX_x|ab\rangle
&=\sum_{u,v}\delta_{bu}\F{X_1}{b}{X_i}{b}{a}{v}(X_x)_{uv,ab}\\
&=\sum_v\F{X_1}{b}{X_i}{b}{a}{v}(X_x)_{bv,ab},\\
\langle ab|T^2X_x|ab\rangle
&=\sum_{v,w}\F{X_1}{b}{X_i}{b}{a}{v}
\F{X_1}{v}{X_i}{v}{b}{w}(X_x)_{vw,ab}.
\end{aligned}
\label{eq:B_transportproducts}
\end{equation}
Thus each transport replaces the ordered vacuum pair by the next pair along the sequence $a_0=a,a_1=b,a_2,\ldots$. Substituting~Eq.~\eqref{eq:B_Dvac} into these products and then into~Eq.~\eqref{eq:B_weightprojection} gives general expression
\begin{equation}
\begin{aligned}
|U_{ab,(p,q)}|^2
={}&\frac{2S_{0p}}
{(1+\delta_{p,k/4})\displaystyle\prod_{r\in\mathcal J_p\setminus\{q\}}(\tau_{pq}-\tau_{pr})}\\
&\times\sum_{E\subseteq\mathcal J_p\setminus\{q\}}
(-1)^{|\mathcal J_p|-1-|E|}
\left[\prod_{r\in\mathcal J_p\setminus(\{q\}\cup E)}\tau_{pr}\right]\\
&\times\sum_{x,c,a_2,\ldots,a_{|E|+1}}
S_{xp}^*R^{xX_1}_c
\left[\prod_{t=0}^{|E|-1}
\F{X_1}{a_{t+1}}{X_i}{a_{t+1}}{a_t}{a_{t+2}}\right]\\
&\qquad\times\F{x}{a}{X_i}{a_{|E|+1}}{a_{|E|}}{b}
\F{x}{X_1}{b}{a_{|E|}}{c}{a}
\F{X_1}{x}{b}{a_{|E|}}{c}{a_{|E|+1}}.
\end{aligned}
\label{eq:Ugeneral}
\end{equation}
Here $a_0=a$ and $a_1=b$ are the specified vacuum labels; $x,c,a_2,\ldots$ run over integer spins from $0$ to $J$, with only admissible fusion vertices included. An empty product equals one and an empty list of vacuum indices requires no sum. The transport eigenvalues are $\tau_{pr}=\exp\left(-2\pi i\frac{p(p+1)-r(r+1)}{(k+2)}\right)$.

In the ordered phase of $\mathcal M_{5,6}$, $k=4$ and $\alpha=X_1$. We use the row order $(01,10,12,21,11)$ and the column order $(0_A,0_B,\tfrac12,+\tfrac13,-\tfrac13)$ of Section~\ref{subsec:M56}. In this order, $\xi=(p,q)$ takes the values $(1,1)$, $(\tfrac12,\tfrac12)$, $(\tfrac12,\tfrac32)$, $(0,1)$ and $(1,0)$. The momentum-basis matrix of $X_1$ is
\begin{equation}
X_1=\operatorname{diag}(-1,0,0,2,-1).
\label{eq:B_M56Dmomentum}
\end{equation}
The shifts are $(0,0,\tfrac12,\tfrac13,-\tfrac13)$, so the two zero-momentum channels are separated by their $X_1$ eigenvalues. Equation~\eqref{eq:Ugeneral} gives the squared amplitudes $|U_{ab,\xi}|^2$ in this order,
\begin{equation}
|U_{ab,\xi}|^2
=\begin{pmatrix}
\frac16&\frac14&\frac14&\frac16&\frac16\\[3pt]
\frac16&\frac14&\frac14&\frac16&\frac16\\[3pt]
\frac16&\frac14&\frac14&\frac16&\frac16\\[3pt]
\frac16&\frac14&\frac14&\frac16&\frac16\\[3pt]
\frac13&0&0&\frac13&\frac13
\end{pmatrix}.
\label{eq:B_M56weights}
\end{equation}

\endgroup

\section{Mass degeneracy of quiver representation}
\label{app:numerics}

At fixed circumference $L$, we extrapolate the periodic gaps and the $X_1$ defect gaps of $\mathcal M_{4,5}$ linearly in $1/N_{\mathrm{cut}}$. For the $X_1$ defect in both phases of $\mathcal M_{5,6}$ and in $\mathcal M_{6,7}$, we fit the rest masses quadratically. The two fit forms are
\begin{align*}
\Delta E(L,N_{\mathrm{cut}}) &= \Delta E^{(\infty)}(L) + \frac{a}{N_{\mathrm{cut}}}\,,\\
m_\xi(L,N_{\mathrm{cut}}) &= m_\xi^{(\infty)}(L) + \frac{b_\xi}{N_{\mathrm{cut}}} + \frac{c_\xi}{N_{\mathrm{cut}}^2}\,.
\end{align*}
All fits use unweighted least squares, with $N_{\mathrm{cut}}=10,12,14,16$ for $\mathcal M_{4,5}$ and $N_{\mathrm{cut}}=8,10,12,14,15$ for the other three deformations. Equation~\eqref{eq:mest} gives the rest mass from the gap above the defect vacuum. We apply it to the extrapolated gap in $\mathcal M_{4,5}$ and at each cutoff before fitting the masses in the other three cases.

We first obtain the cutoff-extrapolated channel masses $m_\xi^{(\infty)}$ in the momentum basis, then use the measured weights $\widehat w_{ab,\xi}$ from Eq.~\eqref{eq:reconstructedweights} in Eq.~\eqref{eq:vacuumweightmass} to obtain $M_{ab}$ in the circle basis. We measure the weights at $N_{\mathrm{cut}}=5$, using $L=100$ for $\mathcal M_{4,5}$ and $L=60$ for the other three deformations. For each deformation we use the same weights at every $L$.

\begin{table}[H]
\centering
\small
\setlength{\tabcolsep}{4pt}
\renewcommand{\arraystretch}{0.94}
\begin{tabular}{lclrrr}
\toprule
Model and coupling & Multiplet & Component or level & Mass & Exact mass & Mass spread \\
\midrule
$\mathcal M_{4,5}$, $\lambda=-0.02$ & $R_1$ & $01,10$ & $0.15213351$ & $0.152215$ & $3.46\times10^{-6}$ \\
 & & $11$ & $0.15213005$ & & \\
\addlinespace[3pt]
$\mathcal M_{5,6}$, $\lambda=-0.02$ & $R_1$ & $01,10,12,21$ & $0.17269915$ & $0.172910$ & $2.77\times10^{-5}$ \\
 & & $11$ & $0.17272681$ & & \\
\addlinespace[3pt]
$\mathcal M_{5,6}$, $\lambda=+0.02$ & $R_1$ & $00$ & $0.17268885$ & $0.172910$ & $6.13\times10^{-5}$ \\
 & & $01,10$ & $0.17275019$ & & \\
 & & $11$ & $0.17269175$ & & \\
\addlinespace[3pt]
$\mathcal M_{6,7}$, $\lambda=-0.02$ & $R_1$ & $01,10$ & $0.18809550$ & $0.188322$ & $4.31\times10^{-5}$ \\
 & & $12,21$ & $0.18813099$ & & \\
 & & $11$ & $0.18811446$ & & \\
 & & $22$ & $0.18808792$ & & \\
\addlinespace[3pt]
$\mathcal M_{6,7}$, $\lambda=-0.02$ & $R_0$ & $n=5$ & $0.371112572$ & $0.374582$ & $7.77\times10^{-3}$ \\
 & & $n=6$ & $0.374088567$ & & \\
 & & $n=7$ & $0.378885502$ & & \\
\bottomrule
\end{tabular}
\caption{Kink and particle masses in $\mathcal M_{k+1,k+2}+\lambda\,\phi_{2,1}$. For $R_1$, each entry is the mass $M_{ab}$ from Eq.~\eqref{eq:vacuumweightmass}. For $R_0$, the entries are the cutoff-extrapolated periodic gaps $E_n-E_0$ at $L=60$. The last column gives the difference between the largest and smallest entries in each multiplet.}
\label{tab:quiver_component_masses}
\end{table}

\begin{figure}[H]
\centering
\includegraphics[width=\linewidth]{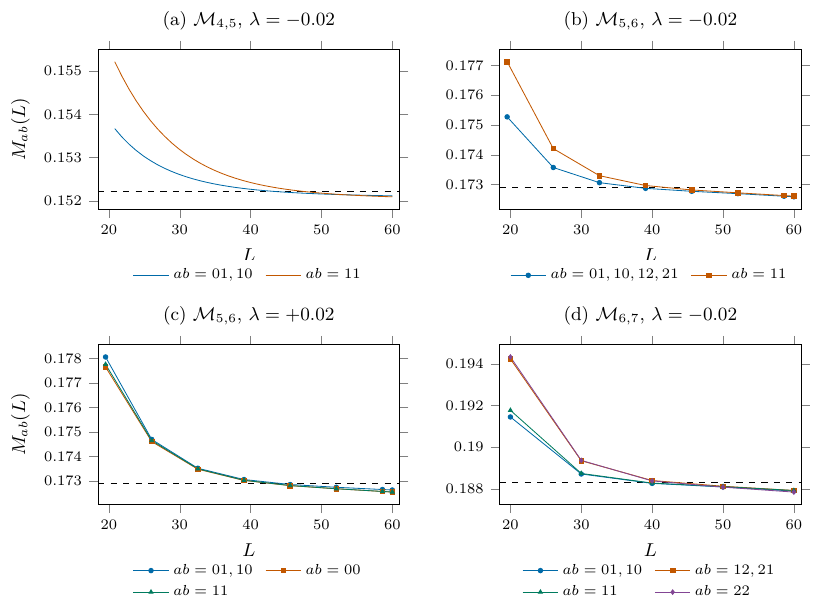}
\caption{Masses $M_{ab}(L)$ of the light $R_1$ multiplets in the vacuum-pair basis for the four $\phi_{2,1}$ deformations. The solid curves are obtained by combining the cutoff-extrapolated momentum-channel masses with the numerical weights $\widehat w_{ab,\xi}$ in Eq.~\eqref{eq:vacuumweightmass}. Vacuum pairs with coincident curves share a legend entry. The dashed line is the exact light mass from Eqs.~\eqref{eq:exact_light_mass} and \eqref{eq:mass_coupling}.}
\label{fig:quiver_mass_degeneracy}
\end{figure}

Figure~\ref{fig:quiver_mass_degeneracy} shows the masses $M_{ab}(L)$ of the light $R_1$ multiplet in the vacuum-pair basis. For Table~\ref{tab:quiver_component_masses}, we use $L=50,55,60$ for $\mathcal M_{4,5}$, $L=52.100073$ for both phases of $\mathcal M_{5,6}$ and $L=50$ for the light multiplet of $\mathcal M_{6,7}$. The exact light mass $m_L$ is given in Eqs.~\eqref{eq:exact_light_mass} and~\eqref{eq:mass_coupling}. At the volumes used for Table~\ref{tab:quiver_component_masses}, the largest mass difference within each light multiplet is less than $0.04\%$ of the corresponding exact mass. In $\mathcal M_{5,6}$, the change of vacuum module replaces the four oriented kinks and one particle of the ordered phase with two oriented kinks and two particles in the disordered phase, at the same exact mass.

For the heavy $R_0$ of $\mathcal M_{6,7}$, the table gives the three zero-momentum gaps from the periodic spectrum at $L=60$ after cutoff extrapolation. Their mean is close to the exact mass obtained from Eq.~\eqref{eq:sm_breather_mass}. The spread is larger than for the light multiplet, consistent with the residual volume and cutoff dependence discussed in Section~\ref{subsec:m67_quiver_reconstruction}.

\section{Exact spectra from integrable scattering matrices}
\label{app:sm}

We compare the multiplets found by TCSA with the exact restricted solid on
solid (RSOS) scattering description of the massive $\phi_{2,1}$
deformations~\cite{Smirnov:1991uw,Colomo:1991gw,TakacsWatts2002RSOS,MussardoTakacs2009}.
Its elementary excitations $K_{ab}$ have a common mass $m_L$ and interpolate
between vacua $a$ and $b$. Thus $a\ne b$ describes a kink, while $a=b$
describes a particle. The allowed transitions follow the fusion adjacency
of the object $X_1$ of spin one and therefore realize the $R_1$ quiver
on the appropriate vacuum module. For the $\phi_{2,1}$ perturbation of $\mathcal M_{k+1,k+2}$ the mass--coupling relation is~\cite{Fateev:1993av,Fateev:1997yg}
\begin{align}
 m_L &= C_k|\lambda|^{\frac{2(k+1)}{3k}}\,,\label{eq:exact_light_mass}\\
 C_k &= \frac{
  2^{\frac{k-3}{3k}}\sqrt{3}\,
  \Gamma\!\left(\frac{1}{3}\right)
  \Gamma\!\left(\frac{k+2}{3k}\right)
  }{
  \pi\,\Gamma\!\left(\frac{2(k+1)}{3k}\right)
  }
  \left(
  \frac{
  \pi^2\,
  \Gamma^2\!\left(\frac{3k+2}{4(k+1)}\right)
  \Gamma\!\left(\frac{1}{2}-\frac{1}{k+1}\right)
  }{
  \Gamma^2\!\left(\frac{k+2}{4(k+1)}\right)
  \Gamma\!\left(\frac{1}{2}+\frac{1}{k+1}\right)
  }
  \right)^{\frac{k+1}{3k}}\,.
 \label{eq:mass_coupling}
\end{align}
For $k=3,4,5$, the regular modules give, respectively, two kinks and one
particle, four kinks and one particle, and four kinks and two particles.
The $k=4$ regular module describes the ordered phase at $\lambda<0$. The disordered
phase at $\lambda>0$ instead has RSOS vacuum labels $\{\tfrac12,\tfrac32\}$. These label the two vacua of the module $\mathcal M_H$ in Section~\ref{subsec:draft2_m56_two_vacua}, giving two kinks and two particles in the light multiplet~\cite{MussardoTakacs2009}.

Additional stable excitations are determined by the bound state poles of
the scattering amplitudes. Let $\theta$ be the rapidity difference of two
light excitations, and write $S_{KK}(\theta)=\sum_j s_j(\theta)P_j$ in a
fusion basis, where $P_j$ projects onto the total fusion channel $X_j$.
The allowed channels are $j=0,1$ for $k=3$ and $j=0,1,2$ for $k=4,5$.
In particular, the amplitude in the unit channel is
\begin{align}
 s_0(\theta)
 &=-\sigma(\theta)
 \frac{\cosh(\Lambda\theta-3i\alpha)}
      {\cosh(\Lambda\theta+3i\alpha)}\,,
 &\Lambda&=\frac{3k}{2(k+2)}\,,\qquad
 \alpha=\frac{\pi}{k+2}\,,
 \label{eq:sm_singlet}
\end{align}
where $\Lambda$ is a dimensionless scattering parameter. The scalar factor
$\sigma(\theta)$ is analytic and nonzero for
$|\operatorname{Im}\theta|<\frac{\pi}{3}$, and $\sigma(iu)>0$ for real
$u$ in this interval. Its integral expression is given in Eq.~(2.26)
of~\cite{DoreyPocklingtonTateo2003}.
The denominator in~Eq.~\eqref{eq:sm_singlet} vanishes at
\begin{align}
 \theta=iu_H\,,\qquad
 u_H=\pi\left(1-\frac{1}{\Lambda}\right)
     =\frac{\pi(k-4)}{3k}\,.
 \label{eq:sm_breather_pole}
\end{align}
For $k=5$, this is a bound state pole in the physical strip
$0<\operatorname{Im}\theta<\pi$. The other fusion channels are regular
there, and the full amplitude has the residue
\begin{align}
 S_{KK}(\theta)
 &\underset{\theta\to iu_H}{\sim}
 \frac{i g_H^2}{\theta-iu_H}P_0\,,
 &g_H^2&=\frac{\sigma(iu_H)\sin(6\alpha)}{\Lambda}>0\,,
 \label{eq:sm_breather_residue}
\end{align}
Here $g_H$ is the real coupling of two light excitations to the heavy bound state. Since $P_0$ selects the unit fusion channel, the bound states form the heavy multiplet $R_0$, with one particle above each vacuum. In $\mathcal M_{6,7}$ this gives three degenerate heavy particles. Relativistic kinematics at the pole fixes their mass to~\cite{DoreyPocklingtonTateo2002II}
\begin{align}
 m_H^2&=2m_L^2(1+\cos u_H)\,, &
 \frac{m_H}{m_L}&=2\cos\left(\frac{u_H}{2}\right)
 =2\cos\left(\frac{\pi}{30}\right)
 \simeq1.98904\,.
 \label{eq:sm_breather_mass}
\end{align}
For $k=3$, $u_H<0$ lies outside the physical strip. For $k=4$, the apparent
singularity at $u_H=0$ is removable: the ratio of hyperbolic cosines in
Eq.~\eqref{eq:sm_singlet} equals $-1$, so $s_0=\sigma$. Consequently, neither
model has this additional bound state~\cite{DoreyPocklingtonTateo2002II}.

The bootstrap constructs the scattering amplitudes of the heavy particles from the amplitudes of their two light constituents at rapidity difference $\ii u_H$. It closes on the light excitations for $k=3,4$ and includes the heavy multiplet for $k=5$. The remaining poles are accounted for by exchange or on-shell rescattering of these excitations~\cite{DoreyPocklingtonTateo2003,DoreyPocklingtonTateo2002II,TakacsWatts2002RSOS}.
The stable multiplets are therefore $R_1$ for $k=3,4$ and
$R_1\oplus R_0$ for $k=5$, in agreement with Table~\ref{tab:spectrum-summary}.

Scattering states are formed by sequences of these excitations with real
rapidities. A constituent of mass $m_i$ and rapidity $\vartheta_i$ has
energy $m_i\cosh\vartheta_i$ and momentum $m_i\sinh\vartheta_i$.
On a periodic circle the vacuum sequence must close, and the eigenphases
acquired by moving an excitation around the circle determine the allowed
rapidities through Bethe--Yang quantization~\cite{Klassen:1992qy,TakacsWatts2002RSOS}. At large circumference their
energies above the vacuum approach the sum of the constituent energies.
At fixed Bethe quantum numbers, the levels of two light excitations approach
the threshold $2m_L$ as the circumference grows, whereas the infinite volume
mass of the heavy $R_0$ particles at $k=5$ lies below it.

\bibliographystyle{unsrt}
\bibliography{references}

@article{CHO,
    author = "Cordova, Clay and Holfester, Nicholas and Ohmori, Kantaro",
    title = "{Representation theory of solitons}",
    eprint = "2408.11045",
    archivePrefix = "arXiv",
    primaryClass = "hep-th",
    doi = "10.1007/JHEP06(2025)001",
    journal = "JHEP",
    volume = "06",
    pages = "001",
    year = "2025"
}

@article{CLSWY,
  author       = {Chang, C.-M. and Lin, Y.-H. and Shao, S.-H. and Wang, Y. and Yin, X.},
  title        = {Topological Defect Lines and Renormalization Group Flows in Two Dimensions},
  journal      = {JHEP},
  volume       = {01},
  pages        = {026},
  year         = {2019},
  eprint       = {1802.04445},
  archivePrefix= {arXiv},
  url          = {https://arxiv.org/abs/1802.04445v3}
}

@article{YZ,
    author = "Yurov, V. P. and Zamolodchikov, A. B.",
    title        = {Truncated conformal space approach to scaling Lee--Yang model},
    reportNumber = "ITEP-89-161",
    doi = "10.1142/S0217751X9000218X",
    journal = "Int. J. Mod. Phys. A",
    volume = "5",
    pages = "3221--3246",
    year = "1990"
}

@article{LMC,
    author = "Lassig, Michael and Mussardo, Giuseppe and Cardy, John L.",
    title = "{The scaling region of the tricritical Ising model in two-dimensions}",
    reportNumber = "UCSBTH-90-28",
    doi = "10.1016/0550-3213(91)90206-D",
    journal = "Nucl. Phys. B",
    volume = "348",
    pages = "591--618",
    year = "1991"
}

@article{CFTCSA,
    author = "Horvath, D. X. and Hodsagi, K. and Takacs, G.",
    title = "{Chirally factorised truncated conformal space approach}",
    eprint = "2201.06509",
    archivePrefix = "arXiv",
    primaryClass = "hep-th",
    doi = "10.1016/j.cpc.2022.108376",
    journal = "Comput. Phys. Commun.",
    volume = "277",
    pages = "108376",
    year = "2022"
}

@article{MM,
    author = "Mossa, Alessandro and Mussardo, Giuseppe",
    title = "{Analytic properties of the free energy: The Tricritical Ising model}",
    eprint = "0710.0991",
    archivePrefix = "arXiv",
    primaryClass = "hep-th",
    doi = "10.1088/1742-5468/2008/03/P03010",
    journal = "J. Stat. Mech.",
    volume = "0803",
    pages = "P03010",
    year = "2008"
}

@article{DLY,
    author = "Bajnok, Zolt{\'a}n and Holl{\'o}, L{\'a}szl{\'o} and Watts, Gerard",
    title = "{Defect scaling Lee{\textendash}Yang model from the perturbed DCFT point of view}",
    eprint = "1307.4536",
    archivePrefix = "arXiv",
    primaryClass = "hep-th",
    reportNumber = "ITP-BUDAPEST-REPORT-661",
    doi = "10.1016/j.nuclphysb.2014.06.019",
    journal = "Nucl. Phys. B",
    volume = "886",
    pages = "93--124",
    year = "2014"
}

@article{Mong,
  author       = {Mong, R. S. K. and others},
  title        = {Universal topological quantum computation from a superconductor/Abelian quantum Hall heterostructure},
  journal      = {Phys. Rev. X},
  volume       = {4},
  pages        = {011036},
  year         = {2014},
  eprint       = {1307.4403},
  archivePrefix= {arXiv},
  url          = {https://arxiv.org/abs/1307.4403}
}

@article{GKO,
  author       = {Goddard, P. and Kent, A. and Olive, D.},
  title        = {Unitary Representations of the Virasoro and Super-Virasoro Algebras},
  journal      = {Commun. Math. Phys.},
  volume       = {103},
  pages        = {105--119},
  year         = {1986},
  doi          = {10.1007/BF01464283}
}

@article{DG,
    author = "Dittrich, Bianca and Geiller, Marc",
    title = "{Quantum gravity kinematics from extended TQFTs}",
    eprint = "1604.05195",
    archivePrefix = "arXiv",
    primaryClass = "hep-th",
    doi = "10.1088/1367-2630/aa54e2",
    journal = "New J. Phys.",
    volume = "19",
    number = "1",
    pages = "013003",
    year = "2017"
}

@article{Muger,
  author       = {M{"u}ger, M.},
  title        = {From Subfactors to Categories and Topology II. The quantum double of tensor categories and subfactors},
  journal      = {J. Pure Appl. Algebra},
  volume       = {180},
  pages        = {159--219},
  year         = {2003},
  eprint       = {math/0111205},
  archivePrefix= {arXiv},
  url          = {https://arxiv.org/abs/math/0111205}
}

@article{Bischoff,
    author = "Bischoff, Marcel",
    title = "{A Remark on CFT Realization of Quantum Doubles of Subfactors: Case Index ${ < 4}$}",
    eprint = "1506.02606",
    archivePrefix = "arXiv",
    primaryClass = "math-ph",
    doi = "10.1007/s11005-016-0816-z",
    journal = "Lett. Math. Phys.",
    volume = "106",
    number = "3",
    pages = "341--363",
    year = "2016"
}

@article{DoreyPocklingtonTateo2003,
  author = {Dorey, Patrick and Pocklington, Andrew and Tateo, Roberto},
  title = {Integrable aspects of the scaling q-state Potts models I:
           bound states and bootstrap closure},
  journal = {Nucl. Phys. B},
  volume = {661},
  pages = {425--463},
  year = {2003},
  doi = {10.1016/S0550-3213(03)00181-0},
  eprint = {hep-th/0208111},
  archivePrefix = {arXiv}
}

@article{PozsgayTakacs2008FF,
  author = {Pozsgay, B. and Tak{\'a}cs, G.},
  title = {Form factors in finite volume {I}: Form factor bootstrap and truncated conformal space},
  journal = {Nucl. Phys. B},
  volume = {788},
  pages = {167--208},
  year = {2008},
  doi = {10.1016/j.nuclphysb.2007.06.027},
  eprint = {0706.1445},
  archivePrefix = {arXiv}
}

@article{FZ1985,
  author = {Zamolodchikov, A. B. and Fateev, V. A.},
  title = {Nonlocal (parafermion) currents in two-dimensional conformal quantum field theory and self-dual critical points in {$Z_N$}-symmetric statistical systems},
  journal = {Sov. Phys. JETP},
  volume = {62},
  pages = {215--225},
  year = {1985},
  url = {https://jetp.ras.ru/cgi-bin/dn/e_062_02_0215.pdf}
}

@article{KZ1984,
  author = {Knizhnik, V. G. and Zamolodchikov, A. B.},
  title = {Current algebra and {Wess--Zumino} model in two dimensions},
  journal = {Nucl. Phys. B},
  volume = {247},
  pages = {83--103},
  year = {1984},
  doi = {10.1016/0550-3213(84)90374-2}
}

@article{ARS1999,
  author = {Alekseev, Anton Yu. and Recknagel, Andreas and Schomerus, Volker},
  title = {Non-commutative world-volume geometries: branes on {$SU(2)$} and fuzzy spheres},
  journal = {JHEP},
  volume = {09},
  pages = {023},
  year = {1999},
  eprint = {hep-th/9908040},
  archivePrefix = {arXiv},
  doi = {10.1088/1126-6708/1999/09/023}
}

@article{TakacsWatts2002RSOS,
    author = "Takacs, G. and Watts, G.",
    title = "{RSOS revisited}",
    eprint = "hep-th/0203073",
    archivePrefix = "arXiv",
    reportNumber = "KCL-MTH-02-04, ITP-BUDAPEST-579",
    doi = "10.1016/S0550-3213(02)00676-4",
    journal = "Nucl. Phys. B",
    volume = "642",
    pages = "456--482",
    year = "2002"
}

@article{MussardoTakacs2009,
    author = "Mussardo, G. and Takacs, G.",
    title = "{Effective potentials and kink spectra in non-integrable perturbed conformal field theories}",
    eprint = "0901.3537",
    archivePrefix = "arXiv",
    primaryClass = "hep-th",
    reportNumber = "SISSA-04-2009-EP, ITP-BUDAPEST-REPORT-NO-643",
    doi = "10.1088/1751-8113/42/30/304022",
    journal = "J. Phys. A",
    volume = "42",
    pages = "304022",
    year = "2009"
}

@article{DoreyPocklingtonTateo2002II,
    author = "Dorey, Patrick and Pocklington, Andrew and Tateo, Roberto",
    title = "{Integrable aspects of the scaling q state Potts models. 2. Finite size effects}",
    eprint = "hep-th/0208202",
    archivePrefix = "arXiv",
    reportNumber = "DCPT-02-53",
    doi = "10.1016/S0550-3213(03)00182-2",
    journal = "Nucl. Phys. B",
    volume = "661",
    pages = "464--513",
    year = "2003"
}

@article{CGH,
  author = {C{\'o}rdova, Clay and Garc{\'i}a-Sep{\'u}lveda, Diego and Holfester, Nicholas},
  title = {Particle-Soliton Degeneracies from Spontaneously Broken Non-Invertible Symmetry},
  journal = {JHEP},
  volume = {07},
  year = {2024},
  pages = {154},
  doi = {10.1007/JHEP07(2024)154},
  eprint = {2403.08883},
  archivePrefix = {arXiv},
  primaryClass = {hep-th}
}

@article{Gaiotto:2014kfa,
    author = "Gaiotto, Davide and Kapustin, Anton and Seiberg, Nathan and Willett, Brian",
    title = "{Generalized Global Symmetries}",
    eprint = "1412.5148",
    archivePrefix = "arXiv",
    primaryClass = "hep-th",
    doi = "10.1007/JHEP02(2015)172",
    journal = "JHEP",
    volume = "02",
    pages = "172",
    year = "2015"
}

@article{McGreevy:2022oyu,
    author = "McGreevy, John",
    title = "{Generalized Symmetries in Condensed Matter}",
    eprint = "2204.03045",
    archivePrefix = "arXiv",
    primaryClass = "cond-mat.str-el",
    doi = "10.1146/annurev-conmatphys-040721-021029",
    journal = "Ann. Rev. Condensed Matter Phys.",
    volume = "14",
    pages = "57--82",
    year = "2023"
}

@inproceedings{Shao:2023gho,
    author = "Shao, Shu-Heng",
    title = "{What's Done Cannot Be Undone: TASI Lectures on Non-Invertible Symmetries}",
    booktitle = "{Theoretical Advanced Study Institute in Elementary Particle Physics 2023}: {Aspects of Symmetry}",
    eprint = "2308.00747",
    archivePrefix = "arXiv",
    primaryClass = "hep-th",
    reportNumber = "YITP-SB-2023-19",
    month = "8",
    year = "2023"
}

@article{Luo:2023ive,
    author = "Luo, Ran and Wang, Qing-Rui and Wang, Yi-Nan",
    title = "{Lecture notes on generalized symmetries and applications}",
    eprint = "2307.09215",
    archivePrefix = "arXiv",
    primaryClass = "hep-th",
    doi = "10.1016/j.physrep.2024.02.002",
    journal = "Phys. Rept.",
    volume = "1065",
    pages = "1--43",
    year = "2024"
}

@article{Frohlich:2004ef,
    author = "Frohlich, Jurg and Fuchs, Jurgen and Runkel, Ingo and Schweigert, Christoph",
    title = "{Kramers-Wannier duality from conformal defects}",
    eprint = "cond-mat/0404051",
    archivePrefix = "arXiv",
    reportNumber = "HU-EP-04-19",
    doi = "10.1103/PhysRevLett.93.070601",
    journal = "Phys. Rev. Lett.",
    volume = "93",
    pages = "070601",
    year = "2004"
}

@article{Choi:2021kmx,
    author = "Choi, Yichul and Cordova, Clay and Hsin, Po-Shen and Lam, Ho Tat and Shao, Shu-Heng",
    title = "{Noninvertible duality defects in 3+1 dimensions}",
    eprint = "2111.01139",
    archivePrefix = "arXiv",
    primaryClass = "hep-th",
    reportNumber = "MIT-CTP/5359",
    doi = "10.1103/PhysRevD.105.125016",
    journal = "Phys. Rev. D",
    volume = "105",
    number = "12",
    pages = "125016",
    year = "2022"
}

@article{Bhardwaj:2022yxj,
    author = "Bhardwaj, Lakshya and Bottini, Lea E. and Schafer-Nameki, Sakura and Tiwari, Apoorv",
    title = "{Non-invertible higher-categorical symmetries}",
    eprint = "2204.06564",
    archivePrefix = "arXiv",
    primaryClass = "hep-th",
    doi = "10.21468/SciPostPhys.14.1.007",
    journal = "SciPost Phys.",
    volume = "14",
    number = "1",
    pages = "007",
    year = "2023"
}

@article{Kaidi:2022cpf,
    author = "Kaidi, Justin and Ohmori, Kantaro and Zheng, Yunqin",
    title = "{Symmetry TFTs for Non-invertible Defects}",
    eprint = "2209.11062",
    archivePrefix = "arXiv",
    primaryClass = "hep-th",
    doi = "10.1007/s00220-023-04859-7",
    journal = "Commun. Math. Phys.",
    volume = "404",
    number = "2",
    pages = "1021--1124",
    year = "2023"
}

@article{Kaidi:2023maf,
    author = "Kaidi, Justin and Nardoni, Emily and Zafrir, Gabi and Zheng, Yunqin",
    title = "{Symmetry TFTs and anomalies of non-invertible symmetries}",
    eprint = "2301.07112",
    archivePrefix = "arXiv",
    primaryClass = "hep-th",
    doi = "10.1007/JHEP10(2023)053",
    journal = "JHEP",
    volume = "10",
    pages = "053",
    year = "2023"
}

@article{Bhardwaj:2017xup,
    author = "Bhardwaj, Lakshya and Tachikawa, Yuji",
    title = "{On finite symmetries and their gauging in two dimensions}",
    eprint = "1704.02330",
    archivePrefix = "arXiv",
    primaryClass = "hep-th",
    reportNumber = "IPMU-17-0049",
    doi = "10.1007/JHEP03(2018)189",
    journal = "JHEP",
    volume = "03",
    pages = "189",
    year = "2018"
}

@article{Thorngren:2019iar,
    author = "Thorngren, Ryan and Wang, Yifan",
    title = "{Fusion category symmetry. Part I. Anomaly in-flow and gapped phases}",
    eprint = "1912.02817",
    archivePrefix = "arXiv",
    primaryClass = "hep-th",
    reportNumber = "PUPT-2603",
    doi = "10.1007/JHEP04(2024)132",
    journal = "JHEP",
    volume = "04",
    pages = "132",
    year = "2024"
}

@article{Bhardwaj:2023idu,
    author = {Bhardwaj, Lakshya and Bottini, Lea E. and Pajer, Daniel and Sch{\"a}fer-Nameki, Sakura},
    title = "{Gapped phases with non-invertible symmetries: (1+1)d}",
    eprint = "2310.03784",
    archivePrefix = "arXiv",
    primaryClass = "hep-th",
    doi = "10.21468/SciPostPhys.18.1.032",
    journal = "SciPost Phys.",
    volume = "18",
    number = "1",
    pages = "032",
    year = "2025"
}

@misc{Chen:2025qub,
    author = "Chen, Jin and Duan, Zhihao and Jia, Qiang and Lee, Sungjay",
    title = "{Fermionic Non-invertible Symmetry Behind Supersymmetric ADE Solitons}",
    eprint = "2511.22129",
    archivePrefix = "arXiv",
    primaryClass = "hep-th",
    reportNumber = "USTC-ICTS/PCFT-25-54",
    month = "11",
    year = "2025",
    howpublished = "{\href{https://arxiv.org/abs/2511.22129}{arXiv:2511.22129}}"
}

@misc{Chen:2026wyn,
    author = "Chen, Jin and Duan, Zhihao and Jia, Qiang and Lee, Sungjay",
    title = "{Superstrip Algebras and Nonperturbative Spectra in Fermionic Theories}",
    eprint = "2609.12750",
    archivePrefix = "arXiv",
    primaryClass = "hep-th",
    reportNumber = "USTC-ICTS/PCFT-26-59, KIAS-P26044",
    month = "9",
    year = "2026",
    howpublished = "{\href{https://arxiv.org/abs/2609.12750}{arXiv:2609.12750}}"
}

@article{Colomo:1991gw,
    author = "Colomo, F. and Koubek, A. and Mussardo, G.",
    title = "{On the S matrix of the subleading magnetic deformation of the tricritical Ising model in two-dimensions}",
    eprint = "hep-th/9108024",
    archivePrefix = "arXiv",
    reportNumber = "NORDITA-91-47, SISSA-94-91-EP",
    doi = "10.1142/S0217751X92002416",
    journal = "Int. J. Mod. Phys. A",
    volume = "7",
    pages = "5281--5306",
    year = "1992"
}

@article{Ellem:1997vz,
    author = "Ellem, R. M. and Bazhanov, V. V.",
    title = "{Thermodynamic Bethe ansatz for the subleading magnetic perturbation of the tricritical Ising model}",
    eprint = "hep-th/9703026",
    archivePrefix = "arXiv",
    doi = "10.1016/S0550-3213(97)00748-7",
    journal = "Nucl. Phys. B",
    volume = "512",
    pages = "563--580",
    year = "1998"
}

@article{Yurov:1991my,
    author = "Yurov, V. P. and Zamolodchikov, Alexei B.",
    title = "{Truncated fermionic space approach to the critical 2-D Ising model with magnetic field}",
    doi = "10.1142/S0217751X91002161",
    journal = "Int. J. Mod. Phys. A",
    volume = "6",
    pages = "4557--4578",
    year = "1991"
}

@article{Martins:1991jj,
    author = "Martins, Marcio Jose",
    title = "{The Off critical behavior of the multicritical Ising models}",
    reportNumber = "UCSBTH-91-45, SISSA-73-92-EP",
    doi = "10.1142/S0217751X92003513",
    journal = "Int. J. Mod. Phys. A",
    volume = "7",
    pages = "7753--7774",
    year = "1992"
}

@article{Konik:2007cb,
    author = "Konik, Robert M. and Adamov, Yury",
    title = "{A Numerical Renormalization Group for Continuum One-Dimensional Systems}",
    eprint = "cond-mat/0701605",
    archivePrefix = "arXiv",
    doi = "10.1103/PhysRevLett.98.147205",
    journal = "Phys. Rev. Lett.",
    volume = "98",
    number = "14",
    pages = "147205",
    year = "2007"
}

@article{James:2017cpc,
    author = "James, Andrew J. A. and Konik, Robert M. and Lecheminant, Philippe and Robinson, Neil J. and Tsvelik, Alexei M.",
    title = "{Non-perturbative methodologies for low-dimensional strongly-correlated systems: From non-Abelian bosonization to truncated spectrum methods}",
    eprint = "1703.08421",
    archivePrefix = "arXiv",
    primaryClass = "cond-mat.str-el",
    doi = "10.1088/1361-6633/aa91ea",
    journal = "Rept. Prog. Phys.",
    volume = "81",
    number = "4",
    pages = "046002",
    year = "2018"
}

@article{Dorey:1997yg,
    author = "Dorey, Patrick and Pocklington, Andrew and Tateo, Roberto and Watts, Gerard",
    title = "{TBA and TCSA with boundaries and excited states}",
    eprint = "hep-th/9712197",
    archivePrefix = "arXiv",
    reportNumber = "DTP-97-67, KCL-MTH-97-71, SACLAY-SPH-T-97-163",
    doi = "10.1016/S0550-3213(98)00339-3",
    journal = "Nucl. Phys. B",
    volume = "525",
    pages = "641--663",
    year = "1998"
}

@article{Feverati:1998va,
    author = "Feverati, G. and Ravanini, F. and Takacs, G.",
    title = "{Truncated conformal space at c = 1, nonlinear integral equation and quantization rules for multi - soliton states}",
    eprint = "hep-th/9803104",
    archivePrefix = "arXiv",
    reportNumber = "DFUB-98-04",
    doi = "10.1016/S0370-2693(98)00543-7",
    journal = "Phys. Lett. B",
    volume = "430",
    pages = "264--273",
    year = "1998"
}

@article{Lepori:2009ip,
    author = "Lepori, Luca and Toth, Gabor Zsolt and Delfino, Gesualdo",
    title = "{Particle spectrum of the 3-state Potts field theory: A Numerical study}",
    eprint = "0909.2192",
    archivePrefix = "arXiv",
    primaryClass = "hep-th",
    reportNumber = "SISSA-55-2009-EP",
    doi = "10.1088/1742-5468/2009/11/P11007",
    journal = "J. Stat. Mech.",
    volume = "0911",
    pages = "P11007",
    year = "2009"
}

@article{Lencses:2014tba,
    author = "Lencs{\'e}s, M. and Tak{\'a}cs, G.",
    title = "{Excited state TBA and renormalized TCSA in the scaling Potts model}",
    eprint = "1405.3157",
    archivePrefix = "arXiv",
    primaryClass = "hep-th",
    doi = "10.1007/JHEP09(2014)052",
    journal = "JHEP",
    volume = "09",
    pages = "052",
    year = "2014"
}

@article{Kormos:2009sk,
    author = "Kormos, Marton and Runkel, Ingo and Watts, Gerard M. T.",
    title = "{Defect flows in minimal models}",
    eprint = "0907.1497",
    archivePrefix = "arXiv",
    primaryClass = "hep-th",
    reportNumber = "KCL-MTH-09-06",
    doi = "10.1088/1126-6708/2009/11/057",
    journal = "JHEP",
    volume = "11",
    pages = "057",
    year = "2009"
}

@article{Klassen:1992qy,
    author = "Klassen, Timothy R. and Melzer, Ezer",
    title = "{Kinks in finite volume}",
    eprint = "hep-th/9202034",
    archivePrefix = "arXiv",
    reportNumber = "CLNS-92-1130, ITP-SB-92-01",
    doi = "10.1016/0550-3213(92)90656-V",
    journal = "Nucl. Phys. B",
    volume = "382",
    pages = "441--485",
    year = "1992"
}

@article{Petkova:2000ip,
    author = "Petkova, V. B. and Zuber, J. B.",
    title = "{Generalized twisted partition functions}",
    eprint = "hep-th/0011021",
    archivePrefix = "arXiv",
    reportNumber = "UNN-SCM-M-00-07, CERN-TH-2000-322",
    doi = "10.1016/S0370-2693(01)00276-3",
    journal = "Phys. Lett. B",
    volume = "504",
    pages = "157--164",
    year = "2001"
}

@article{Colomo:1992nr,
    author = "Colomo, F. and Koubek, A. and Mussardo, G.",
    title = "{The subleading magnetic deformation of the tricritical Ising model in 2-D as RSOS restriction of the Izergin-Korepin model}",
    eprint = "hep-th/9203003",
    archivePrefix = "arXiv",
    reportNumber = "SISSA-36-92-EP",
    doi = "10.1016/0370-2693(92)92000-7",
    journal = "Phys. Lett. B",
    volume = "274",
    pages = "367--373",
    year = "1992"
}

@article{Lencses:2021igo,
    author = "Lencs{\'e}s, M. and Mussardo, G. and Tak{\'a}cs, G.",
    title = "{Confinement in the tricritical Ising model}",
    eprint = "2111.05360",
    archivePrefix = "arXiv",
    primaryClass = "hep-th",
    doi = "10.1016/j.physletb.2022.137008",
    journal = "Phys. Lett. B",
    volume = "828",
    pages = "137008",
    year = "2022"
}

@article{Ostrik:2001xnt,
    author = "Ostrik, Viktor",
    title = "{Module categories, weak Hopf algebras and modular invariants}",
    eprint = "math/0111139",
    archivePrefix = "arXiv",
    doi = "10.1007/s00031-003-0515-6",
    journal = "Transform. Groups",
    volume = "8",
    number = "2",
    pages = "177--206",
    year = "2003"
}

@article{Huang:2021zvu,
    author = "Huang, Tzu-Chen and Lin, Ying-Hsuan and Seifnashri, Sahand",
    title = "{Construction of two-dimensional topological field theories with non-invertible symmetries}",
    eprint = "2110.02958",
    archivePrefix = "arXiv",
    primaryClass = "hep-th",
    reportNumber = "CALT-TH-2021-035, YITP-SB-2021-19",
    doi = "10.1007/JHEP12(2021)028",
    journal = "JHEP",
    volume = "12",
    pages = "028",
    year = "2021"
}

@article{Petkova:2001ag,
    author = "Petkova, V. B. and Zuber, J. B.",
    title = "{The Many faces of Ocneanu cells}",
    eprint = "hep-th/0101151",
    archivePrefix = "arXiv",
    reportNumber = "ESI-975-2000, UNN-SCM-M-00-10, CERN-TH-2000-355",
    doi = "10.1016/S0550-3213(01)00096-7",
    journal = "Nucl. Phys. B",
    volume = "603",
    pages = "449--496",
    year = "2001"
}

@article{Fuchs:2002cm,
    author = "Fuchs, Jurgen and Runkel, Ingo and Schweigert, Christoph",
    title = "{TFT construction of RCFT correlators 1. Partition functions}",
    eprint = "hep-th/0204148",
    archivePrefix = "arXiv",
    reportNumber = "PAR-LPTHE-02-25",
    doi = "10.1016/S0550-3213(02)00744-7",
    journal = "Nucl. Phys. B",
    volume = "646",
    pages = "353--497",
    year = "2002"
}

@article{Tanizaki:2019rbk,
    author = {Tanizaki, Yuya and {\"U}nsal, Mithat},
    title = "{Modified instanton sum in QCD and higher-groups}",
    eprint = "1912.01033",
    archivePrefix = "arXiv",
    primaryClass = "hep-th",
    doi = "10.1007/JHEP03(2020)123",
    journal = "JHEP",
    volume = "03",
    pages = "123",
    year = "2020"
}

@misc{Chen:2026cqz,
    author = "Chen, Jin and Jia, Qiang",
    title = "{On general background of quantum non-invertible symmetry in 2D}",
    eprint = "2609.09550",
    archivePrefix = "arXiv",
    primaryClass = "hep-th",
    month = "9",
    year = "2026",
    howpublished = "{\href{https://arxiv.org/abs/2609.09550}{arXiv:2609.09550}}"
}

@article{Feverati:2006ni,
    author = "Feverati, Giovanni and Graham, Kevin and Pearce, Paul A. and Toth, Gabor Zs. and Watts, Gerard",
    title = "{A Renormalisation group for the truncated conformal space approach}",
    eprint = "hep-th/0612203",
    archivePrefix = "arXiv",
    reportNumber = "KCL-MTH-06-17",
    doi = "10.1088/1742-5468/2008/03/P03011",
    journal = "J. Stat. Mech.",
    volume = "0803",
    pages = "P03011",
    year = "2008"
}

@misc{Giokas:2011ix,
    author = "Giokas, Philip and Watts, Gerard",
    title = "{The renormalisation group for the truncated conformal space approach on the cylinder}",
    eprint = "1106.2448",
    archivePrefix = "arXiv",
    primaryClass = "hep-th",
    reportNumber = "KCL-MTH-11-10",
    month = "6",
    year = "2011",
    howpublished = "{\href{https://arxiv.org/abs/1106.2448}{arXiv:1106.2448}}"
}

@article{Pozsgay:2007gx,
    author = "Pozsgay, B. and Takacs, G.",
    title = "{Form factors in finite volume. II. Disconnected terms and finite temperature correlators}",
    eprint = "0706.3605",
    archivePrefix = "arXiv",
    primaryClass = "hep-th",
    reportNumber = "ITP-BUDAPEST-635",
    doi = "10.1016/j.nuclphysb.2007.07.008",
    journal = "Nucl. Phys. B",
    volume = "788",
    pages = "209--251",
    year = "2008"
}

@article{Smirnov:1991uw,
    author = "Smirnov, F. A.",
    title = "{Exact S matrices for phi(1,2) perturbated minimal models of conformal field theory}",
    doi = "10.1142/S0217751X91000745",
    journal = "Int. J. Mod. Phys. A",
    volume = "6",
    pages = "1407--1428",
    year = "1991"
}

@article{Fateev:1993av,
    author = "Fateev, V. A.",
    title = "{The Exact relations between the coupling constants and the masses of particles for the integrable perturbed conformal field theories}",
    reportNumber = "PAR-LPTHE-93-47",
    doi = "10.1016/0370-2693(94)00078-6",
    journal = "Phys. Lett. B",
    volume = "324",
    pages = "45--51",
    year = "1994"
}

@article{Fateev:1997yg,
    author = "Fateev, Vladimir and Lukyanov, Sergei L. and Zamolodchikov, Alexander B. and Zamolodchikov, Alexei B.",
    title = "{Expectation values of local fields in Bullough-Dodd model and integrable perturbed conformal field theories}",
    eprint = "hep-th/9709034",
    archivePrefix = "arXiv",
    reportNumber = "CLNS-97-1510, RU-97-74",
    doi = "10.1016/S0550-3213(98)00002-9",
    journal = "Nucl. Phys. B",
    volume = "516",
    pages = "652--674",
    year = "1998"
}

@article{Hellerman:2006zs,
    author = "Hellerman, Simeon and Henriques, Andre and Pantev, Tony and Sharpe, Eric and Ando, Matt",
    title = "{Cluster decomposition, T-duality, and gerby CFT's}",
    eprint = "hep-th/0606034",
    archivePrefix = "arXiv",
    doi = "10.4310/ATMP.2007.v11.n5.a2",
    journal = "Adv. Theor. Math. Phys.",
    volume = "11",
    number = "5",
    pages = "751--818",
    year = "2007"
}

@article{Seiberg:2010qd,
    author = "Seiberg, Nathan",
    title = "{Modifying the Sum Over Topological Sectors and Constraints on Supergravity}",
    eprint = "1005.0002",
    archivePrefix = "arXiv",
    primaryClass = "hep-th",
    doi = "10.1007/JHEP07(2010)070",
    journal = "JHEP",
    volume = "07",
    pages = "070",
    year = "2010"
}

@unpublished{Chen:ExceptionalMassiveQuivers,
    author = {Chen, Ting-Kai and Chen, Jin},
    title = {{Exceptional Massive Quivers}},
    note = {In preparation}
}
\end{document}